 \documentclass[aps,pra,letterpaper,superscriptaddress,twocolumn,showpacs,floatfix,10pt]{revtex4-1}
\date{\today}

\usepackage{amsfonts}
\usepackage{amsmath}
\usepackage{amssymb}
\usepackage{graphicx}
\usepackage{epstopdf}
\usepackage{epsfig}
\usepackage{color}
\usepackage{mathtools}
\usepackage{hyperref}
\usepackage{tabularx}

\def\bra#1{\langle#1\vert}
\def\ket#1{\vert#1\rangle}

\begin{document}

\title{Emergence of the XY-like phase in the deformed spin-3/2 AKLT systems 
}

\author{Ching-Yu Huang}
 \affiliation{C. N. Yang Institute for Theoretical Physics and Department of Physics and Astronomy, State University of New York at Stony Brook, NY 11794-3840, United States}  
 
\author{Maximilian Anton Wagner}
 \affiliation{C. N. Yang Institute for Theoretical Physics and Department of Physics and Astronomy, State University of New York at Stony Brook, NY 11794-3840, United States} 
 
\author{Tzu-Chieh Wei}
 \affiliation{C. N. Yang Institute for Theoretical Physics and Department of Physics and Astronomy, State University of New York at Stony Brook, NY 11794-3840, United States}  

\vfill
\begin{abstract}
Affleck, Kennedy, Lieb and Taski (AKLT) constructed an exemplary spin-3/2 valence-bond solid (VBS) state on the hexagonal lattice, which is the ground state of an isotropic quantum antiferromagnet and possesses no spontaneous magnetization but finite correlation length. This is distinct from the N\'eel ordered state of the spin-3/2 Heisenberg model on the same lattice. Niggemann, Kl\"umper and Zittartz then generalized the AKLT Hamiltonian to one family  invariant under spin rotation about the z-axis. The ground states of this family can be parameterized by a single parameter that deforms the AKLT state, and  this system exhibits a quantum phase transition between the VBS and N\'eel phases, as the parameter increases from the AKLT point to large anisotropy.  We investigate the opposite regime when the parameter decreases from the AKLT point and find that there appears to be a Berezinskii-Kosterlitz-Thouless-like transition from the VBS phase to an  XY phase. Such a transition also occurs in the deformation of other types of AKLT states with triplet-bond constructions on the same lattice. However, we do not find such an XY-like phase in the deformed AKLT models on other trivalent lattices, such as square-octagon, cross and star lattices. On the star lattice, the deformed family of AKLT states remain in the same phase as the isotropic AKLT state throughout the whole region of the parameter. However, for two triplet-bond generalizations, the triplet VBS phase is sandwiched between  two ferromagnetic phases (for large and small deformation parameters, respectively), which are characterized by spontaneous magnetizations along different axes. Along the way, we also discuss how various deformed AKLT states can be used for the purpose of universal quantum computation.

\end{abstract}

\maketitle

\section{ introduction}

Quantum magnetism~\cite{auerbach2012interacting}, since the invention of Heisenberg model  and later Bethe's solution in 1931~\cite{Bethe1931}, remains a very  active research field with  various fascinating phenomena emerging, such as quantum phase transitions~\cite{sachdev2001quantum}, spin liquids~\cite{Balents2010}, and topological order~\cite{Wen_TO1990,Kitaev20032,Yan2011,Stefan_SL2012}, etc.  
The Heisenberg model has been studied in higher dimensions and exhibits N\'eel order for various spin magnitudes on various bi-partite lattices. The spin-3/2 Heisenberg model on the hexagonal lattice is such an example. In contrast, Affleck, Kennedy, Lieb and Taski (AKLT) constructed an exemplary spin-3/2 valence-bond solid (VBS) state on the hexagon lattice, which is the ground state of a quantum antiferromagnet whose Hamiltonian is also invariant under spin rotation, and possesses a finite correlation length but no spontaneous magnetization~\cite{AKLT_1988}. Such a VBS phase is an generalization of the 1D Haldane phase to two dimensions, and is also of interest given recent development in symmetry-protected topological phases~\cite{Pollmann2012,ChenSPT2013,Chen1604}. 
Additionally, from the viewpoint of quantum computation, the spin-3/2 AKLT state on the hexagonal lattice~\cite{WeiQCR2011,Miyake20111656}, as well as other trivalent lattices~\cite{2Dbeyond_wei_2013},  has been shown to provide the necessary entanglement for implementing universal measurement-based quantum computation~\cite{Raussendeorf2001}.

Niggemann, Kl\"umper and Zittartz (NKZ) generalized the AKLT Hamiltonian to one family that is invariant under spin rotation about the z-axis~\cite{Hexagon_Niggemann}. The ground states of this family can be parameterized by a single parameter (denoted by $a$ below) that deforms the AKLT state.  They found that the system exhibits a quantum phase transition between the VBS and N\'eel phases, as the parameter $a$ increases from the isotropic AKLT point ($a=\sqrt{3}$) to large anisotropy  $a$. The transition can be obtained accurately by mapping to a classical eight-vertex model  and agrees with their Monte Carlo study, yielding a critical point at $a_{c2}\approx 2.54$. Incidentally, a recent work by Darmawan, Brennen and Bartlett showed that, as regards to quantum computation, in addition to the utility at the AKLT point~\cite{WeiQCR2011,Miyake20111656}, this family of deformed AKLT states can be used as a universal resource for $a\ge1$ in the VBS phase up to the same VBS-N\'eel transition~\cite{Darmawan_2012}. 

For small $a$ parameter ($a<1$), the deformed family is less explored, and the classical model by NKZ is no longer valid, so we are motivated to examine this regime further. 
When the parameter $a$ decreases from the AKLT point and we find, via the numerical tensor-network methods, that there appears to be a Berezinskii-Kosterliz-Thouless  transition at $a_{c1}\approx 0.42$ from the VBS phase to an quantum XY-like phase. Such a transition also occurs in the deformation of other types of AKLT states with triplet-bond constructions (more details below). In contrast, on other trivalent lattices, such as square-octagon, cross and star lattices, we do not find such an XY phase in the deformed spin-3/2 AKLT models. On the star lattices, the deformed family of AKLT states (with the singlet-bond construction) remains in the same phase as the isotropic AKLT state throughout the whole region of the parameter considered. Moreover, for two triplet-bond generalizations, the triplet VBS phase is sandwiched between  two ferromagnetic phases (for large and small parameters, respectively), which are characterized by spontaneous magnetizations along different axes. The deformed AKLT states on various trivalent lattices, therefore, provide a rich variety of phases.

For the various deformed AKLT families, we also discuss whether they can be used for universal quantum computation. The situation at the exact AKLT point was previously studied in Refs.~\cite{WeiQCR2011,Miyake20111656} on the hexagonal lattice and  in Ref.~\cite{2Dbeyond_wei_2013} on other trivalent lattices. In particular, we find that the resourcefulness of the deformed AKLT families, for all four types of bond states, persists (from $a=1$) up to the transition of the VBS to the ordered phase on the square-octagon and cross lattices,  extending the  results of Darmawan, Brennen and Bartlett on the hexagonal lattice~\cite{Darmawan_2012}. The loss of the capability for universal quantum computation for these deformed AKLT states is due to the growth of the size of an effective qubit to a macroscopic size, and this is consistent with the percolation of ferromagnetic or antiferromagnetic domains at the transition. However,
on the star lattice, regardless of the type of bonds, the deformed AKLT states are found not to have the capability for universal quantum computation.

The paper is organized as follows. 
In Sec.~\ref{sec:States},  we describe the spin-3/2 AKLT states with various bond generalization and local deformation. In Sec.~\ref{sec:GSTNS} we discuss how to express these various states in terms of  tensor network or tensor product states~\cite{PEPS_2004}.  
We also review the method of the tensor renormalization group (TRG)~\cite{Levin_TRG2007}, which is used to find the overlap of the wave functions in a polynomially efficient way. 
In Sec.~\ref{sec:AKLThexagon},  we first briefly  review  a mapping from the deformed AKLT state, $| \Psi(a,\omega) \rangle $,  to the 2D classical eight-vertex model, from which the critical point between the disordered phase and an ordered phase can be obtained analytically. 
We  then discuss how the deformation can be applied to the AKLT model via the tensor product states and present the phase diagram of the deformed AKLT state on the hexagon lattice by evaluating the order parameter, a scale invariant quantity (which we call the Chen-Gu-Wen X-ratio) and the correlation function. 
The transition point from valence bond state to ordered phases agrees with analytical results via mapping to classical models.
In Sec.~\ref{sec:AKLTOther}, we focus on other trivalent lattices, such square-octagon, cross and star lattices and study the respective phase diagrams. 
In Sec.~\ref{sec:MBQC}, we discuss how various deformed AKLT states can be used for universal quantum computation, in particular on square-octagon and cross lattices.
We conclude in Sec.~\ref{sec:conclude} with a summary and some discussion.
In Appendix~\ref{App:classicalvertex}, we briefly review the classical vertex models~\cite{mccoy2011advanced,baxter2013exactly}. 
The exact solution of free fermion model is given in Appendix~\ref{App:freefermion}. 
In Appendix~\ref{App:Isingasunionjack}, we review the classical Ising model on the union jack and checkerboard lattices, and use the property to explain to critical point from disordered phase to ordered phase on the square-octagon lattice as well as the behavior of the spontaneous magnetization.

\section{AKLT states, their deformation and generalization with triplet bonds}\label{sec:States}

\noindent {\bf The AKLT state}. 
The original spin-3/2 AKLT state is constructed as follows~\cite{AKLT_1988}. 
One splits a spin-3/2 particle into the three virtual spin-1/2 degrees of freedom and each virtual spin-1/2 particle forms, with its partner virtual spin-1/2 particle on the adjacent site, a singlet state, $|\uparrow \downarrow \rangle - | \downarrow \uparrow \rangle$.
The on-site Hilbert space of three virtual spin-1/2 particles is then mapped to that of a physical spin-3/2 particle by projecting onto the symmetric part to the  tensor product of three spin-1/2 spaces, as shown in Fig.~\ref{fig:three_bond} (a).
The AKLT state has a closed form in terms of tensor product state (TPS), with a local rank-4 tensor at each site, where three virtual indices represent the degrees of freedom the three virtual spin-1/2 particles, as shown in Fig.~\ref{fig:three_bond} (a).

The AKLT state on the hexagonal lattice is a valence bond solid state  and it is an exact and unique ground state of the following two-body interacting Hamiltonian~\cite{AKLT_1988},
\begin{align}
&H_\text{AKLT} =  \sum_{\langle i,j\rangle} (h_\text{AKLT})_{ij}   \notag\\
&= \sum_{\langle i,j\rangle} \Big[   (\vec{S_i}  \cdot	 \vec{S_j} )+\frac{116}{243}    (\vec{S_i}  \cdot	 \vec{S_j} )^2 +   \frac{16}{243}    (\vec{S_i}  \cdot	 \vec{S_j} )^3     +\frac{55}{108}  \Big], 
\label{AKLT_H}
\end{align} 
where the sum runs over all nearest neighbor pairs of lattice sites and $\vec{S_i}  $ is the spin-3/2 operator. From the construction of the AKLT state, it is easy to see that $\langle S^{\alpha}\rangle=0$ for $\alpha=x,y,z$.

In contrast, the Heisenberg model on the hexagonal lattice exhibits N\'eel order and hence is not in the same phase as the AKLT model. In fact, the phase diagram has been explored by allowing the second and third terms to have varying coupling constants $J_2$ and $J_3$, respectively, while preserving the rotational symmetry~\cite{HuangSPT2013},
\begin{align}
H = \sum_{\langle i,j\rangle}   \Big[  (\vec{S_i}  \cdot	 \vec{S_j} )+ J_2  (\vec{S_i}  \cdot	 \vec{S_j} )^2 +   J_3   (\vec{S_i}  \cdot	 \vec{S_j} )^3  \Big].
\label{AKLT_Iso}
\end{align} 
 A phase transition between the disordered VBS phase and N\'eel phase can occur for sufficiently large $J_3$.

\smallskip\noindent {\bf Deformation}. 
In this paper we focus on the family of spin-3/2 AKLT ground states parameterized by a single parameter $a$ as follows~\cite{Hexagon_Niggemann}, 
\begin{align}
|\Psi (a) \rangle \propto D(a)^{ \otimes N}  |\Psi_{AKLT} \rangle, 
\end{align} 
where  $D(a)^{ \otimes N}$ is to apply operator $D(a)$ to each physical particles, where $D(a) = diag(\frac{a}{\sqrt{3}},1,1, \frac{a}{\sqrt{3}})$ in the $S^z$ basis and $N$ denote the total number of sites.
This ground state at a fixed $a$ is a unique ground state of a 5-parameter Hamiltonian constructed in Ref.~\cite{Hexagon_Niggemann}, whose detailed form is not repeated here.  But a simplified parent Hamiltonian such that $|\Psi(a)\rangle$ is the ground state can written alternatively as 
\begin{align}
H(a)  =  \sum_{\langle i,j \rangle}   \mathbb{D}_{ij}(a) (h_\text{AKLT} )_{ij} \mathbb{D}_{ij}(a),
\end{align} 
where  we have defined, for convenience, $ \mathbb{D}_{ij}(a) \equiv   \big[ D(a)^{-1}_i \otimes D(a)^{-1}_j \big]$.
For $a=\sqrt{3}$, the Hamiltonian reduces to Eq.~(\ref{AKLT_H}) and the ground state is the AKLT state.

By examining the properties of the ground states, NKZ found that there is a phase transition at $a_c\approx2.5416$ from the disordered VBS to a N\'{e}el ordered phase as $a$ increases~\cite{Hexagon_Niggemann}.
The emergence of the N\'eel order can be understood in the large deformation limit $a\to \infty$, where there are two possible configuration of neighboring sites:   (i) $| \Uparrow \rangle $  (i.e. $S^z  | \Uparrow \rangle  =\frac{3}{2} | \Uparrow \rangle $) on the first sublattice and a $| \Downarrow   \rangle $ (i.e. $S^z  | \Downarrow  \rangle = -\frac{3}{2} | \Downarrow  \rangle $) on the second one and (ii) vice versa. The resulting ground state is a superposition of two possible N\'{e}el states: $| \Psi (a\to \infty) \rangle =  |  \Uparrow  \Downarrow   \Uparrow  \Downarrow   ... \Uparrow  \rangle +  |  \Downarrow  \Uparrow   \Downarrow   \Uparrow  ...\Downarrow   \rangle $.
Even though it is of the Schr\"{o}dinger-cat form, small fluctuations will break the $\mathbb{Z}_2$ symmetry and the ground state will select spontaneously either  $ |  \Uparrow  \Downarrow   \Uparrow  \Downarrow   ... \Uparrow  \rangle$ or $ |  \Downarrow  \Uparrow   \Downarrow   \Uparrow  ...\Downarrow   \rangle  $, displaying long-range order.  Thus, a transition between VBS and N\'eel phases is expected.

In the other limit $a\to 0$, the basis states  $| \Uparrow \rangle $ and $| \Downarrow  \rangle$ are suppressed and the resulting ground state is essentially composed of spin-1/2 particles (except that $\pm 3/2$ components of $S^x$ and $S^y$ are allowed), and one expects that the $U(1)$ symmetry on the x-y plane might give rise to an XY phase.  
A major question to be addressed in the one-parameter deformed AKLT state is whether an XY phase actually emerges or whether there might be any other phase that is not previously identified by NKZ and if so where the transition is.

It turns out that it is particularly helpful to use the tensor network representation of the state and see the phase transition by tuning the parameter. 
In particular we will employ  a  renormalization scheme of the tensors, coined the tensor renormalization group (TRG), much in the spirit of the spin blocking scheme used in the coarse graining of the Kadanoff real-space renormalization group (RG) approach.  
For example, we confirm the phase transition from valence bond solid state to N\'{e}el phase by calculating directly the magnetization.
For the small $a$ region in the deformed AKLT model on the hexagonal lattice,  we  find that as $a$ decreases, the VBS phase makes a transition to the XY phase  via a Berezinskii-Kosterlitz-Thouless (BKT) transition~\cite{Berezinskii1970,KT_1973,2DXYmodel_1977} first revealed in a classical XY model, and later studied in quantum spin-1/2 models, see e.g.~\cite{Ding_1990,Ding_1992,1995_Cuccoli}. This conclusion is reached by characterizing the scaling behavior of the correlation length.

\smallskip
\noindent {\bf Generalization with triplet bonds\/}.
The original AKLT states, including the one-dimensional (1D) spin-1 AKLT state  and the two-dimensional (2D) spin-3/2 AKLT on  hexagon lattice,  were constructed using the virtual qubit pairs forming a singlet bond, i.e. $|\psi^{-}\rangle\equiv  |01\rangle - |10 \rangle$,
where  $\sigma ^z |0 \rangle =  |0 \rangle$ and
$\sigma ^z |1 \rangle = - |1 \rangle$.
However,  there are three other  orthogonal maximally entangled quantum states of two spins in the form of triplet bonds, which can also be used in the general construction,
\begin{align}
|\phi^{\pm} \rangle = |00\rangle \pm |11\rangle   \notag \\
|\psi^{+} \rangle =  |01\rangle +|10 \rangle.
\end{align}
It is worth mentioning that the 1D AKLT states built up from different bond states $| \omega \rangle,  \quad   \omega  \in \{ \phi^{\pm}, \psi^{\pm} \}$ belong to different symmetry protected topological ordered phases with different 1D representations. 

A natural question is what difference will be for these 2D deformed AKLT states, labeled by $| \Psi(a,\omega) \rangle $, where $a$ is the same deformation parameter used before.
In general, the valence bond state construction  can be used on any lattices, and with regards to spin-3/2 cases, on lattices with coordination number three, such as 
the hexagonal, square-octagon, cross and star lattices, all of which will be considered below.

As a brief summary, we find a phase transition at $a_c =2.5425 $ on the hexagonal lattice and $a_c = 2.6547$ on the square-octagon lattice and $a_c=2.7280$ on the cross lattice, separating the disordered VBS phase from an ordered phase (antiferromagnetic phase for $ \omega = \psi^{\pm}  $  and ferromagnetic phase for  $ \omega = \phi^{\pm}  $).
The XY phase only appears in the hexagonal lattice and the BKT-like transition is located at $a_{c}\approx 0.42$. 
In particular, the star lattice is geometrically frustrated  and the frustration effect has important consequences on the ground-state properties, especially bonds with $\omega=\psi^\pm$.  Namely, the deformed AKLT states constructed with these bond states remain disordered and in the same phase for all parameters $a\ge 0$.
In contrast, for the deformed AKLT states, $| \Psi(a,\omega= \phi^{\pm} )  \rangle $,  on the star lattice with ferromagnetic bond state $ |\phi^{\pm} \rangle $, we find a phase transition from the disordered phase to an ordered ferromagnetic phase at $a_c= 3.0243$, as well as another transition at $a_c=0.58$ to another ferromagnetic phase.

\section{ The ground states in terms of tensor network states} \label{sec:GSTNS}
In this section, we discuss the tensor-network form of the various ground states, parameterized by parameter $a$.
To begin with, we consider the original AKLT state under deformation $D(a)$,
\begin{align}
|\Psi \rangle \propto D(a)^{ \otimes N}  |\psi_\text{AKLT} \rangle, 
\end{align} 
where $D(a) = diag(\frac{a}{\sqrt{3}},1,1, \frac{a}{\sqrt{3}})$ in the $S^z$ basis.
Locally the physical degree of freedom can be obtained by applying a  map $D'(a) P$ to the three spin-1/2 particles at each site, where $P$ is the projector that map the virtual space to the physical one, 
\begin{align}
P=& |\Uparrow \rangle  \langle 000| + | \uparrow \rangle \big( \langle 001|  +  \langle 010|  + \langle 100| \big) +  \notag \\
    &|\downarrow \rangle \big(   \langle 011|  +  \langle 110|  + \langle 101| \big)  +  |\Downarrow  \rangle  \langle 111| , 
\end{align} 
and the $D'(a)$ is a local deformation as 
\begin{align}
D'(a)= a & \big( |\Uparrow  \rangle  \langle \Uparrow    |   +  |\Downarrow \rangle  \langle \Downarrow | \big) +\notag \\
             &         |\uparrow \rangle  \langle \uparrow |   +  | \downarrow \rangle  \langle \downarrow|. 
\end{align} 
Hence, the AKLT wave function with deformation can be written as 
\begin{align}
|\Psi \rangle = \bigotimes_{v\in V} \Big( D'(a) P  \Big)_{v} \bigotimes_{l\in L} | \psi^- \rangle_{l}, 
\end{align} 
where the bond states are placed on every link $l$ of lattice and the projectors $D'(a) P$ map the virtual space at each vertex $v$ to physical space. 
In general, we can place any bond state $|\omega \rangle$ as shown in Table~\ref{table:bond_state} on each edge of lattice.   
\begin{table}[h!]
\centering
\begin{tabularx} { 0.45\textwidth}{ c c c c } 
\hline  \hline 
 $\omega $  &  z base & x base & y base  \\  \hline 
$\phi^+ \quad$ & $|00\rangle+|11\rangle \quad$ &  $|0_x0_x\rangle+|1_x1_x\rangle \quad$  &  $ |0_y1_y\rangle+|1_y0_y\rangle \, $\\  
$\phi^-  \quad$ & $|00\rangle-|11\rangle  \quad$ &   $|0_x1_x\rangle+|1_x0_x\rangle \quad$  &  $ |0_y0_y\rangle+|1_y1_y\rangle \,$\\  
$\psi^+ \quad$ &$|01\rangle+|10\rangle  \quad$ &  $|0_x0_x\rangle-|1_x1_x\rangle \quad$   & $ |0_y0_y\rangle-|1_y1_y\rangle \,$ \\  
$\psi^- \quad$ & $|01\rangle-|10\rangle  \quad$ &   $|0_x1_x\rangle-|1_x0_x\rangle \quad$  &  $ |0_y1_y\rangle-|1_y0_y\rangle \,$ \\  \hline  \hline 
\end{tabularx}
 \caption{ The representations of  bond states with different basis.  }
 \label{table:bond_state}
\end{table}
For tensor multiplication, it is natural  that two neighboring virtual spins of two different physical sites are in the same  state, i.e $| \phi^+ \rangle= |00\rangle+|11\rangle$. 
In particular, these four  two-qubit orthogonal basis can be transformed each other by local operator. 
For example, 
\begin{align}
& | \phi^- \rangle = |00\rangle-|11\rangle =  I \otimes \sigma^z | \phi^+ \rangle  \notag \\
& | \psi^+ \rangle = |01\rangle+|10\rangle =  I \otimes \sigma^x | \phi^+ \rangle  \notag \\
& | \psi^- \rangle = |01\rangle-|10\rangle =  I \otimes i \sigma^y | \phi^+ \rangle,
\end{align} 
where $I$ denotes $2\times 2$ identity matrix, and $\sigma^i, i \in \{ x,y,z\}$ are Pauli matrices.
The hexagonal lattice is bipartitioned into $A$ and $B$ sublattice sites.  

\begin{figure}[ht]
\includegraphics[width=0.5\textwidth]{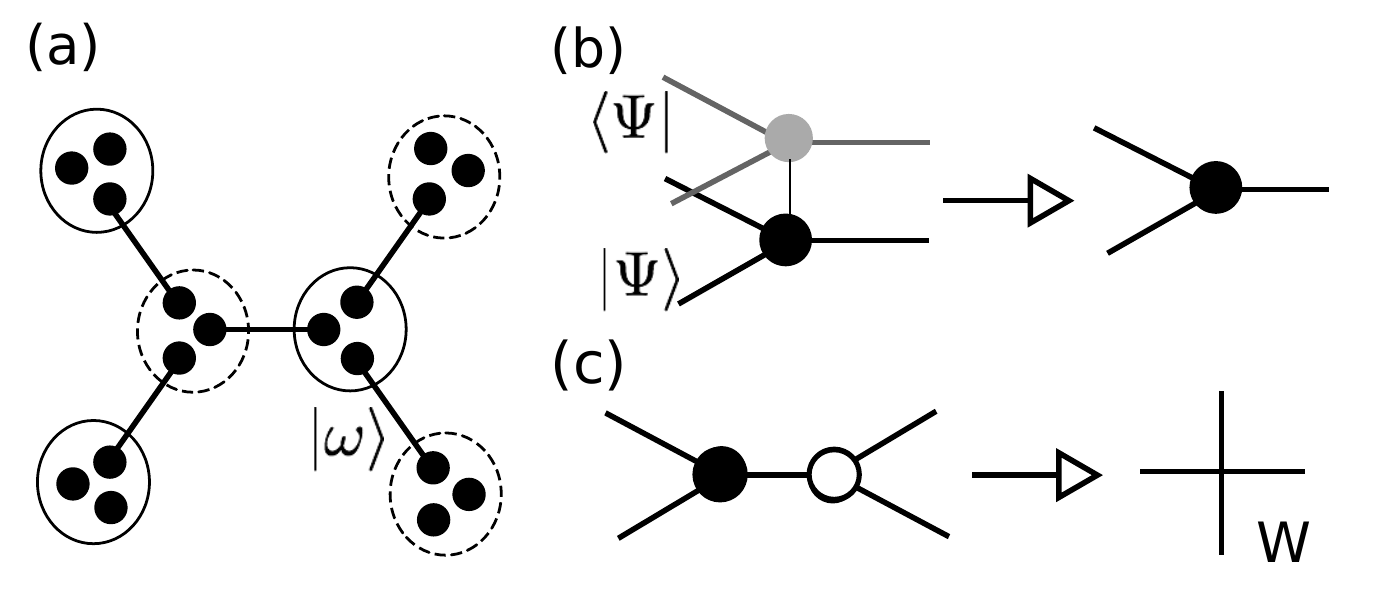}
\caption{ (a)The Valence bond picture of the spin-3/2 AKLT state on the hexagonal lattice: The circles represent the map form virtual to physical spin space, the dots represent the virtual spins, and the lines represent the maximally entangled bond. (b) Contract the physical index of a tensor on a lattice site a with the physical index of its complex conjugated tensor and remove the non-diagonal  term that one double bond shows an anti-parallel  pair to form a single bond tensor.   (c) Horizontal contraction of two lattice sites belonging to the sublattices $A$ and $B$ respectively.}
  \label{fig:three_bond}
\end{figure}

The deformed AKLT wave function with bond state $|\omega \rangle$ can be represented by 
\begin{align}
|\Psi (a,\omega) \rangle = &  \bigotimes_{v\in V} \Big( D'(a) P  \Big)_{v} \bigotimes_{l\in L}  |\omega \rangle_{l} =  \notag \\ 
|\Psi (a,k) \rangle = & \bigotimes_{v\in V}\Big( D'(a) P  \Big)_{v} \bigotimes_{l\in L}  \sigma^k |\phi^+ \rangle_{l} \notag \\ 
                      =& \!  \bigotimes_{v\in V_A} \!  \Big( \! D'(a) P  \Big)_{v} \!    \bigotimes_{v\in V_B} \!  \Big(  \!  D'(a) P (\sigma^k)^{\otimes3} \!   \Big)_{v} \!   \bigotimes_{l\in L} \!  |\phi^+ \rangle_{l},
\end{align} 
where $ \sigma^k,   k\in \{0,x,y,z\}$ are Pauli matrices and $ \sigma^0=  \mathbb{I}$. Graphically this construction is shown in Fig.~\ref{fig:three_bond}(a).
The deformed AKLT state can thus be represented as a tensor-network representation on the hexagonal lattice with bond state $|\omega \rangle$ (of bond dimension $\chi=2$) given by 
\begin{align}
|\Psi  (a, \omega)  \rangle \!  =\!  \sum_{\! s_1,s_2,....s_N \! }  \!  tTr ( A^{\! s_1}B^{ s_2}...A^{s_{\! N\! -\! 1\! }}B^{ s_N})  | s_1s_1...s_n\rangle,
\end{align} 
where $A_{\alpha_i \beta_i \gamma_i}^{s_i}$ and $B_{\alpha_j \beta_j \gamma_j}^{s_j}$  are rank-4 tensors with physical index $s_i$ and $s_j$, respectively, and internal indices $\alpha_i \beta_i \gamma_i$ and $\alpha_j \beta_j \gamma_j$, respectively. The $s_i \in \{ \Uparrow,\uparrow, \downarrow, \Downarrow   \}$ corresponds to the basis of $S=3/2$ states.
The ``tTr'' denotes tensor contraction of all the connected inner indices according to the underlying lattice structure. 
For example, the nonzero elements of the tensors with deformation $a$ and bond state $|\phi^- \rangle$ are 
\begin{align}
&A^{\Uparrow }_{000} = a,  \quad   A^{\uparrow }_{100}= A^{\uparrow }_{010} =A^{\uparrow }_{001} =1 \notag \\
&A^{\Downarrow }_{111} = a, \quad  A^{\downarrow }_{101}= A^{\downarrow }_{110} =A^{\downarrow }_{011} =1,
\end{align} 
\begin{align}
&B^{\Uparrow }_{111} = a ,  \quad  B^{\uparrow }_{101}= B^{\uparrow }_{110} =B^{\uparrow }_{011} =-1 \notag \\
& B^{\Downarrow }_{000} = -a,  \quad B^{\downarrow }_{001}= B^{\downarrow }_{010} =B^{\downarrow }_{001} =1.
\end{align} 
It is straightforward to write down the tensors of the AKLT states with other bond states $\omega$.

\begin{figure}[ht]
\includegraphics[width=0.5\textwidth]{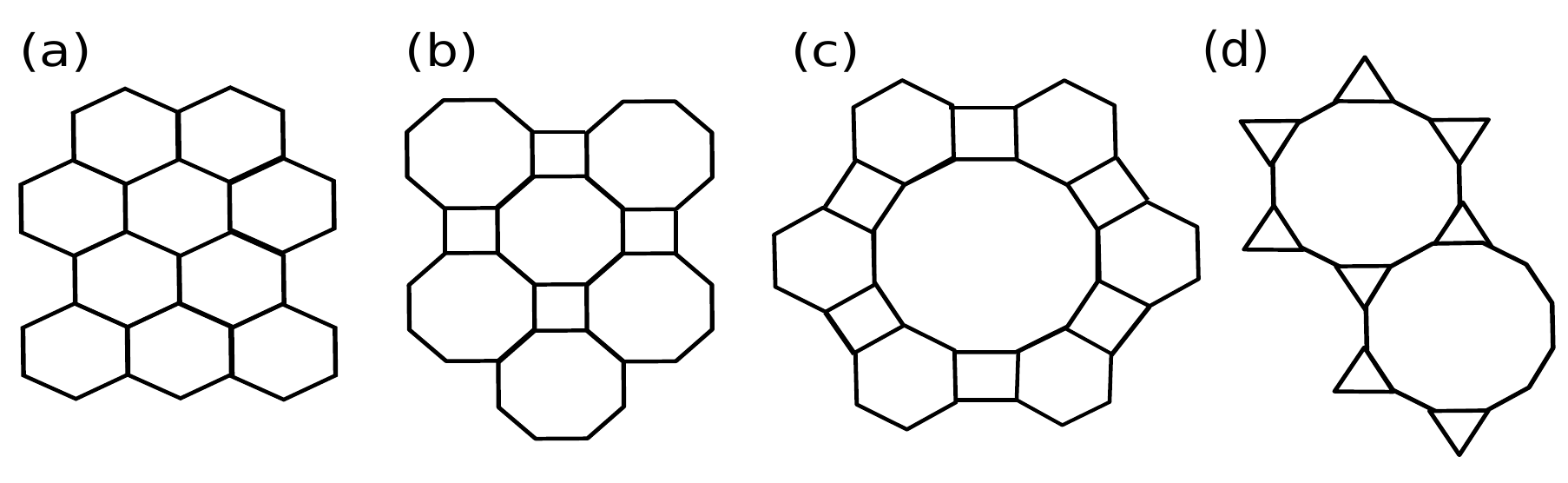}
\caption{ The lattice structure (a) hexagon, (b) square octagon, (c) cross, (d) star }
  \label{fig:Lattice}
\end{figure}

\subsection{Other lattices }
Below we discuss the deformed AKLT states with various bond states and on various other trivalent lattices.

\begin{figure}[ht]
 \includegraphics[width=0.5\textwidth] {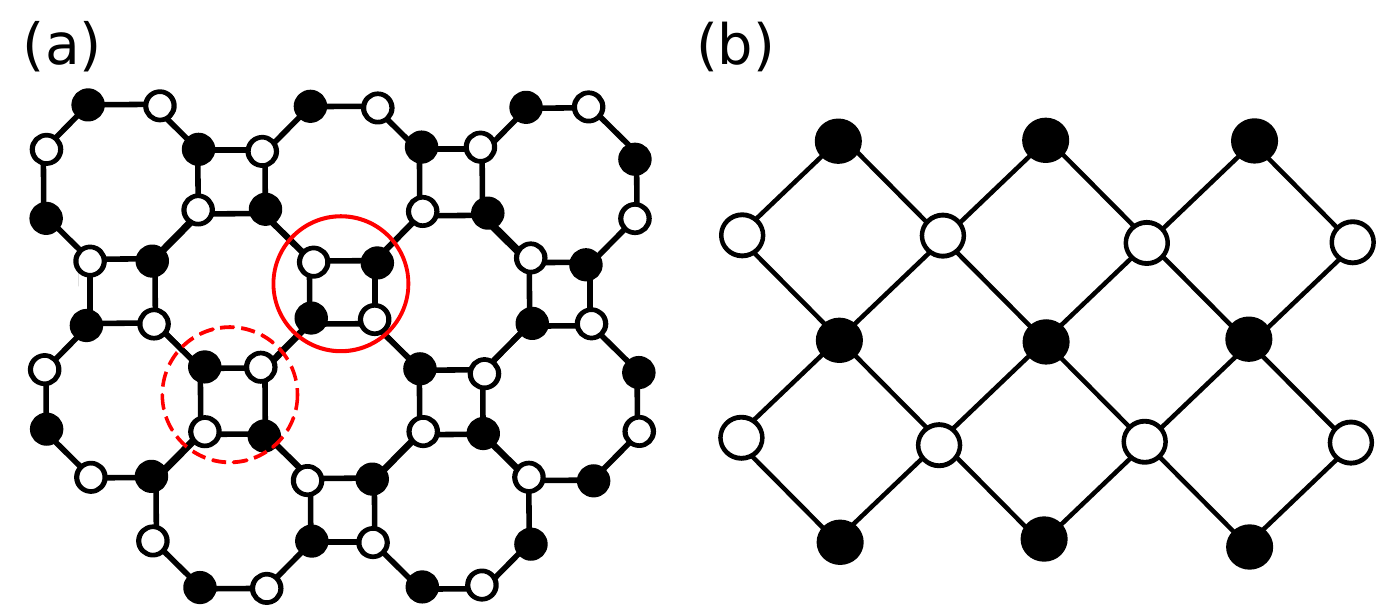}
\caption{(a) The tensor network representation for square octagon lattice. (b)   Combining the four tensors in a square to form a new tensor. The square octagon lattice thus deforms into a square lattice with $T_a$ (white) and $T_b$ (black) tensors on the respective sublattices.   }
  \label{fig:SO_lattice}
\end{figure}

\smallskip\noindent {\bf Square-octagon lattice}. 
The deformed  AKLT wave function on the square-octagon lattice also can be represented by the tensor network state with $A$ (black circle) and $B$ (white circle) sublattice in Fig.~\ref{fig:SO_lattice}. 
It  can be built by maximally entangled state  $| \omega \rangle$ that can transformed by applying Pauli matrices $(\sigma^x, \sigma^y, \sigma^z)$ to $|\phi^+ \rangle =|00 \rangle+|11 \rangle $, as follows: 
\begin{align}
|\Psi (a, \omega) \rangle =& \bigotimes_{v\in V_A}  \Big( (\sigma^k)  D'(a) P  \Big)_{v}    \notag \\
&\bigotimes_{v\in V_B}  \Big( (\sigma^k)^{\otimes2}  D'(a) P  \Big)_{v}  \bigotimes_{l\in L} |\phi^+ \rangle_{l},
\label{eq:AKLT_TPS}
\end{align} 
where $k \in \{ 0,x,y,z \}$ and $\sigma^0 = 1$. The local tensors can then be written down easily.

We resort to the TRG method to calculate the physical quantities and describe the steps, for example, on a square-octagon lattice starting with the original local double tensors $\mathbb{T}_A$ (or $\mathbb{T}_B$), which  can be formed by merging two layers tensors $A$ (or $B$) and $A^*$ (or $B^*$) with the physical indices contracted.
From this we  build a new rank-3 tensor as shown in Fig.~\ref{fig:SO_lattice}(b) by combining the four tensors around a square to form a new tensor   $\mathbb{T}'_A$  and $\mathbb{T}'_B$.
The original square-octagon lattice tensor network is now mapped to a square-lattice model, for which one can then  apply the usual TRG method.

\begin{figure}[ht]
\includegraphics[width=0.5\textwidth]{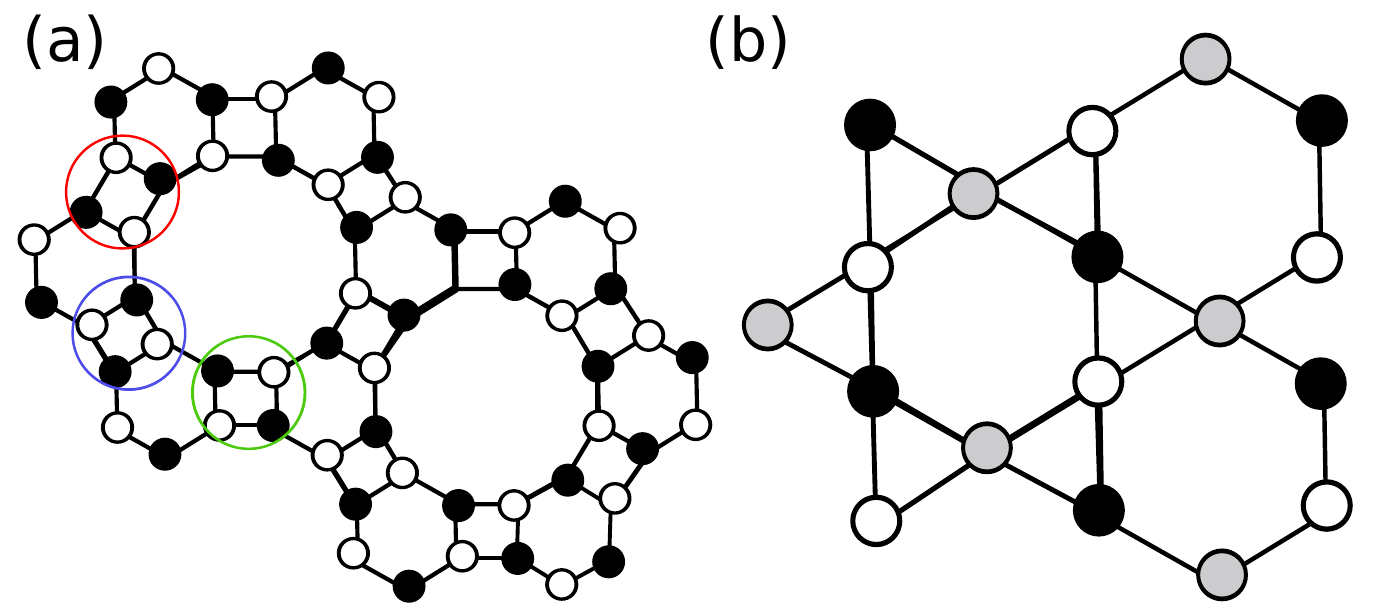}
\caption{ (a) The tensor network representation for cross lattice. (b)   Combining the four tensors in a square to form a new tensor. The cross lattice  deforms into a kagom\'e lattice with three tensors on the respective sublattices.}
  \label{fig:cross_lattice}
\end{figure}

\smallskip\noindent {\bf Cross lattice}.
The deformed AKLT family wave function on the cross lattice also can be represented by a tensor network state with $A$ (black circle) and $B$ (white circle) on respective sublattices in Fig.~\ref{fig:cross_lattice}. The wave function is given formally by Eq.~(\ref{eq:AKLT_TPS}), except on the cross lattice.  The local tensors can then be written down easily. To compute physical quantities with TRG, it is useful to merge the local tensors into to a tensor network on the kagome lattice. 

\begin{figure}[ht]
 \includegraphics[width=0.5\textwidth]{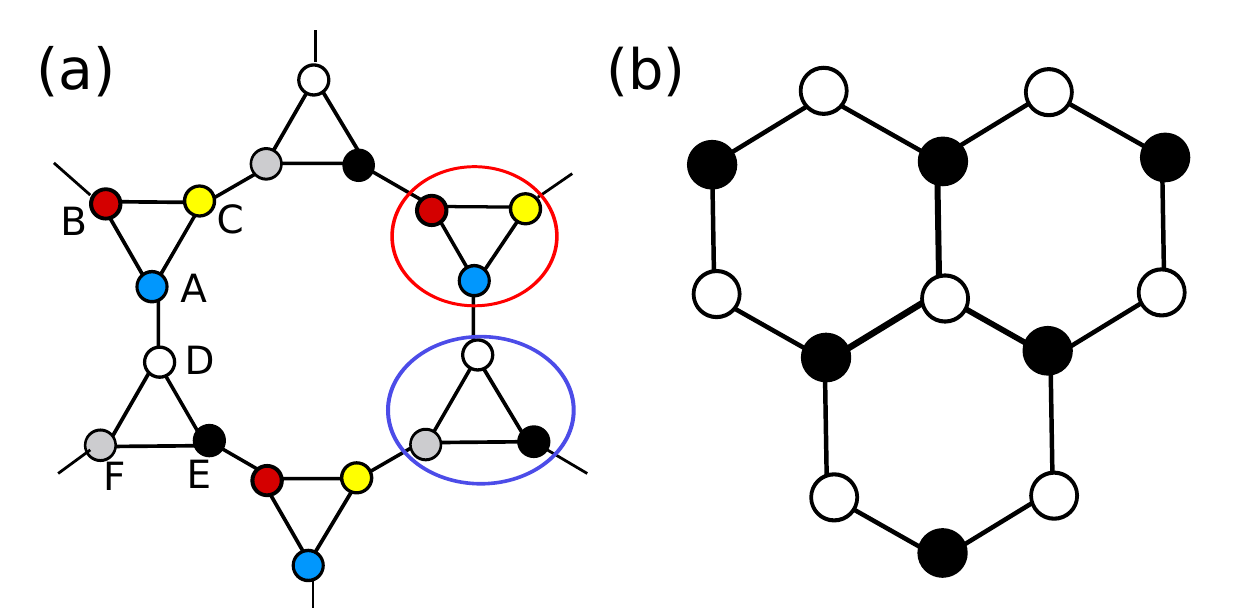}
\caption{ (a) The tensor network representation for star lattice with six sublattices. (b)   Combining the three tensors in a triangular to form a new tensor. The star lattice  deforms into a hexagon lattice with two tensors on the respective sublattices.}
  \label{fig:star_lattice}
\end{figure}

\smallskip\noindent {\bf Star lattice}.
To construct the ground state, we use the tensor network and valence bond solid construction once again. 
The wave function with six sublattice as shown in Fig.~\ref{fig:star_lattice} (a) can be given by 
\begin{align}
|\Psi (a, \omega) =& \bigotimes_{v\in V_A}  \Big( (\sigma^k)^{\otimes3}   D'(a) P  \Big)_{v}    \bigotimes_{v\in V_B}  \Big( (\sigma^k)^{\otimes2} D'(a) P  \Big)_{v}  \notag\\
& \bigotimes_{v\in V_C}  \Big(   D'(a) P  \Big)_{v}    \bigotimes_{v\in V_D}  \Big(  D'(a) P  \Big)_{v}  \notag\\ 
& \bigotimes_{v\in V_E}  \Big( \sigma^k  D'(a) P  \Big)_{v}    \bigotimes_{v\in V_F}  \Big( (\sigma^k)^{\otimes3} D'(a) P  \Big)_{v} \notag\\
& \bigotimes_{l\in L}|\phi^+ \rangle_{l}.
\end{align} 
The local tensors can then be written down easily.
To compute the physical quantities by using TRG, we can merge nearby local tensors to form a tensor network on a hexagonal lattice.

\subsection{Numerical methods}
In a two-dimensional system, it is difficult to calculate the tensor trace (tTr) exactly since all indices on the connected links in the network need to be summed over. 
Several approximation schemes have been proposed as solutions in this context such as the iPEPS algorithm~\cite{iPEPS_2009}, the corner transfer matrix renormalization group (CTMRG) method~\cite{CTMRG1997}, and the tensor renormalization approach~\cite{Levin_TRG2007,Xiang_TRG2008} which tackle this problem essentially by using trucation to scale the computational effort down to the polynomial level.
In this paper, we use the tensor renormalization group (TRG) approach which is akin to the real space renormalization in the way that, at each step, the RG is structured by merging sites (by contracting respective tensors) and truncating the bond dimension according to the relevance of the eigenvalues in the Schmidt decomposition of the old tensors. 
Each step of the TRG approach reduces the number of sites by half. 
Eventually, the entire network collapses to only a few sites and the double tensor trace appearing in the expectation value can be calculated easily.


\section{The deformed AKLT family  on the hexagonal lattice} \label{sec:AKLThexagon}

\subsection{Mapping to a classical vertex model}

In this section, we study the phase transitions of a family of quantum spin-3/2 models on the  trivalent lattice such as  a hexagonal lattice, a square octagon and a cross lattice analytically and numerically.
In the work of NKZ~\cite{Hexagon_Niggemann}, they found the phase transition of between the VBS phase to a N\'eel phase in the one-parameter family of the deformed AKLT state on the hexagonal lattice. 
They showed that the ground state properties can be obtained from the exact solution of a corresponding classical eight vertex model, and the resulting transition point agreed with their Monte Carlo simulations. 
We briefly review their classical mapping. First, they look at the norm $\langle \Psi | \Psi \rangle$ of the ground state with bond state $| \psi^-\rangle$ (and $|\psi^+\rangle$ as well). 
In graphical language, that means they place a copy of the lattice on top of the first one and connect the vertices which lay exactly on top of each other as shown in Fig.~\ref{fig:three_bond}(b).  
In tensor network language, after contracting physical index, the tensor is called double tensor that means each virtual bond contain two indices such as $\alpha, \alpha' \in \{0,1\}$. 
The double tensor with at least one  unequal pair $(0,1)$ are called  off-diagonal vertices and vice versa.

The classical vertex model consists of the 16 vertex weight $w(i_1,i_2,i_3,i_4)$ in which the $i_k$ can take value $0$ for left/down arrow and $1$ for  right/left arrow  (see Appendix \ref{App:classicalvertex}). 
The first step is to treat the two-valued state variable of double tensor as single bonds. 
This step is we called ``diagonal approximation''. 
Interestingly,  for the hexagonal lattice, it was shown in \cite{Hexagon_Niggemann}, that the probability of finding an unequal pair $(0,1)$ for each bond vanishes exponentially as $a$ increases. 
The off-diagonal vertices become negligible in the regime of the phase transition between VBS and an ordered phase. 
Thus, the double tensor can be treated as a  the diagonal vertices, with double indices reduced to only one index on each bond. 

The second step is to map the network on the hexagonal lattice  to that on the square lattice. 
This can be done by contracting all horizontal links in Fig.~\ref{fig:three_bond}(c). 
Now the norm of ground state represented by double tensors is a 16-vertex model on the square lattice. 
This is still not a solvable model, then by applying a Hadamard transformation to all bonds, all vertices (in the sense of vertex models) which have an odd number of arrows pointing towards them have a vanishing Boltzmann weight. 
Therefore, a eight-vertex model can be obtained with the following weights of vertices,
\begin{eqnarray}
&& w_1=\frac{1}{2}( a^2+3)^2 \\
& &w_2=w_3=w_4=w_5=w_6=\frac{1}{2}( a^2-1)^2\\
& &w_7=w_8=\frac{1}{2}(a^2+1)^2-2.
\end{eqnarray}

Moreover, the Boltzmann weights of this eight-vertex model satisfy the free fermion condition (see Appendix~\ref{App:freefermion}).
For $a \geq1 $, all vertex weights are non-negative. 
By exploiting such mapping, NKZ~\cite{Hexagon_Niggemann} showed   that for $a < a_{c_2} = \sqrt{3+2 \sqrt{3}} \approx  2.5424$ the original quantum state is in a disordered phase while for $a > a_{c_2}$ the quantum state exhibits N\'eel  order. 
Both the AKLT point $a = \sqrt{3}$ and the infinite temperature point $a = 1$ lie within the disordered phase.

In addition to the results just summarized from NKZ~\cite{Hexagon_Niggemann} we also replace the singlet bond states by the three triplet states.
In tensor network language, this means that we exchange the bond generators by $\sigma^x$  or $\sigma^z$ or the identity respectively. 
It turns out that the norm square from the deformed AKLT with different bond states all get mapped to the same classical model, as the lattice is bipartite and any bond states can be related to one another by a physical local unitary transformation (which cancels in the double-layer contraction).

\subsection{Spontaneous magnetization with tensor network}
As a check and benchmark, we study the magnetization near this transition with the tensor-network method. As explained above, depending on the bond state $|\omega\rangle$, the corresponding VBS phase will make a transition to either antiferromagnetic or ferromagnetic phase, but the transition point in terms of $a$ is the same.

\begin{figure}[ht]
\includegraphics[width=0.5\textwidth] {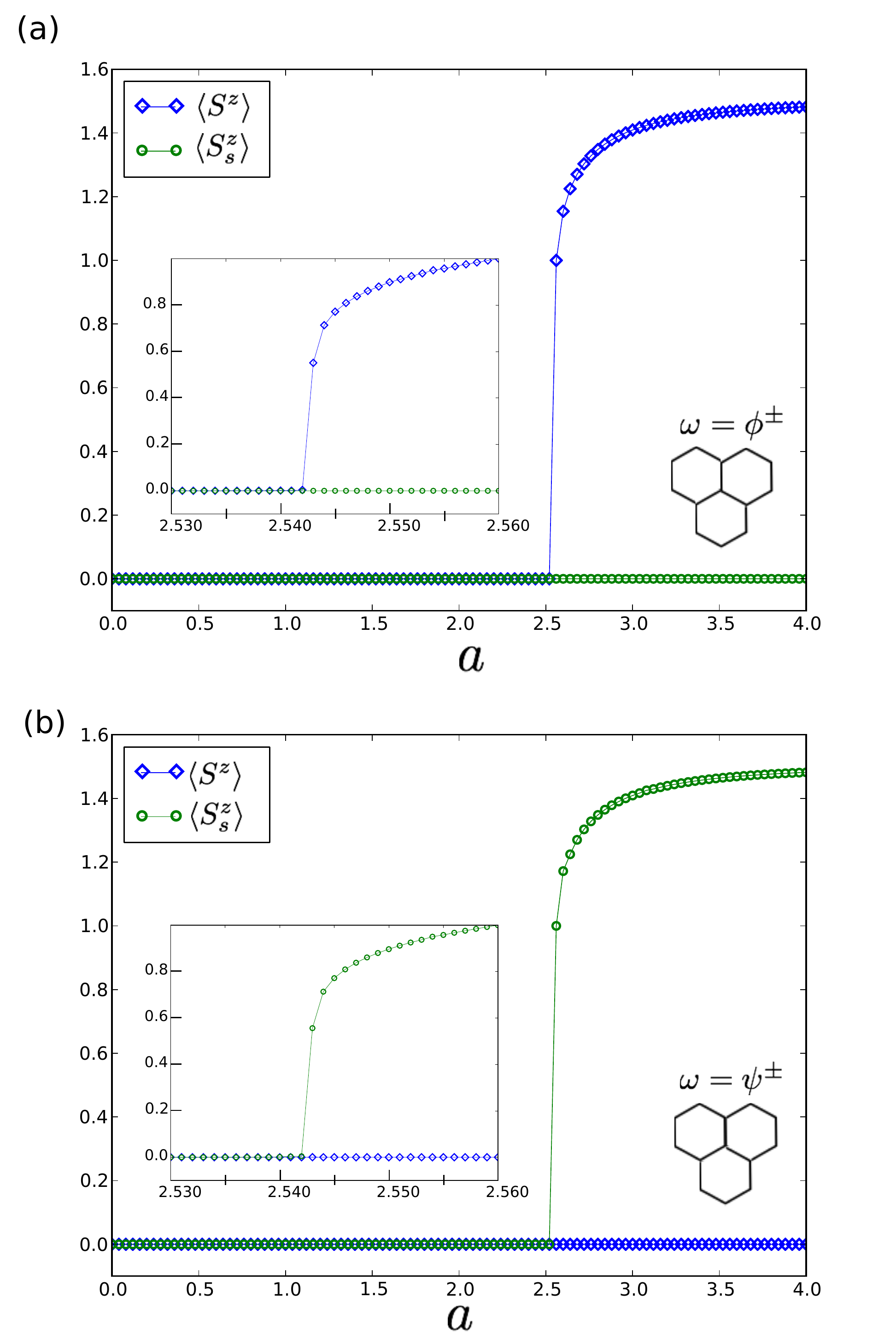}
\caption{ The magnetization  $\langle S^z \rangle$  and  staggered  magnetization  $\langle S^z_s \rangle$  as a function of parameter $a$ using TRG with bond dimension $D_c=24$ on the hexagon lattice. It indicates a transition from VBS phase to  ordered phase at $a_{c_2} = 2.5425$ with (a) $\omega =\phi^{\pm}$ and (b) $\omega =\psi^{\pm}$. However, it does not capture the phase transition from XY phase to VBS for bond state }
  \label{fig:SZ_hexagon}
\end{figure}

As shown in Fig.~\ref{fig:SZ_hexagon}(a), our numerical results verify that the second order transition between the VBS and ferromagnetic phase  with $\omega =\phi^{\pm} $ occurs at $a_{c_2}=2.5425$ and it can be characterized by a nonzero spontaneous magnetization in the FM phase. 
As for the antiferromagnetic case (with $ \omega =\psi^{\pm} $),  the system undergoes a second-order quantum phase transition from the VBS phase to the N\'{e}el phase characterized by staggered magnetization $\langle S^z_s \rangle = \frac{1}{N} \sum_{i=1}^N (-1)^i \langle S^z \rangle$  as shown in Fig.~\ref{fig:SZ_hexagon}(b). 

We remark that if the wave function  in the ferromagnetic phases is a superposition of both possible ordered states, we will have zero magnetization. 
The spin-up ferromagnetic phases will give a positive magnetization and the spin-down one will give negative magnetization. The equally weighted superposition of them gives rise the zero  magnetization.
In order to obtain the spontaneous magnetization we can apply a very tiny symmetry breaking field in practice (and check that the obtained magnetization is independent of the small breaking field). But this is usually done at the level of the Hamiltonian.  However, here we only  have the ground-state wavefunctions. We achieve the effect of symmetry breaking by applying, e.g., an operator 
\begin{align}
{\cal O}(h_z)=\left(\begin{array}{cccc}
1+\frac{3}{2}h_z & 0 & 0 & 0\\
0 & 1+ \frac{1}{2}h_z & 0 & 0\\
0 & 0 & 1-\frac{1}{2}h_z & 0\\
0 & 0 & 0 &  1+\frac{3}{2}h_z\end{array} \right),
\end{align}
to all sites of the ground-state wavefunction before evaluating the magnetization, where $h_z$ is a small numer such as $10^{-5}$. 
For staggered magnetization, we apply ${\cal O}(h_z)$ and ${\cal O}(-h_z)$ to different sublattices, respectively.
The transition from the magnetization (or staggered magnetization) for all cases is found to be at $a_{c_2}= 2.5425$, consistent with results from NKZ.
\subsection{Chen-Gu-Wen X ratio}
 
In the special limit case $a \to \infty $, the effective terms of local tensor representation in $A$ site are $A^{\Uparrow }_{000} = a $,  $A^{\Downarrow }_{111} = a $ and  in $B$ site the effective ones are $B^{\Uparrow }_{000} = a $,  $B^{\Downarrow }_{111} = a $ for bond state $|\phi^{\pm} \rangle$  and  $B^{\Uparrow }_{111} = a $,  $B^{\Downarrow }_{000} = a $ for bond state $|\psi^{\pm} \rangle$.
In this limit the wave function is a cat state, containing two dominant configurations $| \Uparrow , \Downarrow ,\Uparrow ,\Downarrow ,...    \rangle$ and $| \Downarrow , \Uparrow , \Downarrow ,\Uparrow ,...    \rangle$ for the bond state $|\psi^{\pm} \rangle$, and  $| \Uparrow , \Uparrow ,\Uparrow ,\Uparrow ,...    \rangle$ and $| \Downarrow , \Downarrow , \Downarrow ,\Downarrow ,...    \rangle$ for the bond state $|\phi^{\pm} \rangle$. 
So the transition from the disordered VBS to the ordered phase is expected and is exactly what has been found and the transition point is labeled as $a_{c_2}$.

However, for $a<1$ the diagonal approximation is not valid, and thus the classical vertex model cannot be used to describe this region. 
But from arguments presented in Sec.~\ref{sec:States}, we expect that an XY phase might emerge, as the $|\Uparrow\rangle$ and $|\Downarrow\rangle$ components are suppressed. We shall examine this small $a$ regime in detail later.

\begin{figure}[ht]
 \includegraphics[width=0.5\textwidth]{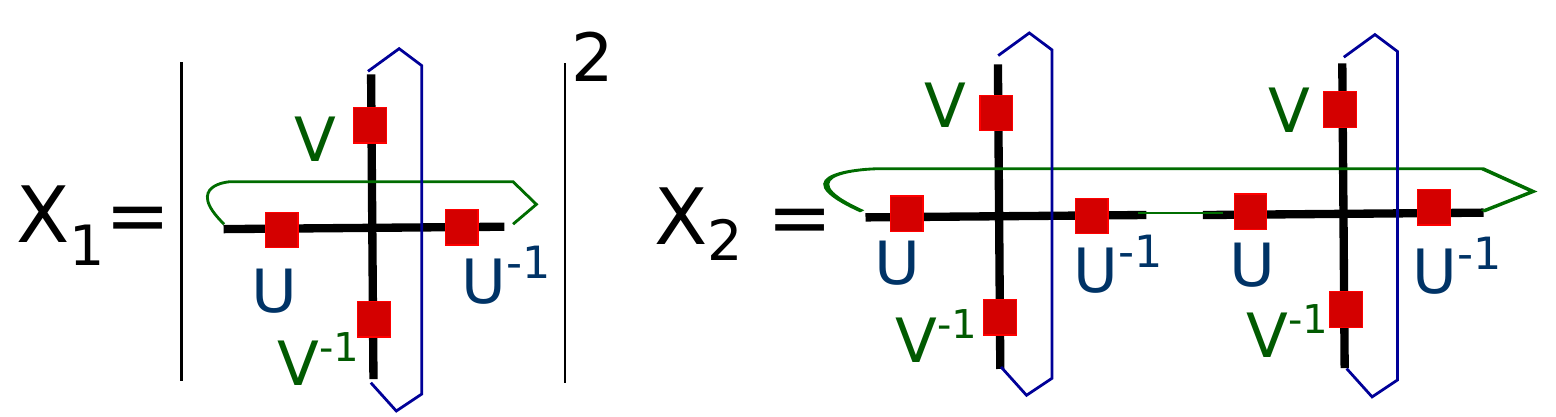}
\caption
{The quantity $X_2/X_1$ obtained by taking the ratio of the contraction value of the double tensor in two different ways. $X_2 / X_1$ is invariant under gauge transformation, such as unitary operators $U$ and $V$. It can be used to distinguish different fixed-point tensors. }
   \label{fig:X2X1}
\end{figure}

Since our ground states are expressed in terms of the tensor network, there is a useful quantity, which we call the X-ratio, introduced by Chen, Gu and Wen~\cite{Xie_LU}, that can be used to probe phase transitions.  
For a local tensor, the X-ratio is given by $X_2/X_1$, where $X_1$ and $X_2$, shown as diagrams in Fig.~\ref{fig:X2X1}, are defined as follows,
\begin{align}
& X_1 =    \left(   \sum _{s,\alpha, \beta, \alpha', \beta'} A^s_{\alpha, \beta, \alpha, \beta}   \times  (A^{s}_{\alpha', \beta', \alpha', \beta'} )^*      \right) ^2, \notag \\
& X_2 =  \sum _{s,s',\alpha, \beta,\gamma,\delta, \alpha', \beta',\gamma',\delta'} 
\Big(A^s_{\alpha, \beta, \gamma, \beta} \times A^{s'}_{\gamma, \delta,  \alpha, \delta}\Big)
\times \notag \\
&\left( (A^s_{\alpha', \beta', \gamma', \beta'})^*   \times  \big( A^{s'}_{\gamma', \delta',  \alpha', \delta'})^*   \right). 
\end{align} 
According to Chen, Gu and Wen, this X-ratio is expected to have the same value in the same phase after sufficient RG coarse-graining, and therefore can be used to detect phase transitions~\cite{Xie_LU}. To compute this, we use the tensor renormalization group (TRG) to flow our wave function to the fixed point (or for large number of coarse-graining steps) and compute the invariant X-ratio  $X_2/X_1$.  A sharp change in the X-ratio signals a phase transition.

\begin{figure}[ht] 
\includegraphics[width=0.5\textwidth]{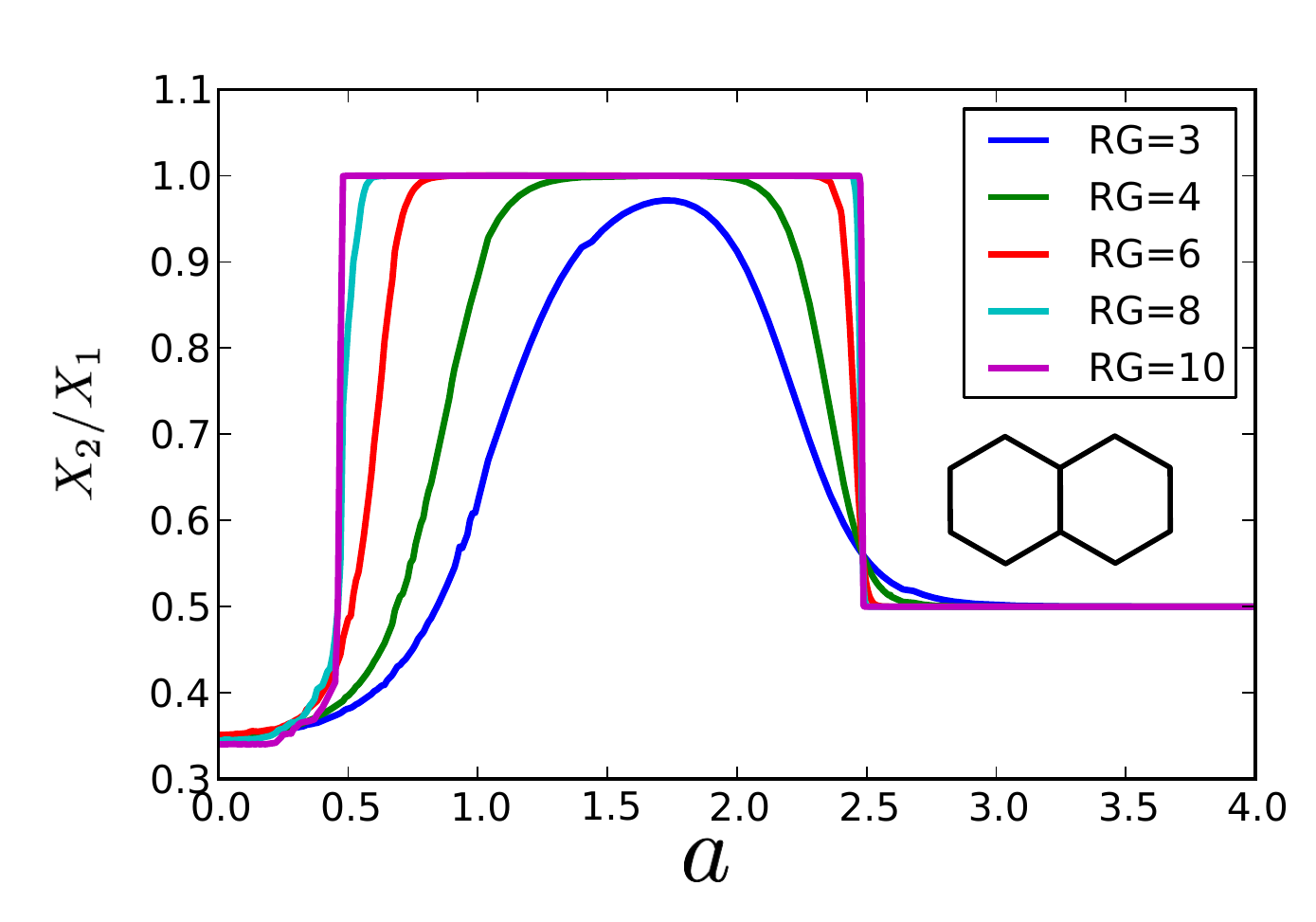}
\caption{  The quantity X ratio $X_2/X_1$ for tensors under the renormalization flow with cut-off $D_c=32$ by tuning a parameter $a$. For $a>1$, it displays a phase transition at   $a_{c_2} =2.5425$ on the hexagon lattice for $\omega = \phi^{\pm},\psi^{\pm}$. The critical exponent $\nu_{a_2} \approx 1.01$.  However, it is not clear for $a<1$.   }
  \label{fig:X12_hexagon}
\end{figure}

The resulting X-ratio vs. parameter $a$ is shown in Fig.~\ref{fig:X12_hexagon} and we monitor its value as the number of RG steps increases. 
We clearly see that a phase transition is identified at $a_{c_2}=2.5425$, which coincides with the transition between the VBS and ordered phases. The value of the X-ratio in the ordered phase can be understood as follows.
As the wave function in the large $a$ limit  is a superposition of $| \Uparrow , \Downarrow ,\Uparrow ,\Downarrow ,...    \rangle$ and $| \Downarrow , \Uparrow , \Downarrow ,\Uparrow ,...    \rangle$,  and   such a cat state has non-trivial $X_2/X_1=1/2$. 
For example,  the state $| \Uparrow , \Downarrow ,\Uparrow ,\Downarrow ,...    \rangle+| \Downarrow , \Uparrow , \Downarrow ,\Uparrow ,...    \rangle$ on the square lattice can be represented by the tensor product state with nonzero components:  $A^{\Uparrow}_{0000}$  and $A^{\Downarrow}_{1111}$, leading to $X_2/X_1=1/2$. 
We remark that as $a$ approaches $a_{c_2}$ , the curves for  $X_2/X_1$  in Fig.~\ref{fig:X12_hexagon} show a crossing, and from this we obtain the critical exponent $\nu_{a_2} \approx 1.01$ using the scaling method in Ref.~\cite{HuangTO2015}.  

Between $0<a<0.5$, there seems to be a non-trivial value of $X_2/X_1$ and it suggests that there is a transition there. 
However, the results are noisy and are likely due to the fact that TRG cannot handle critical states well without using sufficiently bond dimensions, so we must examine this region more carefully with other quantities.

\subsection{XY-like phase: induced magnetization and correlation length}
We now examine the phase diagram by tuning parameter $a$ with bond state $\omega$. 
We find the planar state and (anti)ferromagnetic phases, as well as a valence bond solid (VBS) state between them.  
\begin{figure}[ht]
 \includegraphics[width=0.5\textwidth] {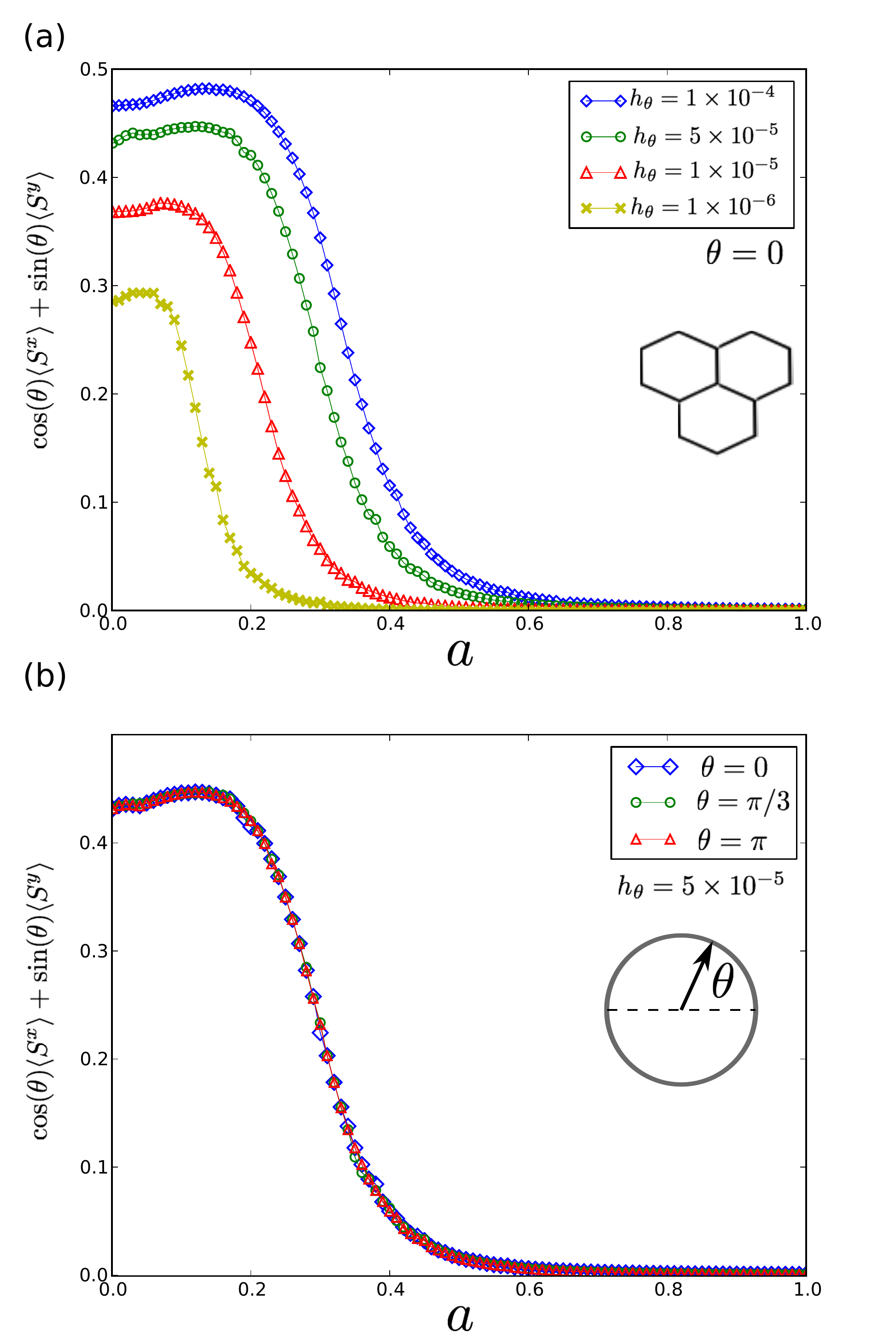}
\caption{The $\langle S^x \rangle $ and  $\langle S^y \rangle $  as a function of parameter $a$ in bond state $\omega = \phi^-$ with small perturbation (a) with different  small field $h_{\theta}$ along $x$ direction  and  (b) in the XY plane along angle $\theta$ with $h_{\theta} = 5\times 10^{-5}$.}
  \label{fig:SXY_hexagon}
\end{figure}
To examine the magnetic properties, we apply a finite small field $h_{\theta}$ along $\hat{\theta}$ direction in XY plane to obtain the induced magnetization. 
To realize this in the wavefunction, we can apply to it with an effective field operator,
\begin{align}
D (h_{\theta}) \! = \! U_{\hat{\theta} \hat{z} }^{\dagger}  \, \text{diag} (  1\!+\! \frac{3}{2}h_{\theta}, 1\!+\! \frac{1}{2}h_{\theta},1\!- \! \frac{1}{2}h_{\theta},1\!-\! \frac{3}{2}h_{\theta} ) U_{\hat{\theta} \hat{z} },
\end{align}
where $U_{\hat{\theta}\hat{z}}=e^{i (\pi/2) S^y} e^{i \theta S^z}$ is an unitary gate that takes $\hat{\theta}$ direction back to $\hat{Z}$ direction.
The induced XY order magnetization depends on the magnitude of the field $h_{\theta}$ as shown in Fig.~\ref{fig:SXY_hexagon}(a), demonstrating that there is no spontaneous magnetization. 
The results, presented in Fig.~\ref{fig:SXY_hexagon}(b), show that the induced XY magnetization $ \cos(\theta)  \langle S^x \rangle +  \sin(\theta)  \langle S^y \rangle$ has a magnitude that is independent of the $\hat {\theta}$ direction. 
This is consistent with the fact that the wavefunction does not have a spontaneous magnetization and has a $U(1)$ symmetry, hence it suggests that this may be an XY phase.

To check the important feature of the XY phase that the correlation length is infinite, we compute the correlation function (in absence of an external field),
\begin{align}
C(r) = \langle \vec{S}(\vec{r_i})  \vec{S}(\vec{r_j}) \rangle  - \langle \vec{S}(\vec{r_i})   \rangle  \langle \vec{S}(\vec{r_j}) \rangle,
\end{align} 
where $r= | \vec{r_i}-\vec{r_j}|$ and $\vec{S} = (S^x, S^y, S^z)$  is spin operators.
\begin{figure}[ht]
\includegraphics[width=0.5\textwidth]{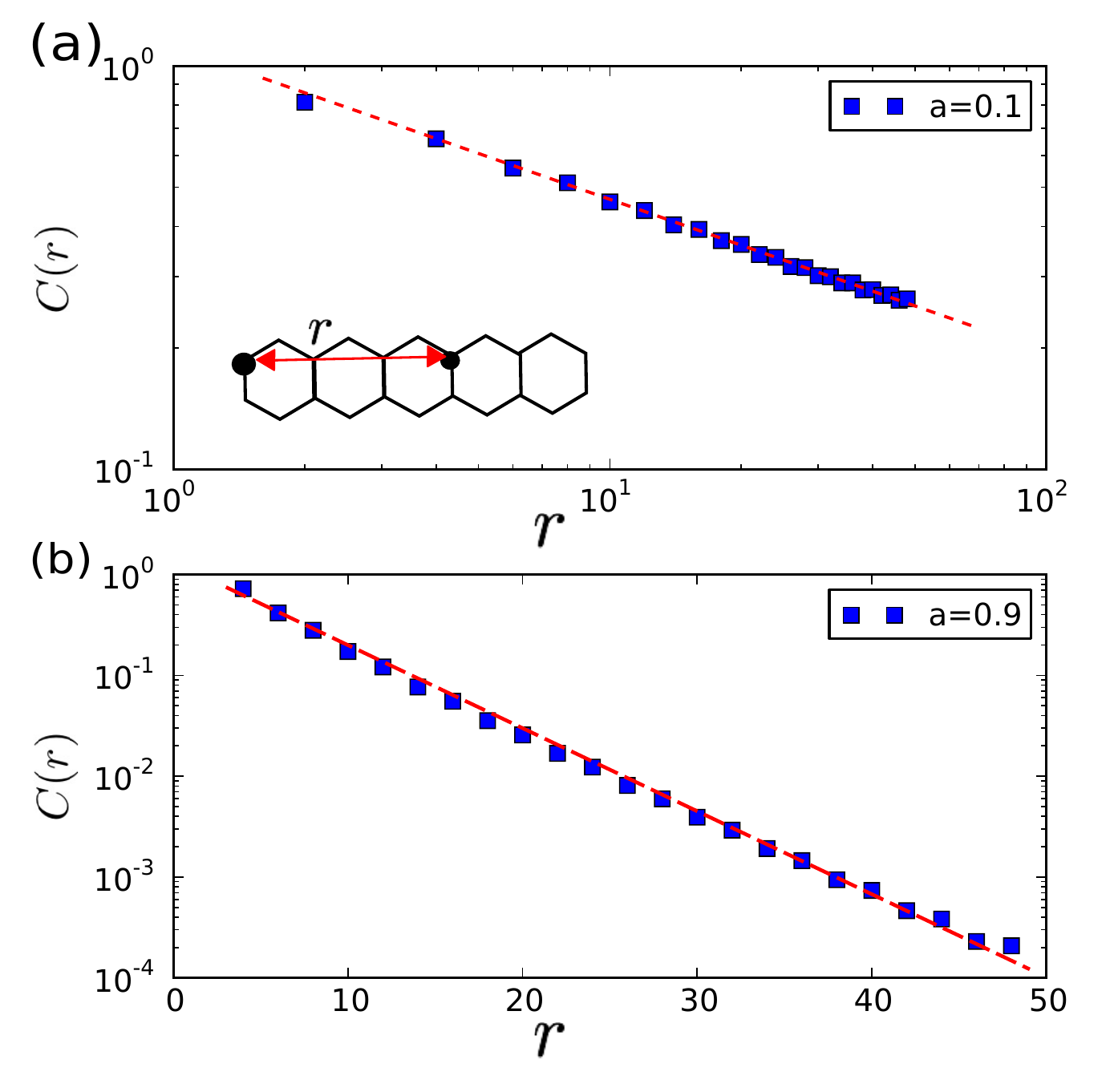}
\caption{ The correlation function under the deformation with parameter  (left)  $a=0.1$, (right) $a=0.9$ on hexagon lattice. }
  \label{fig:DR_hexagon}
\end{figure}
On general grounds, one expects that the correlation function behaves as 
\begin{align}
 C(r) = A\, r^{-\eta} e^{-r/\xi}, 
 \end{align}
where $\eta$ is an algebraic exponent (which is the anomalous exponent at criticality) and  $\xi$ the correlation length. 
For example, at $a=0.1$, we find that  the correlation function displays an algebraic decay, as shown in Fig.~\ref{fig:DR_hexagon}(a). 
This is obtained using the mean-field second renormalization group (SRG)~\cite{Xiang_SRG2009}, which is an improvement over the simple TRG method.  We find that there is an extended gapless region near $a=0$.
In contrast, in Fig.~\ref{fig:DR_hexagon}(b), at $a=0.9$ the correlation function decays to zero exponentially, and is consistent  with a finite correlation length in the VBS phase. 
This suggests that for $a$ small enough, an XY phase emerges, and there should be a phase transition as $a$ decreases from the VBS phase.

\begin{figure}[ht]
   \includegraphics[width=0.5\textwidth]{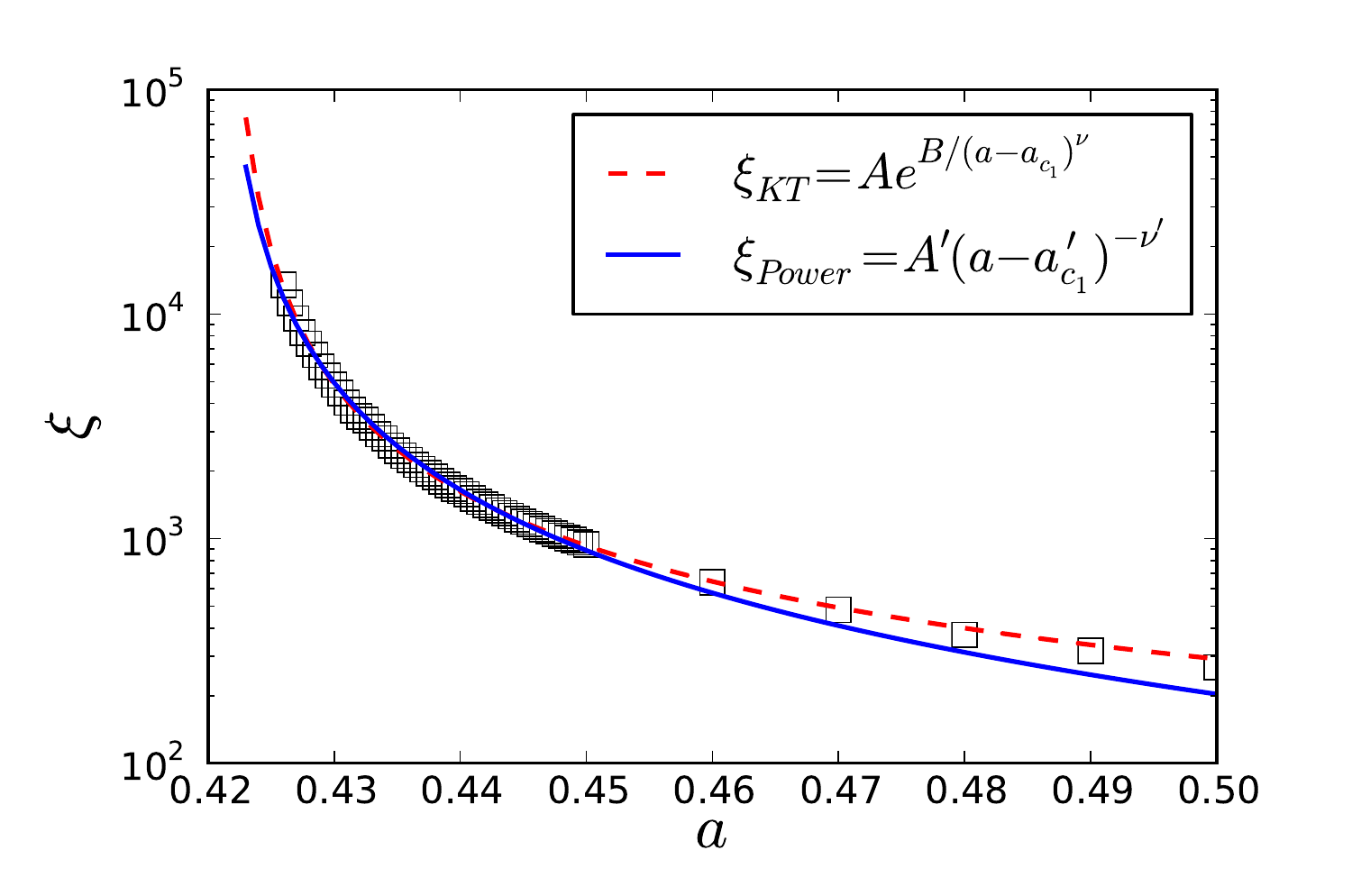}
\caption{ The correlation length under the deformation with parameter  $a$ on hexagon lattice.   The related error ratio of $\sum_a \delta_{power}(a)/\sum_a \delta_{KT}$ for the above range of data is 4.98. }
  \label{fig:KT_scaling}
\end{figure}

To locate the transition and characterize its nature, we examine the dependence of the correlation length as $a$ decreases from the VBS phase. (The correlation length is infinite in the XY phase.)
We find that as $a$ is lowered,  the $\xi$ increases rapidly and it diverges at $a_{c_1}$ (and stays infinite below that critical value), which is the characteristic of continuous transition. Whether the transition is finite-order (e.g. second) or infinite-order can be inferred from the scaling of the correlation.
As shown in Fig.~\ref{fig:KT_scaling} we fit the correlation length to  the essential-singularity form predicted in the Berezinskii-Kosterlitz-Thouless transition~\cite{Kosterlitz1974}, 
\begin{align}
\xi_\text{KT} (a) = A e^{ B/ (a-a_{c_1})^{\nu} }. 
\label{fn:ktscaling}
 \end{align}
The fit  gives 
\begin{align}
A = 1.23 ; \quad B =2.85; \quad  \nu =0.185    ; \quad  a_{c_1} = 0.421.
 \end{align}
 We note that in the original BKT transition tuned by temperature, the exponent $\nu$  is $1/2$~\cite{Kosterlitz1974}. However, as our parameter $a$ is not a temperature, we allow the exponent $\nu$ to be determined by the fit and it needs not be identical to $1/2$.
To compare with the typical continuous transition,  on the other hand,  we also fit the correlation length to the typical power-law dependence
\begin{align}
\xi_\text{Power} (a) = A' (a-a'_{c_1})^{-\nu'}, 
\label{fn:powerscaling}
 \end{align}
resulting in 
$A' = 1.23 ; \quad  \nu' =1.47    ; \quad  a'_{c_1} = 0.4205.$
 %
In order to determine which fit function is better, we then evaluate the  deviation of correlation lengths from the corresponding fit, as follows:
\begin{align}
\delta_i = \frac{  | \xi_i(a)- \xi(a) | }{  \xi(a)}, 
 \end{align}
where $i=$  `KT' or `power'  labels the form of the correlation length from Eq.~\ref{fn:ktscaling} or Eq.~\ref{fn:powerscaling}, respectively. 
We find that the related error ratio  $r=\sum_a \delta_{power}(a)/\sum_a \delta_{KT}$ is 4.98 for the range of data in Fig.~\ref{fig:KT_scaling}, and  this shows that the fitting $\xi_\text{KT} (a)$ is better than $\xi_\text{Power} (a)$. 
Thus we conclude that the transition at $a_c\approx 0.421$  in deformed AKLT state on hexagonal lattice is BKT-like as in both the classical and quantum 2D XY models with temperature~\cite{KT_1973,2DXYmodel_1977,Ding_1990,Ding_1992,1995_Cuccoli}. 
The same conclusion holds regardless of the bond state $\omega$ used in the construction of  the deformed AKLT family on the hexagonal lattice.

\section {Other trivalent lattices: square octagon and cross}  \label{sec:AKLTOther}
In this section, we discuss deformed AKLT states on the square-octagon and cross lattices, as their phase diagrams are similar.

\subsection{ The spin-3/2 model on the square-octagon lattice }
Niggemann and Zittartz (NZ) considered the deformation of the AKLT state on the square-octagon lattice, and they found that similar to the case on the hexagonal lattice there is a VBS to N\'eel transition~\cite{SO_Niggemann}. 
They also used the on-site diagonal approximation to construct a solvable 8-vertex model and obtain the approximate transition at $a_{c_2}\approx 2.65158$.
 We will briefly review their construction and derive an improved 8-vertex model, even though not exactly solvable, that gives a closer transition (obtained numerically) to results from direct TRG evaluation of the spontaneous magnetization in the quantum model.

\begin{figure}[ht]
\includegraphics[width=0.5\textwidth]{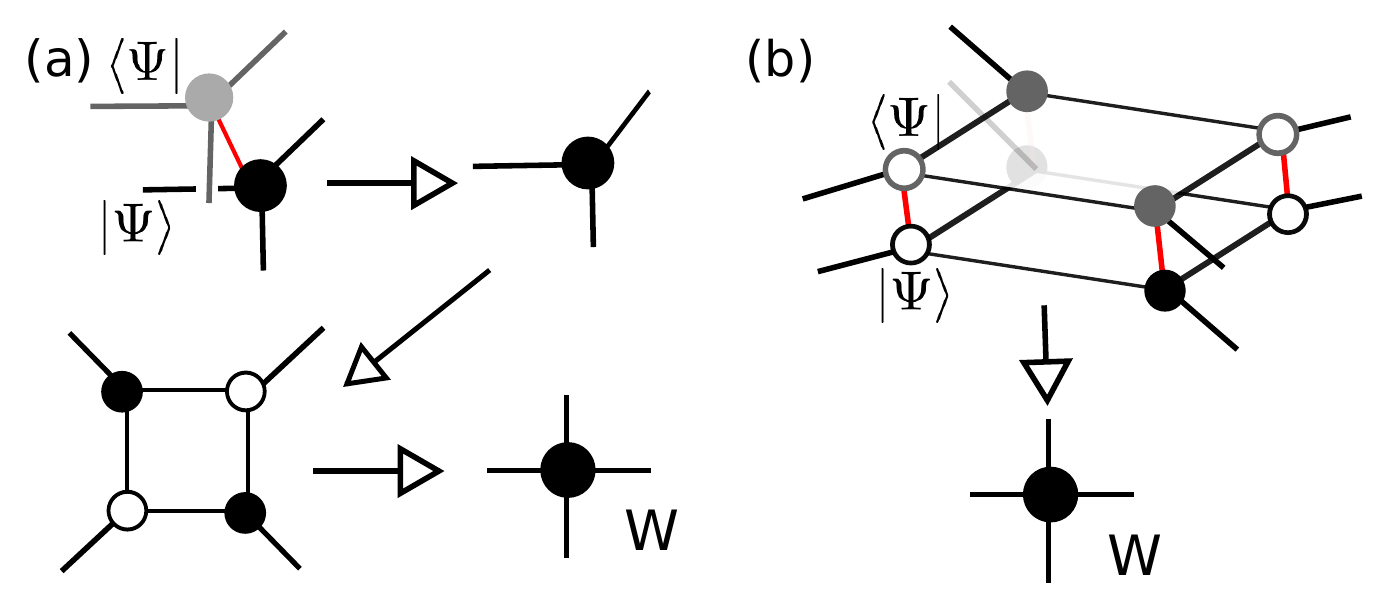}
\caption{ (a) On-site diagonal approximation: merged unit cell for the square octagon lattice. (b) Loop diagonal approximation: unit cells of the new approximation }
  \label{fig:SO_classical}
\end{figure}

\smallskip\noindent {\bf Vertex model by NZ under on-site diagonal approximation.}
Let us consider the deformed spin 3/2-AKLT state (with the single state $|\psi^{-} \rangle$ bond state)  on the  square octagon lattice. 
As in the case of the hexagonal lattice by NKZ, the norm square $\langle \Psi| \Psi \rangle$  of the quantum state will be mapped to the partition function of a vertex model with some approximations, in particular the diagonal approximation.
By the diagonal approximation, one neglects the off-diagonal terms of the double  tensor as in the hexagonal case,  owing to the exponentially decaying probability of unequal bonds as the parameter $a$ increases. 
This approximation is done for each site, i.e., every vertex on the square-octagon lattice. We shall refer to this as the on-site diagonal approximation. The next step is to generate a vertex model on a square lattice by merging several sites (four here) as shown in Fig.~\ref{fig:SO_classical}(a).
This again gives rise to a 16-vertex model, as in the hexagonal case. 
The 16-vertex mode is mapped to a new 8-vertex model via 
 the Hadamard transformation, with weights being
\begin{align}
&w_1 = \frac{1}{2} ( a^8+4a^6+ 30a^4 + 52 a^2  + 41 )   \notag \\
&w_2 = \frac{1}{2}  (a^2-1)^4   \notag \\
&w_3 =  w_4 = - \frac{1}{2} ( a^2-1 )^3    ( a^2+3 )      \notag \\
&w_5 =w_6 =  -\frac{1}{2} (a^2-1)^2(a^4+2a^2+5)   \notag \\
& w_7 =  w_8=  \frac{1}{2} (a^2-1)^2(a^4+2a^2+5).
\end{align}

It turns out that these weights satisfy the free fermion condition (see also Appendix \ref{App:freefermion}), $w_1 w_2+w_3w_4=w_5w_6+w_7w_8$, and thus one can find the transition point at 
\begin{align}
 a_{c_2} = ( \sqrt{2(5+4\sqrt{2})}+ \sqrt{2} +1)^{1/2}   \approx 2.65158.
\end{align} 
Thus under the on-site diagonal approximation, NZ concluded that for $a >a_c$ the quantum state possesses N\'eel order and for $a<a_{c_2}$ the quantum state is in a disordered  VBS phase.

For the deformed AKLT states constructed using  the triplet bond state $|\psi^{+} \rangle = |01 \rangle + |10 \rangle$, we check that the same vertex model is obtained and the same approximate transition point lies between a disordered phase and a N\'eel ordered phase.  
But for the deformed AKLT states from both $|\phi^{\pm} \rangle = |00 \rangle \pm |11 \rangle$  bond states,   the transition is between a disordered phase and a ferromagnetic order phase, a conclusion drawn from almost the same classical eight-vertex model except the signs in $w_3, w_4, w_5, w_6$.

\begin{figure}[ht]
  \includegraphics[width=0.5\textwidth]{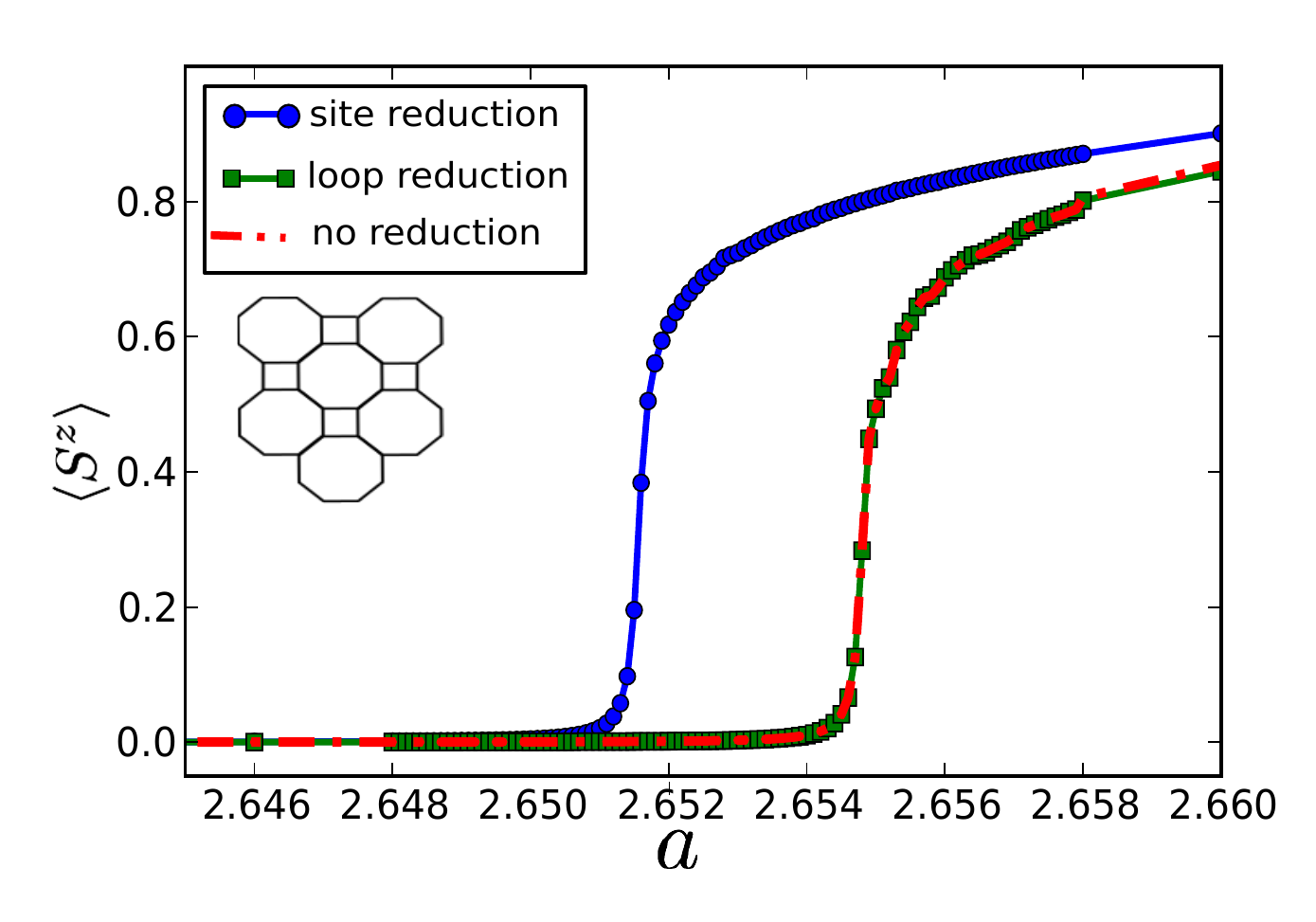}
\caption{ Magnetization of the deformed spin-3/2 AKLT state on the square-octagon lattice 
using the exact tensor network (no reduction), only on-site diagonal double tensors (site reduction), and the  loop diagonal approximation (loop reduction).
 It indicates a transition from VBS phase to N\'eel  phase at $a_{c_2} = 2.6547$ for exact,  at $a_{c_2} = 2.65158$ for site reduction  and  $a_{c_2} = 2.6547$ for  plaquette reduction.  }
  \label{fig:SO_reSZ}
\end{figure}

\smallskip \noindent {\bf Improved vertex model by loop diagonal approximation}.
In the above procedure to derive an effective vertex model on the square lattice (starting from the deformed AKLT states on the square-octagon lattice), four sites from the square-octagon lattice are merged into one site on the resultant square lattice. 
It turns that instead of making the diagonal approximation at each site, we can first merge the four sites in a loop and make diagonal approximation later. 
We call this loop diagonal approximation, in which off-diagonal double tensors are dropped only  after the merging; see Fig.~\ref{fig:SO_classical}(b).  
This means that we allow unequal pairs $(0,1)$ on all contracted double bonds.
This procedure is justified for the square octagon lattice by a result of Niggemann and Zittartz~\cite{SO_Niggemann}, where they found that the probability of finding an unequal pair on a double bond within the merged plaquette is greater than the probability of finding an unequal pair on the free double bonds.
The resulting 16-vertex model can then  be reduced to an eight-vertex models: 
\begin{align}
&w_1 = \frac{1}{2} ( a^8+ 4a^6 + 30 a^4 + 52 a^2 + 57 )     \notag \\
&w_2 =  \frac{1}{2} ( a^2-1 )^4           \notag \\
&w_3 =w_4=  -\frac{1}{2} (a^2-1)^3(a^2+3)       \notag \\
&w_5 =w_6 =  -\frac{1}{2} (a^2-1)^2(a^4+2a^2+5)      \notag \\
&w_7 = w_8 =  \frac{1}{2} (a^2-1)^2(a^4+2a^2+5).  
\end{align} 
We note that this model differs from that of NZ by just one Boltzmann weight $w_1$, with a difference of 8. 
However, the free fermion condition no longer holds. 
It is useful to use the tensor network algorithm to compare these two approximations.
But the transition point predicted from this model, $a_{c_2} = 2.6547$,  turns out to be closer  than that predicted by NZ ($a_{c_2} \approx 2.65158$) to the transition $a_{c_2} = 2.6547$ from the numerical study (see Fig.~\ref{fig:SO_reSZ}) directly on the deformed AKLT family.

\begin{figure}[ht]
\includegraphics[width=0.5\textwidth]{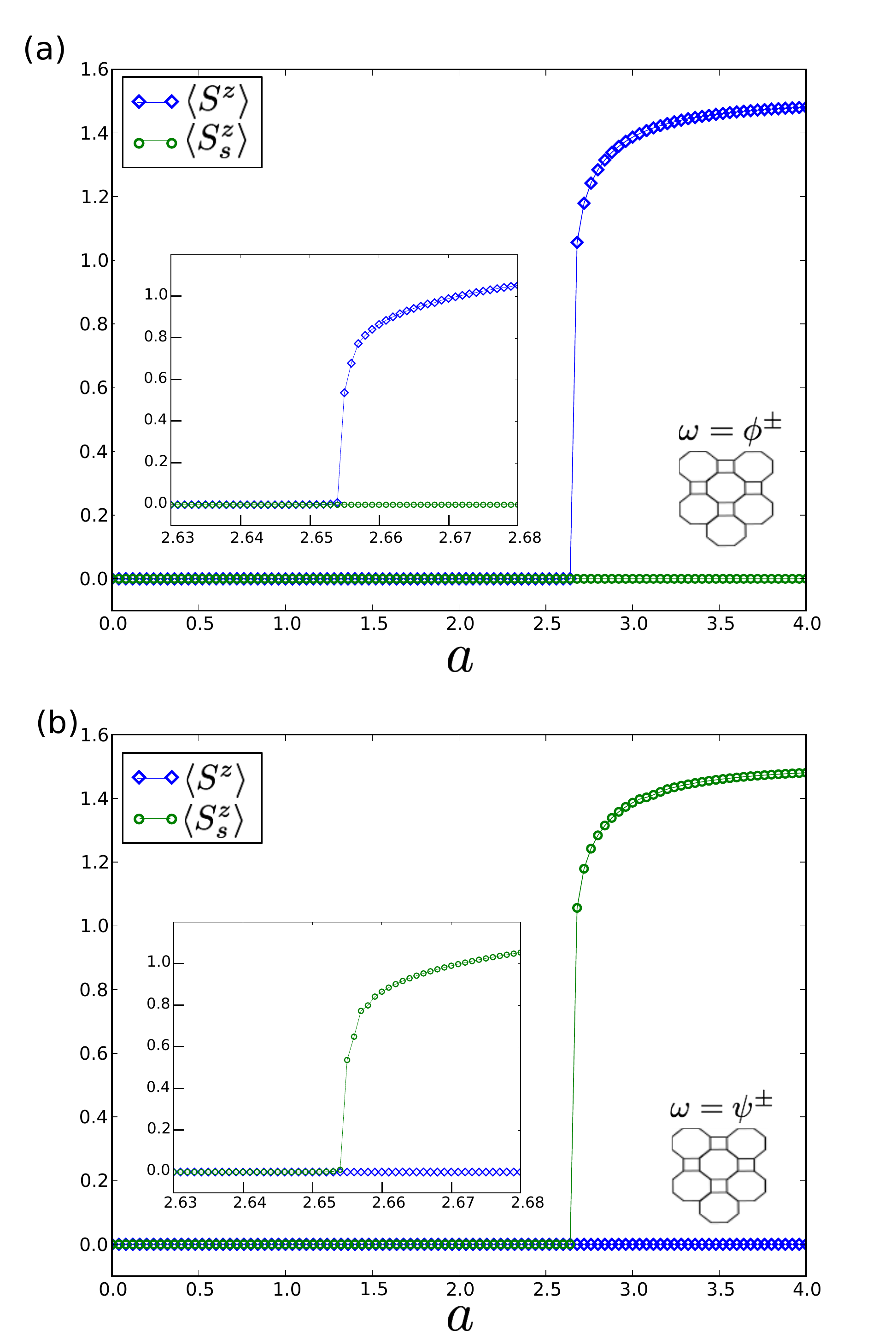}
\caption{ The magnetization  $\langle S^z \rangle$  and  staggered  magnetization  $\langle S^z_s \rangle$  as a function of parameter $a$ using TRG with bond dimension $D_c=24$ on the square octagon lattice. It indicates a transition from VBS phase to  ordered phase at $a_{c_2} = 2.6547$ with (a) $\omega =\phi^{\pm}$ and (b) $\omega =\psi^{\pm}$. }
  \label{fig:SO_SZ}
\end{figure}

\smallskip \noindent {\bf Numerical results with tensor network}.
Using the TRG method we find that the family of the deformed AKLT states  has a transition point located at $a_{c_2}=2.6547$, as captured by behavior of the local order parameter (namely $\langle S^z \rangle$ in bond $\omega = \phi^{\pm}$ and the staggered one $\langle S^z_s \rangle$ in bond $\omega = \psi^{\pm}$) as shown in Fig.~\ref{fig:SO_SZ}.   
This is similar to the large-$a$ side of the phase diagram in the hexagon case. 
The observable of quantum state can be represented by the tensor network and evaluate it by using TRG. 
As mentioned above,  in order to compare with these two approximations, we can reduced the local tensor of observable first. 
The numerical results in Fig.~\ref{fig:SO_reSZ} shows that the transition point from the on-site diagonal approximation  ($a_{c_2} = 2.65158$) is close to the NZ result ($a_{c_2} \approx 2.65158$). 
Then, the magnetization curves of with/without loop diagonal approximation are almost exactly the same.  
This is, on square octagon lattice, the loop diagonal approximation is better than on-site diagonal approximation.

\begin{figure}[ht]
 \includegraphics[width=0.5\textwidth]{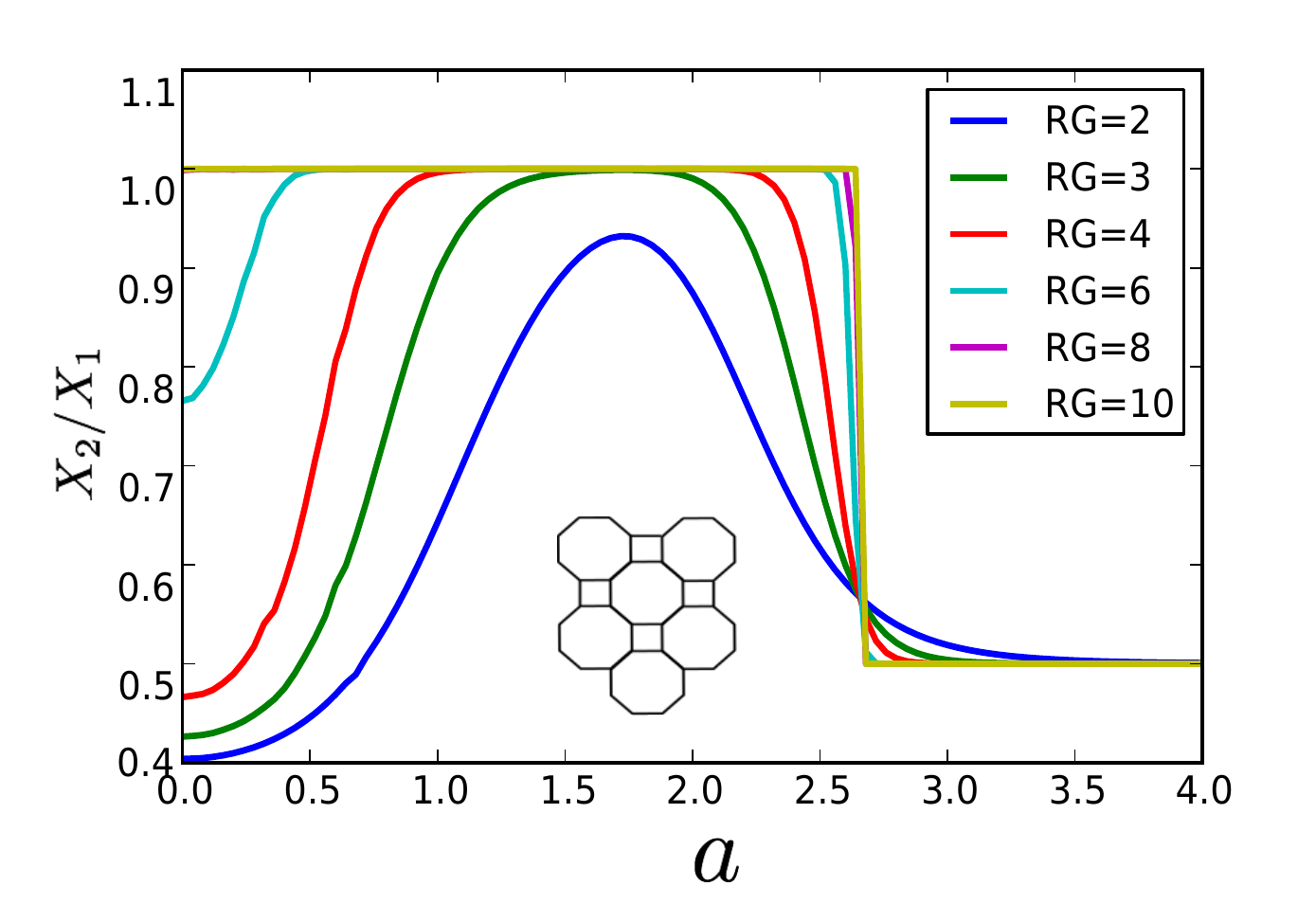}
\caption{  The quantity $X_2/X_1$ for tensors under the renormalization flow by tuning a parameter $a$. As performing more steps of renormalization group, the crossover become sharper and sharper around $a_{c_2}=2.6547$. 
The critical exponent $\nu_{a_2} \approx 1.02$. }
  \label{fig:SO_X12}
\end{figure}

We also find that the  X-ratio $X_2/X_1$ can identify this critical point by their sharp change as shown in Fig.~\ref{fig:SO_X12}.
At $a>1$ region, we extract the critical exponent $\nu$ from the data collapse of the order parameter $X_2/X_1$ under the renormalization flow and find $\nu_{a_2} \approx 1.02$. 
However, unlike the hexagonal case, we do not find any transition for small $a$ values. 
The system remains disordered in the same phase all the way down to $a=0$. 
This can be seen by exponential decaying correlation for small $a$ values. 
We find the finite correlation length at $a=0$ and they are almost the same by increasing the cutoff $D_c$ under SRG processing.

\begin{figure}[ht]
 \includegraphics[width=0.5\textwidth] {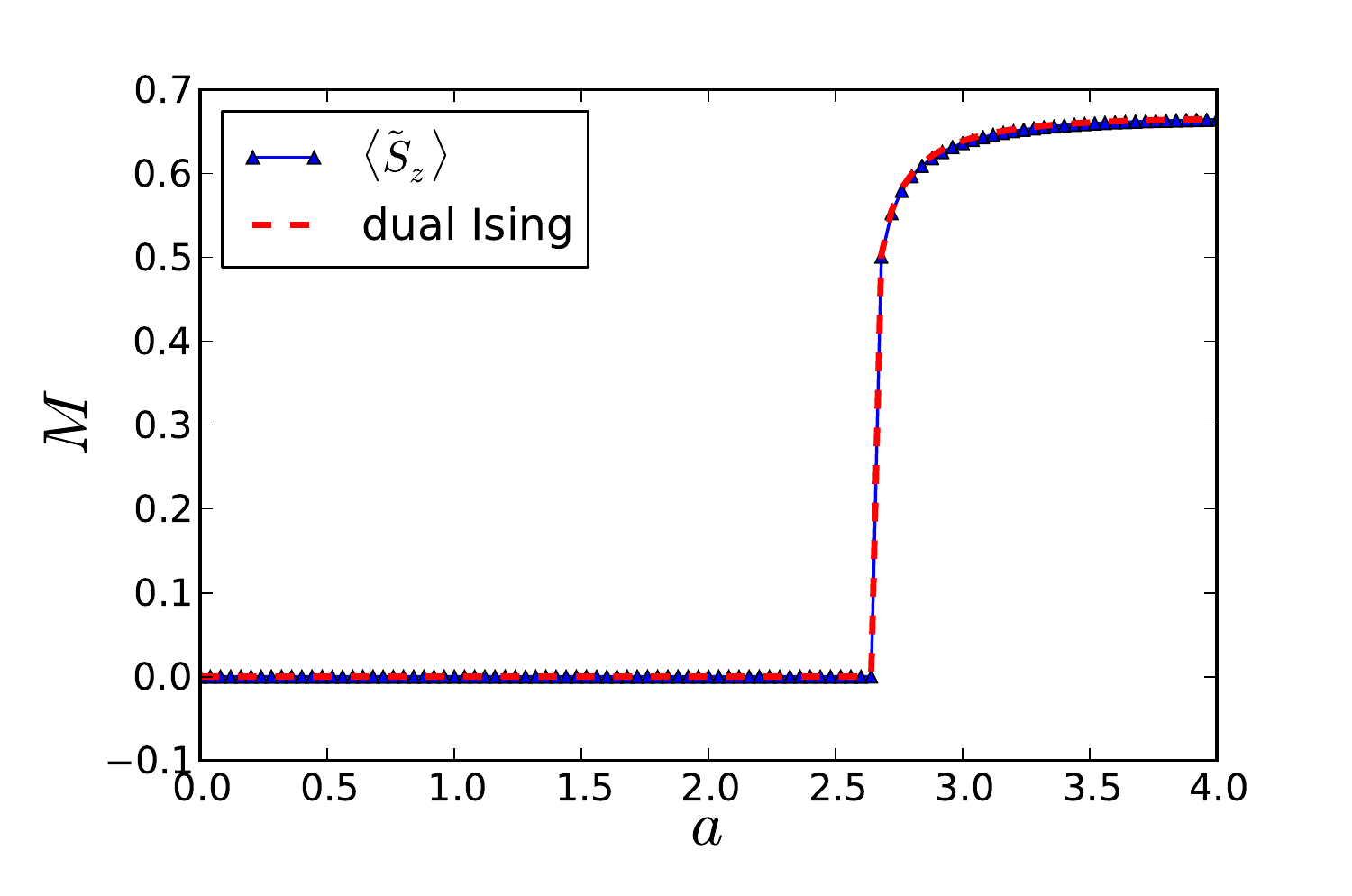}
\caption{ The spontaneous magnetization $M$ of the dual checkerboard Ising model (red) compared to the ground state expectation value of $\tilde{S}_z$ (blue).}
  \label{fig:dual_M}
\end{figure}

\medskip
\noindent {\bf The spontaneous magnetization}. 
Here we discuss how we can obtain spontaneous magnetization in the ordered phase. Since the difference in different bond constructions is whether the resulting ordered phase is ferromagnetic or antiferromagnetic, we shall for simplicity consider the bond state $|\phi^+\rangle$. The free-fermion eight-vertex model using this bond state (as well as $|\phi^-\rangle$),  under the on-site diagonal approximation, gives the following weights,
\begin{align}
&w_1 = \frac{1}{2} ( a^8+4a^6+ 30a^4 + 52 a^2  + 41 )   \notag \\
&w_2 = \frac{1}{2}  (a^2-1)^4   \notag \\
&w_3 =  w_4 =  \frac{1}{2} ( a^2-1 )^3    ( a^2+3 )      \notag \\
&w_5 =w_6 =   \frac{1}{2} (a^2-1)^2(a^4+2a^2+5)  \notag \\
& w_7 =  w_8=  \frac{1}{2} (a^2-1)^2(a^4+2a^2+5).
\end{align} 
It was known that the free fermion model can be mapped onto the Ising model on the union jack lattice~\cite{Baxter1988,Choy1987} (see Appendix~\ref{App:Isingasunionjack} ). 
We find the various interaction energies of the Ising model are given by
\begin{align}
&k = K_1=K_2=K_3=K_4 = \frac{1}{2}\ln \big(  \frac{a^2 -1}{a^2+3} \big)     \notag \\
& \kappa = K=K'=- \ln \big(  \frac{a^2 -1}{a^2+3} \big) \notag \\
&\rho = \frac{1}{2}(a^2-1)^3 (a^2+3).
\end{align}
The Ising model undergoes a phase transition at $\Omega ^2 =1$ and $a_m \approx 2.65158$ exactly the same transition point that we extracted from the free-energy behavior of the eight-vertex model.
However, from this mapping, the ferromagnetic phase lies in $1 < a < a_m$, opposite to the phase diagram of the quantum model.
This is due to the Hadamard transformation used in deriving the 8-vertex model, and as noted in~\cite{Hexagon_Niggemann}, this transformation maps the low temperature region of the quantum state to the high temperature region of the eight-vertex model and vice versa.
But we can reconstruct the original temperature behavior of the classical model using the Kramers-Wannier duality. 
This duality maps the original Ising model on the union jack to an Ising model on the dual lattice, i.e., the square octagon and the low-temperature regime of the original model to the high-temperature regime of the dual model and vice versa~\cite{Kramers1941}, and hence the correct correspondence of the classical magnetization to the quantum magnetization can be identified.  As pointed out by Baxter~\cite{Baxter1986},  the free-fermion vertex model can also be mapped to the Ising model on the checkerboard lattice, which is self-dual. We thus carry out the duality on the checkerboard lattice and we obtain the magnetization that reverts the temperature dependence, 
\begin{align}
 M =
  \begin{cases}
    (1-\Omega ^{-2})^{1/8}      & \quad \text{if }   \Omega^2 >1\\
    0  & \quad  \text{ others},\\
  \end{cases}
\end{align}
where 
\begin{align}
\Omega^2 =  \frac{(w_1^2+w_2^2-w_3^2-w_4^2)^2-4(w_5w_6-w_7w_8)^2}{ 16 w_5 w_6 w_7 w_8 }.
\end{align}
This magnetization behavior is shown in  Fig.~\ref{fig:dual_M} and it can be regarded as the disorder parameter of the free fermion model~\cite{Baxter1988} .
Surprisingly, as shown in Fig.~\ref{fig:dual_M} we find that the behavior of the magnetization matches very well the expectation value of the following effective quantum spin operator
of the deformed AKLT states under the same on-site diagonal approximation
\begin{align}
\tilde{S}_z= 
 \begin{pmatrix}
  1&0 &0&0\\
  0&1 &0&0\\
  0&0 &-1&0\\
  0&0 &0&-1\\
 \end{pmatrix}.
 \end{align}
This seems to hold for the ordered phases in other lattices, such as the hexagon and the cross lattice.
\begin{figure}[ht]
 \includegraphics[width=0.2\textwidth] {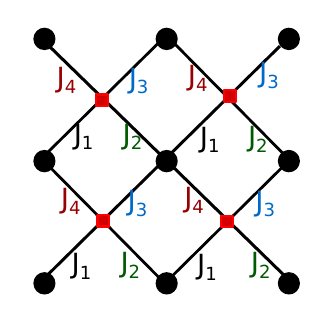}
\caption{ Checker board lattice and the two types of Ising spins associated with its vertices.}
  \label{fig:checker}
\end{figure}

\subsection{ The spin-3/2 on the cross lattice }

The alternative lattice to investigate  in this section is the cross lattice  shown in Fig.~\ref{fig:cross_lattice}.
We adapt the procedure introduced in the previous section. 
The deformation and the AKLT state are applied to the new spatial geometry of the system.
To construct the ground state, we use the tensor network and valence bond solid construction again.

For the mapping to the classical model we carry out the on-site diagonal approximation and arrive at  a 16-vertex model on the kagome lattice with all 16 nonzero vertex weights. 
By employing the same Hadamard transformation and on-site diagonal approximation, we obtain an eight-vertex model, whose  Boltzmann weights for all virtual bonds are given by 
\begin{align}
&w_1 = \frac{1}{2} ( a^8+4a^6+ 30a^4 + 52 a^2  + 41 )   \notag \\
&w_2 = \frac{1}{2}  (a^1-1)^4   \notag \\
&w_3 =  w_4 =  \frac{1}{2} ( a^2-1 )^3    ( a^2+3 )      \notag \\
&w_5 =w_6 =  \frac{1}{2} (a^2-1)^2(a^4+2a^2+5)  \notag \\
& w_7 =  w_8=  \frac{1}{2} (a^2-1)^2(a^4+2a^2+5).
\end{align} 
While the Boltzmann weights from  the loop diagonal approximation almost the same classical eight-vertex model from on-site diagonal approximation expect the constant term of $w_1$, with a difference of 8.
However, we cannot find an analytic solution of the transition.

\begin{figure}[ht]
 \includegraphics[width=0.45\textwidth]{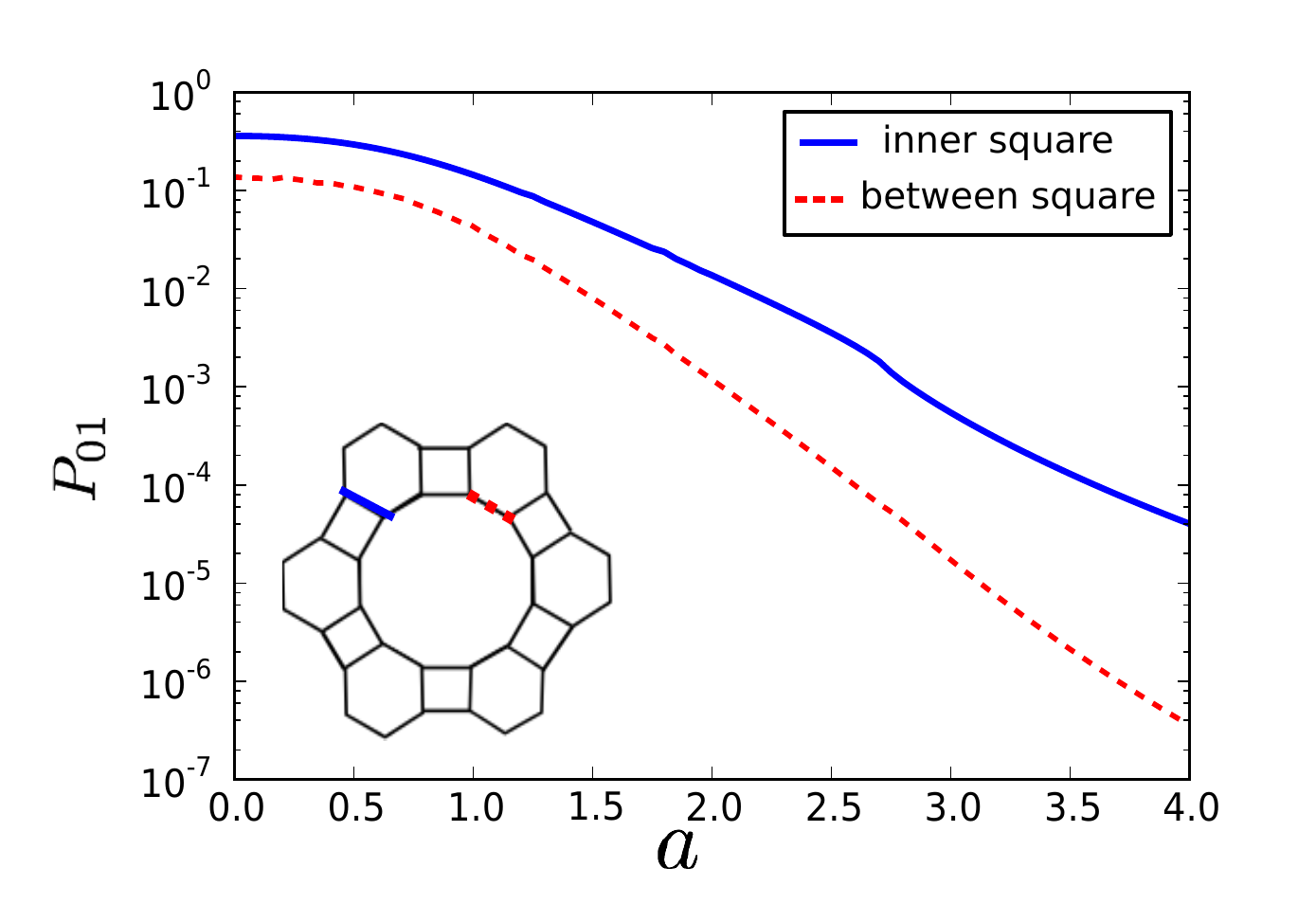}
\caption{ The probability for finding the unequal pair on a bond within in a plaquette (inner) and between two plaquettes (between) on the cross lattice.  }
  \label{fig:cross_error}
\end{figure}
On the other hand,  we found that the probability of finding an unequal pair on a double bond within the merged plaquette is greater than the probability of finding an unequal pair on the free double bonds as shown in Fig.~\ref{fig:cross_error}.
Let us discuss how to obtain the the probability of finding an unequal pair $P_{01}$ numerically.
First, the wave function of deformed AKLT state can be represented by tensor product state, $|\Psi\rangle = tTr(A^{s_1} A^{s_2} A^{s_3}...)|s_1,s_2,s_3,... \rangle$. 
The local double tensor  $\mathbb{T}$ can be formed by merging two layers, tensors $A$ and $A^*$ with only physical indices contracted.
We then prepare a operator $O_{01}$ defined on the two-virtual-particle space, which gives one as two virtual particles are unequal and others are zero. 
The $P_{01}$ can be obtained by determining the expectation value of operator  $O_{01}$ for one bond,  $\langle \Psi | O_{01}^i  | \Psi \rangle = tTr ( \mathbb{T}^1 \mathbb{T}^2 \mathbb{T}^3 O_{01}^i  ....) $, where $i$ is the location of operator $O_{01}$ which could be inner the square or between two squares. 
The result, in Fig.~\ref{fig:cross_error}, shows that the the probability for finding an unequal pair  between two squares is lower than inner the square. 
From it, we can predict that loop (square) diagonal  approximation is better than on-site diagonal approximation. 
The loop diagonal approximation can avoid the truncation error from the reduction of virtual bonds in the inner square. 

\begin{figure}[ht]
\includegraphics[width=0.5\textwidth]{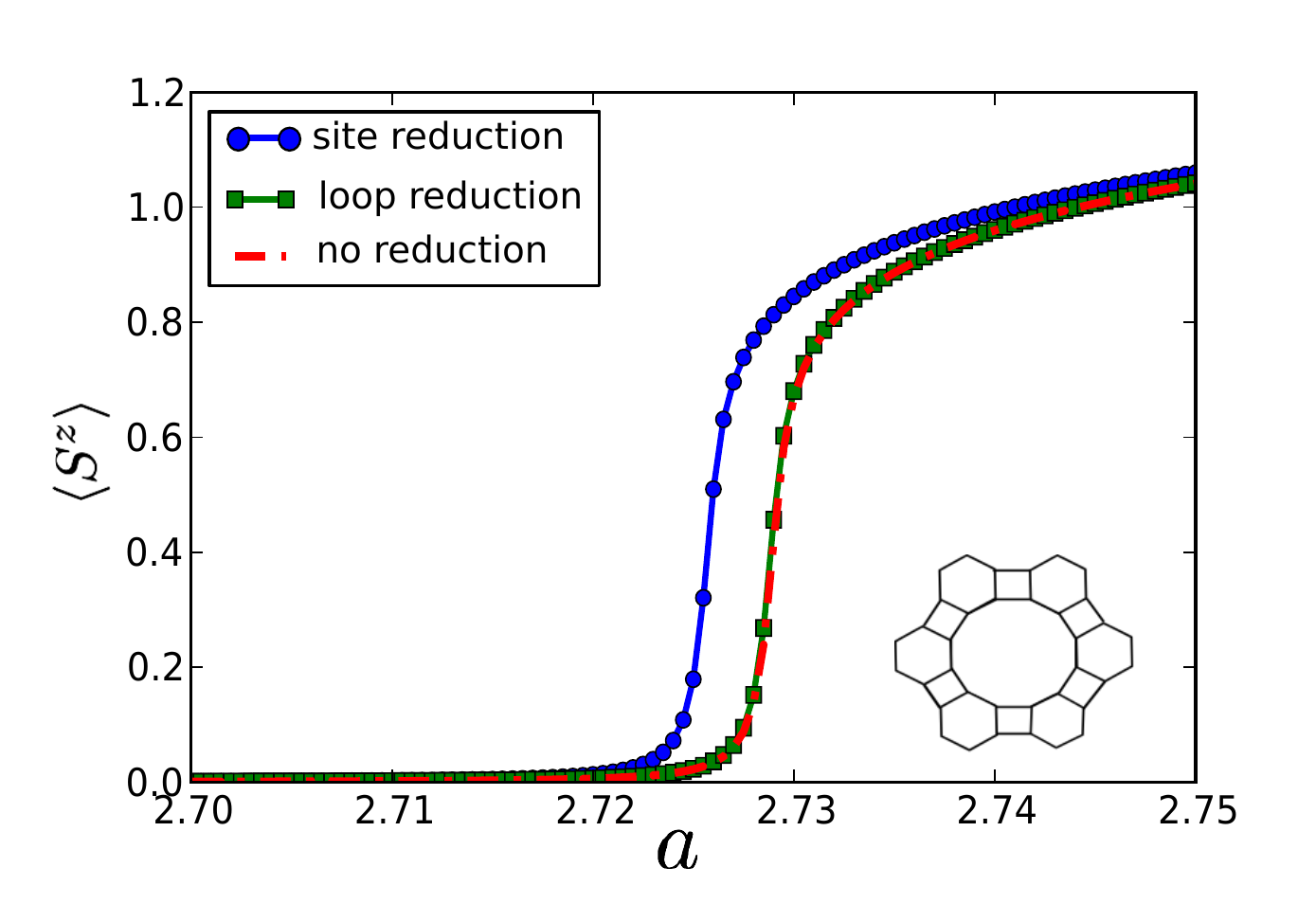}
\caption{ The magnetization  as a function of parameter $a$ with $D_c=24$ for TRG in taking the expectation value. It indicates a transition from VBS phase to N\'eel  phase at $a_{c_2} = 2.7280$ (no reduction). By reducing the off-diagonal terms, we can obtain the transition at $a_{c_2} = 2.7245$ for on-site approximation  (site reduction)  and  $a_{c_2} = 2.7280$ for  loop approximation (loop reduction) .   }
  \label{fig:cross_reSZ}
\end{figure}
\begin{figure}[ht]
 \includegraphics[width=0.5\textwidth]{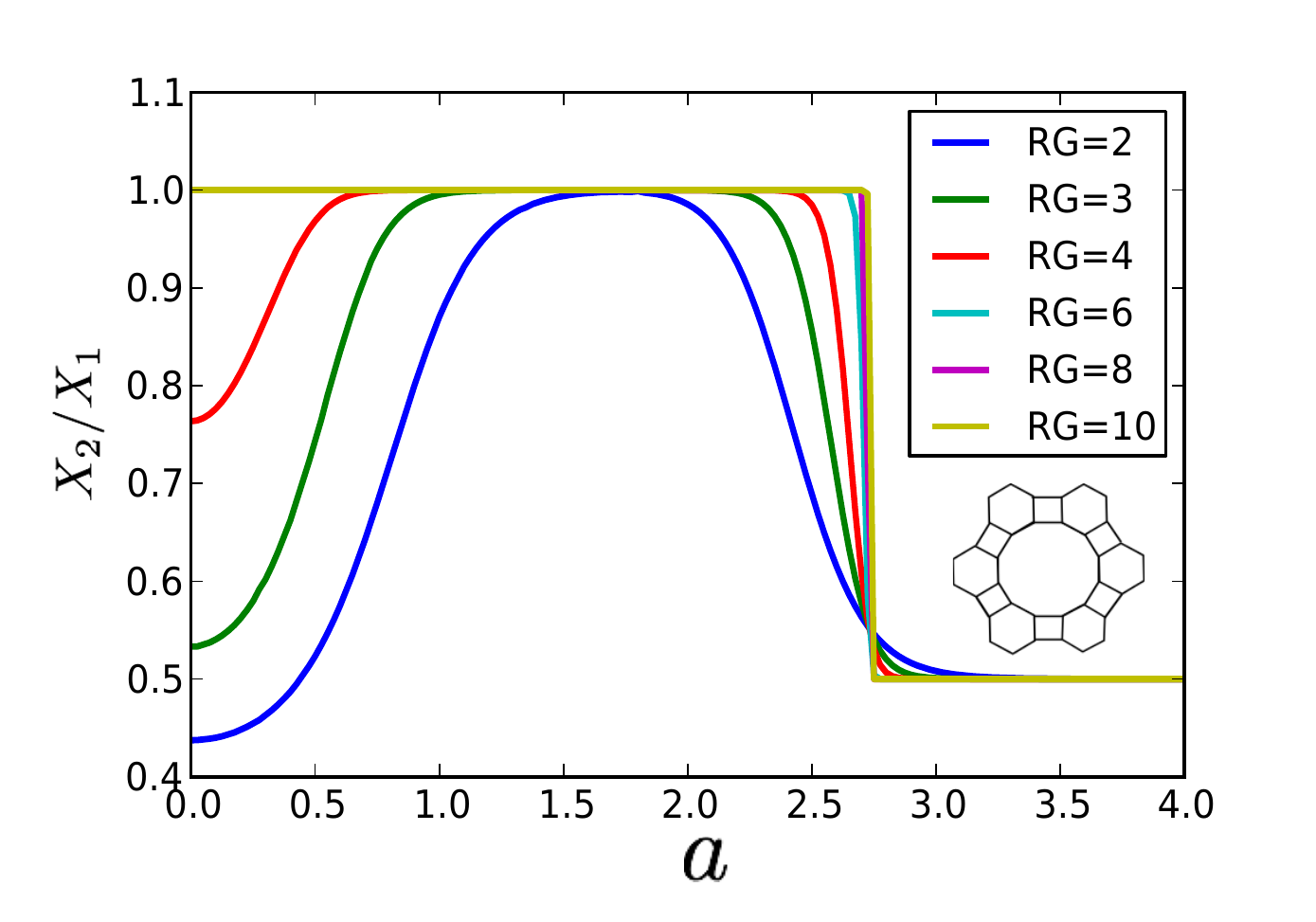}
\caption{  The quantity $X_2/X_1$ for tensors  as a function of parameter $a$ under the renormalization flow.  It indicates a phase transition at $a_{c_2}=2.7280$ on the cross lattice.
The critical exponent $\nu_{a_2} \approx 1.0$. }
  \label{fig:cross_X12}
\end{figure}
\begin{figure}[ht]
 \includegraphics[width=0.5\textwidth] {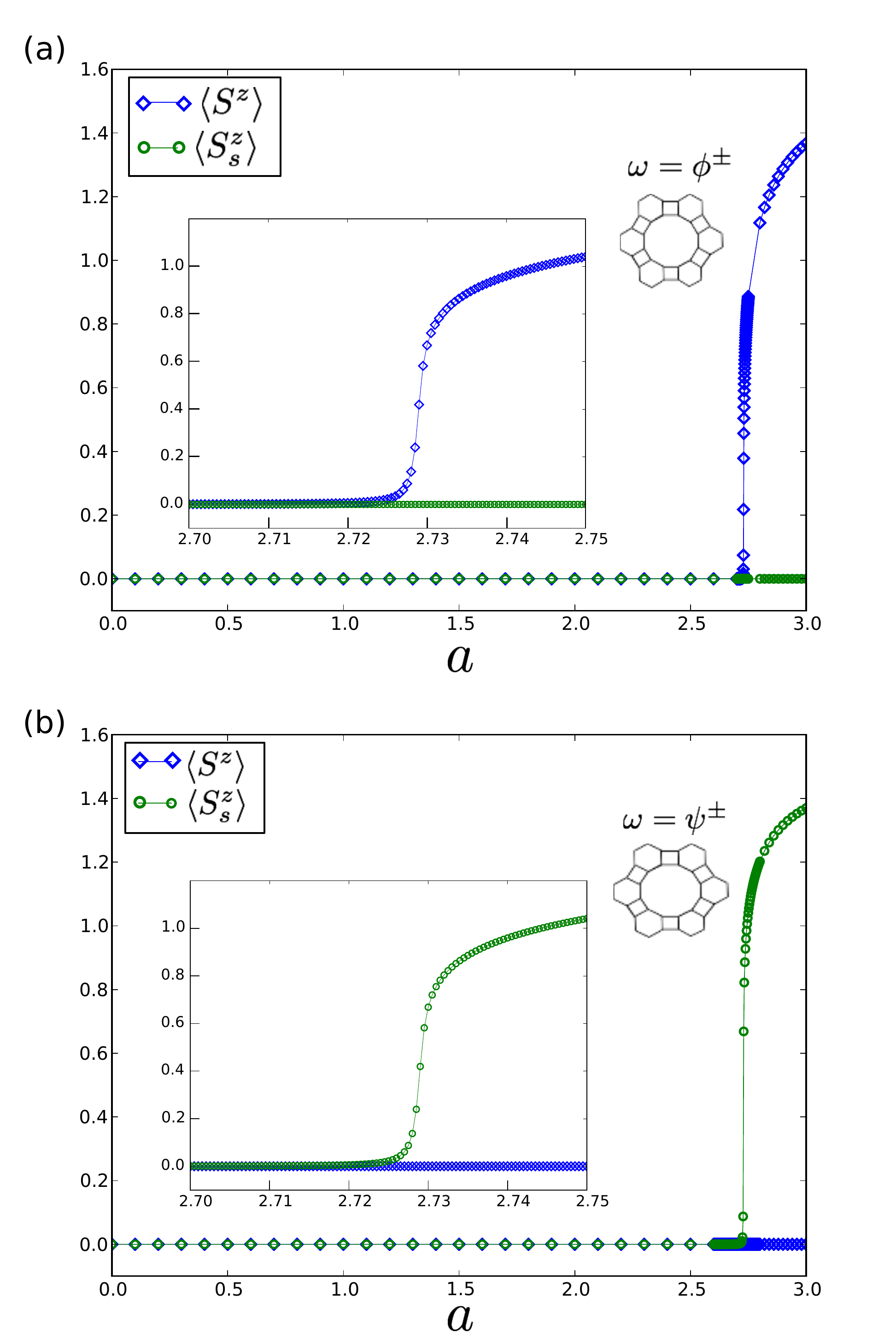}
\caption{ The magnetization  $\langle S^z \rangle$  and  staggered  magnetization  $\langle S^z_s \rangle$  as a function of parameter $a$ using TRG with bond dimension $D_c=24$ on the cross lattice. It indicates a transition from VBS phase to  ordered phase at $a_{c_2}=  2.7280$ with (a) $\omega =\phi^{\pm}$ and (b) $\omega =\psi^{\pm}$. }
  \label{fig:cross_SZ}
\end{figure}

Again, we can obtain the transition point numerically at  $a_{c_2} = 2.7245$ by on-site diagonal approximation  and  at  $a_{c_2} = 2.7280$ for  loop diagonal approximation, as shown in Fig.~\ref{fig:cross_reSZ}. 
The magnetization under the latter approximation matches very well with the magnetization obtained from full quantum calculations using TRG, which gives 
the transition from the VBS to the ordered phase  at  $a_{c_2} = 2.7280$,  as shown in Fig.~\ref{fig:cross_SZ}.
The results of the X-ratio $X_2/X_1$ also confirm the transition point, as shown in Fig.~\ref{fig:cross_X12}. 
Again, as $a$ approaches $a_{c_2}$, the curves for $X_2/X_1$ show a crossing, and from this we obtain the critical exponent $\nu_{a_2} \approx 1.0$.  

The phase diagram of the deformed AKLT family on the cross lattice is similar to that on the square-octagon lattice, where there is a transition from VBS to ordered phase, but no XY phase is found for small $a$. 
This is consistent with our calculations of the correlation length, which never diverges in this parameter region.

\section{ The deformed AKLT states on the star lattice }
The phase diagram of the deformed AKLT states on the star lattice depends on the type of bond states used in the construction. We discuss in turn the antiferromagnetic and ferromagnetic bonds.
\subsection{Antiferromagnetic bonds}
First, we consider the deformed AKLT states constructed from antiferromagnetic bonds $\omega = \psi^\pm$, 
\begin{align}
|\Psi \rangle =  \bigotimes_{v}  \Big( D'(a) P  \Big)_{v}   \bigotimes_{l\in L}|\psi^\pm \rangle_{l}. 
\end{align} 
The wavefunctions can be in turn written as tensor product states. In the limit $a\rightarrow \infty$, the wavefunction is essentially a two-state system of $|\Uparrow\rangle\equiv|000\rangle$ and $|\Downarrow\rangle\equiv111\rangle$. 
Because of the anti-correlation in the $|\psi^\pm\rangle=|01\rangle\pm|10\rangle$,  the spins $\Uparrow$ and $\Downarrow$'s can only arrange in the same way as the classical antiferromagnetic Ising spins on the frustrated star lattice.  
In this limit, the deformed AKLT states with $|\psi^\pm\rangle$ bonds are equivalent to classical spin liquids.

\begin{figure*}[ht]
  \includegraphics[width=1.0\textwidth] {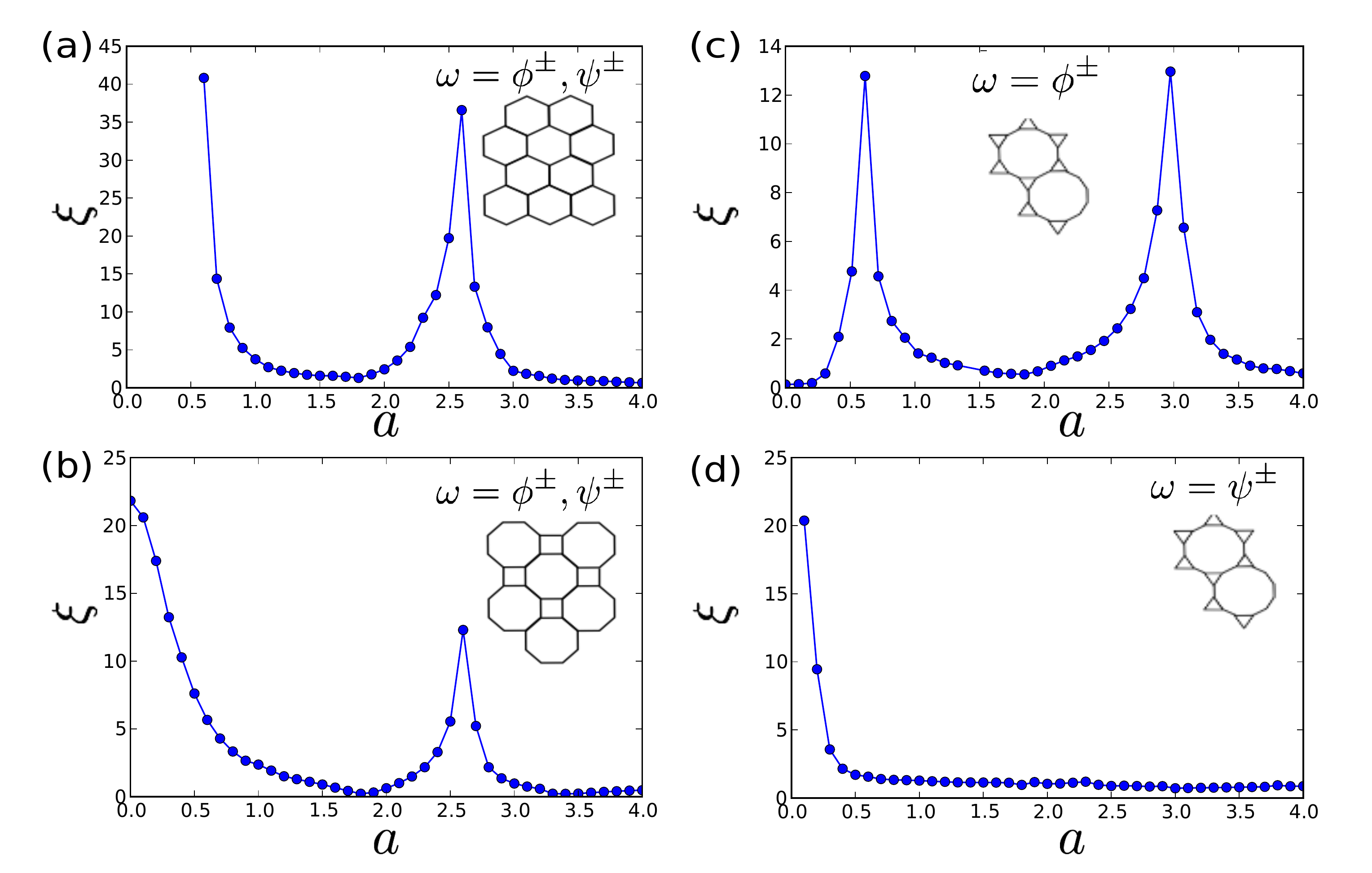}
\caption{ (a) The correlation length for tensors under the renormalization flow by tuning a parameter a. It displays phase transitions at $a_{c_2} = 2.54$ on the hexagon lattice.   
(b) It shows that a transition occur around $a_{c_2}=2.65$ on square-octagon lattice. 
(c)   It indicates  phase transitions at $a_{c_1}=0.58$ and $a_{c_2}=3.02$ on star  lattice with bond state $ \omega =  \phi^{\pm}$.  
(d) There is no singularity with bond state $ \omega  = \psi^{\pm} $. We check that even at $a=0$, the correlation length at cases (b), (c) and (d) is finite and does not increase as the bond dimension is increased. }
  \label{fig:correlation_length}
\end{figure*}

By employing the TRG method, we do not find any transition, as is seen from the finite correlation length throughout the parameter range as shown in Fig.~\ref{fig:correlation_length}(d).
In particular, the VBS AKLT state is in the same phase as the classical spin-liquid state.
The finding that there is no transition is also confirmed by the mapping to classical vertex models. 
We follow the procedure described earlier and construct corresponding 8-vertex models for the deformed AKLT states with $\omega = \psi^{\pm}$ bonds. 
The two 8-vertex models are the same under the loop (one triangle) diagonal approximation, which are given by 
\begin{align}
&w_1 =2 (3a^4+6a^2+7)^2   \notag \\
&w_2 =  w_3 =  w_4 =  2 ( a^2-1 )^4        \notag \\
&w_5 =w_6 =  -2 (a^2-1)^2(3a^4+6a^2+7)  \notag \\
& w_7 =  w_8=  2 ( a^2-1 )^4,
\end{align} 
but are different under the loop (two triangles ) diagonal approximation.
They turn out to satisfy free-fermion condition and are thus solvable, but no transition is found for $a\ge 1$ (in the range where the models are valid).  

\begin{figure}[ht]
 \includegraphics[width=0.5\textwidth]{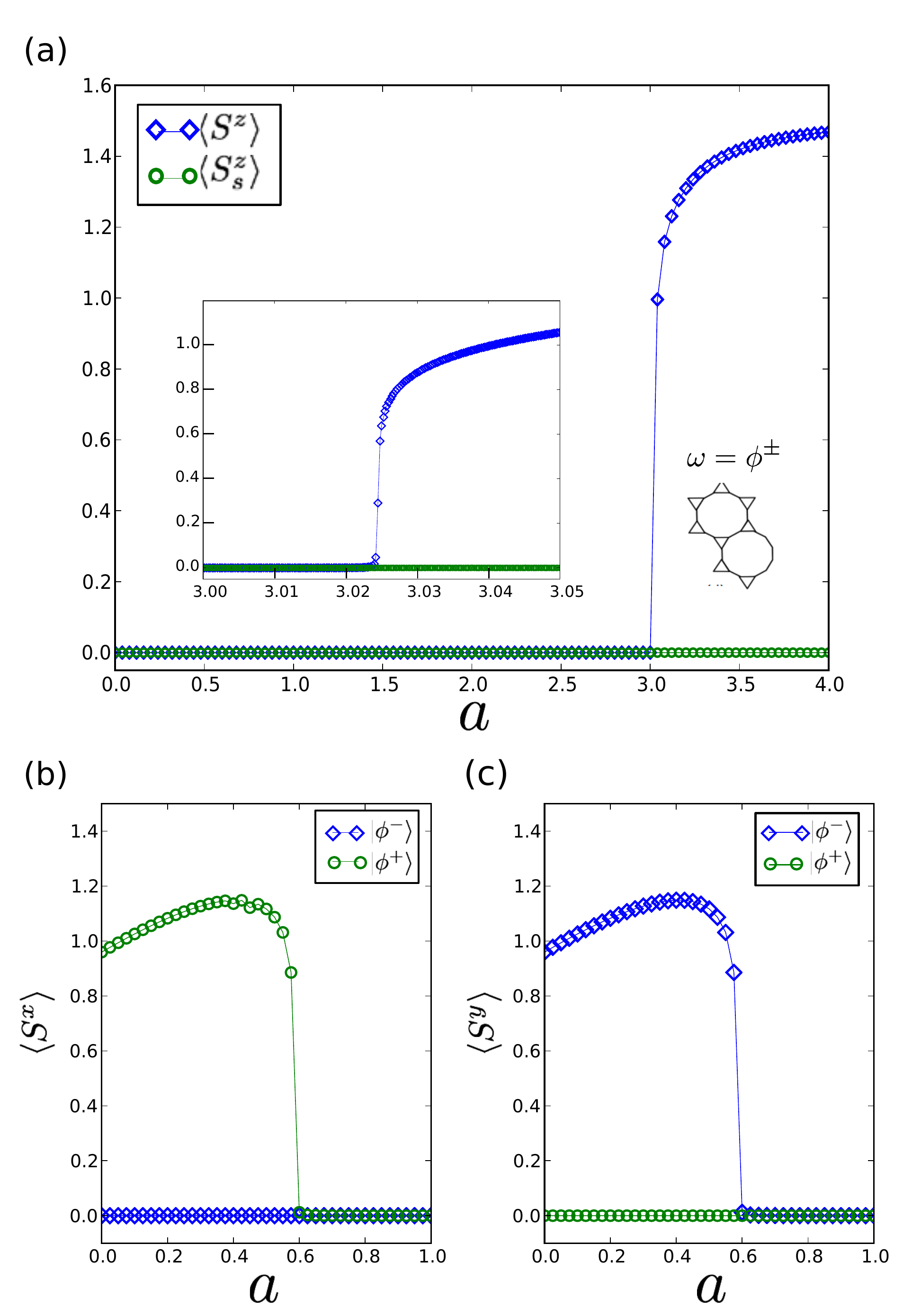}
\caption{ (a) The magnetization  as a function of parameter $a$ with $D_c=24$ for TRG in taking the expectation value. It indicates a transition from VBS phase to ferromagnetic phase at $a_{c_2}= 3.0243$ on star lattice.    
(b) The $\langle S^x\rangle$as a function of parameter $a$ with bond state $\phi^{\pm}$. It shows that a transition from  x-direction ferromagnetic phase to VBS phase at $a_{c_1}=0.5850$ on the hexagon lattice. 
(c) The $\langle S^y\rangle$as a function of parameter $a$ with  bond state $\phi^{\pm}$. It shows that a transition from  y-direction ferromagnetic phase to VBS phase at $a_{c_1}=0.5850$ on the hexagon lattice
}
  \label{fig:star_SZ}
\end{figure}

\subsection{Ferromagnetic bonds}
Next, we consider the ferromagnetic case
\begin{align}
|\Psi \rangle =  \bigotimes_{v}  \Big( D'(a) P  \Big)_{v}   \bigotimes_{l\in L}|\phi^+ \rangle_{l}. 
\end{align} 
The wavefunctions can be in turn written as tensor product states.

\smallskip \noindent {\bf Ferromagnetic phase at large $a$}.
In the limit $a \to \infty $, the effective terms of local tensor representation  are $A^{\Uparrow }_{000} = a $,  $A^{\Downarrow }_{111} = a $. 
The global ground state is therefore a superposition of both possible ferromagnetic states $| \Uparrow , \Uparrow ,\Uparrow ,\Uparrow ,...    \rangle+ | \Downarrow , \Downarrow , \Downarrow ,\Downarrow ,...    \rangle$. 
By increasing the parameter $a$ from the VBS phase, the wave function undergoes a second order quantum phase transition  to a ferromagnetic phase, which can be characterized by the spontaneous magnetization. 
The Fig.~\ref{fig:star_SZ} (a) shows that the magnetization (ordered in the $z$ direction) versus $a$ with $D_c=24$ for TRG method. 
The transition is found to be at $a_{c_2}=3.0243$.

\begin{figure}[ht]
 \includegraphics[width=0.5\textwidth]{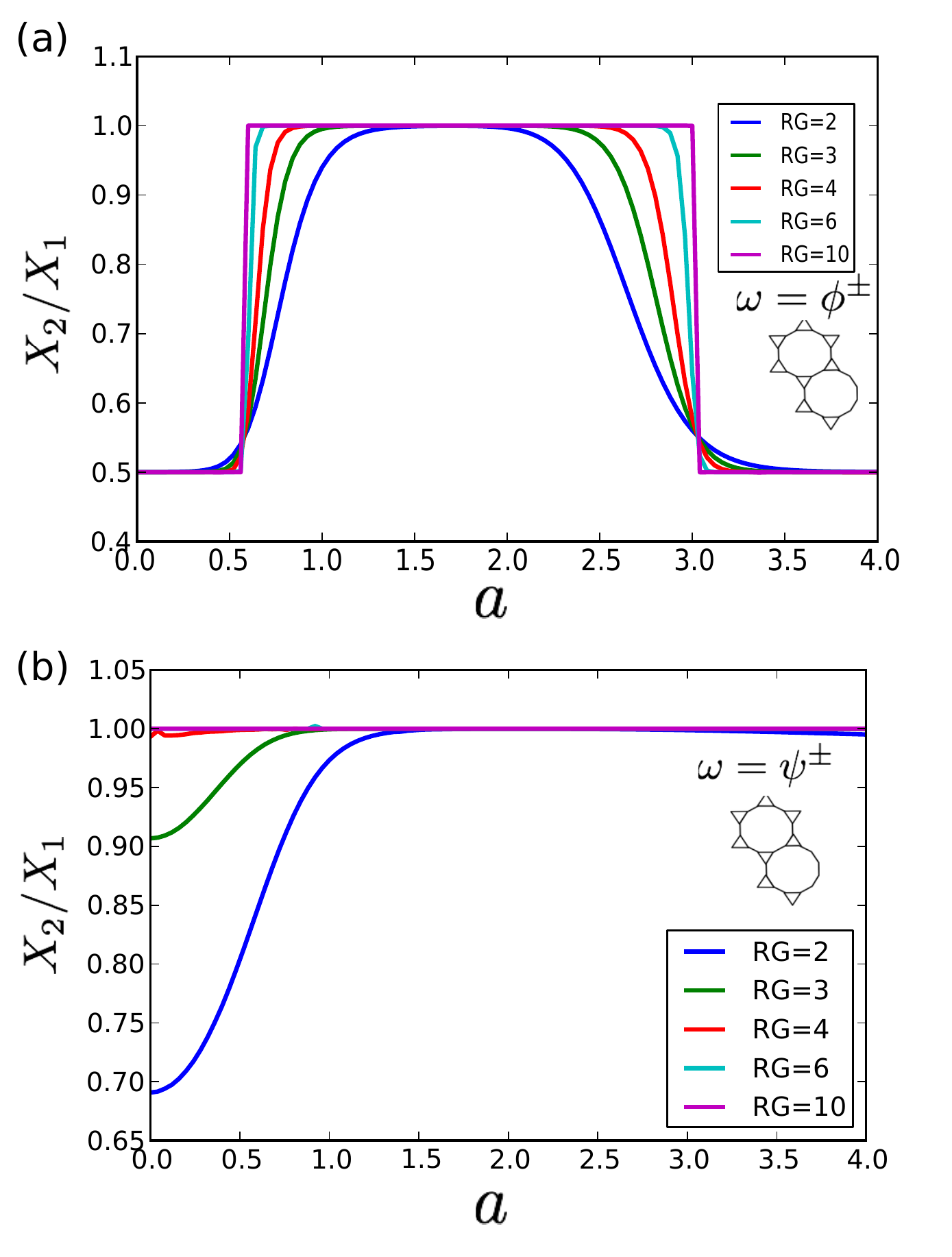}
\caption{The quantity $X_2/X_1$ for tensors  as a function of parameter $a$ under the renormalization flow.  (a) It indicates  phase transitions at $a_{c_1}=0.5850$ and $a_{c_2}=3.0243$ on star lattice with bond state $|\phi^{\pm} \rangle$. The critical exponent $\nu_{a_1} \approx 1.0$ and $\nu_{a_2} \approx 1.0$. (b) The transition do not appear for the deformed AKLT states constructed with   $|\psi^{\pm} \rangle$ bond states.  }
  \label{fig:star_X12}
\end{figure}

\smallskip\noindent {\bf Mapping to an 8-vertex model}. 
By employing the loop (one triangle) diagonal approximation, we merge six sites of two neighboring triangles on the star lattice (see Fig.~\ref{fig:star_lattice}), and arrive at an 8-vertex model on the resultant square lattice,   
\begin{align}
\label{eqn:star8}
&w_1 = \frac{1}{2} ( a^6+ 3a^4 + 15 a^2  + 13 )^2   \notag \\
&w_2 =  w_3 =  w_4 =  \frac{1}{2} ( a^2-1 )^4    ( a^2+1 )^2      \notag \\
&w_5 =w_6 =  \frac{1}{2} (a^4-1)^2(a^4+2a^2+13)      \notag \\
&   w_7 =  w_8=  \frac{1}{2} ( a^2-1 )^4    ( a^2+1 )^2 . 
\end{align} 
It turns out that these Boltzmann weights satisfy the free fermion condition, and thus 
we can locate a transition point (see Appendix \ref{App:freefermion}) at
\begin{align}
a_c = \big(2+\sqrt{3}+ \sqrt{2(6+5\sqrt{3})}   \big)^{\frac{1}{2}} \approx 3.02438.
\end{align} 
This matches the numerically found transition value  $a_{c_2}=3.0243$ from TRG, shown in Fig.~\ref{fig:star_SZ}(b).

\smallskip \noindent {\bf Ferromagnetic phase at small $a$}.
To explore the whole phase diagram, we also compute the Chen-Gu-Wen X-ratio and the results are 
 show in Fig.~\ref{fig:star_X12} (a). 
 We find that in addition to the transition at $a_{c_2}=3.0243$ found by the magnetization, there is a second   transition at $a_{c_1}=0.5850$ from the VBS to a ferromagnetically ordered phase. 
 Again, near transition point, we find the critical exponent $\nu_{a_1} \approx 1.0$ and $\nu_{a_2} \approx 1.0$. 
 We note, however, such a transition does not appear in the deformed AKLT states constructed with   $|\psi^{\pm} \rangle$ bond states as shown in Fig.~\ref{fig:star_X12} (b) .

For $a<1$ regime the diagonal approximation is not valid as the antiparallel configurations cannot be ignored.  We cannot find a classical model for the transition near $a=0.585$.  Instead we confirm the transition by evaluating the spontaneous magnetizations using TRG.
As shown in Fig.~\ref{fig:star_SZ} (b), we see that the ferromagnetic phase is characterized by nonzero x-direction magnetization $\langle S_x \rangle$ that vanishes into the disordered VBS phase for  $|\phi^{+} \rangle$ bond construction. On the other hand, for $|\phi^{-} \rangle$ bond construction, the deformed AKLT states 
display nonzero y-direction magnetization $\langle S_y \rangle$ for small $a$ before they make a transition to the disordered VBS phase, as shown in Fig.~\ref{fig:star_SZ} (c).
The direction in which the magnetization orders in the ferromagnetic phase therefore depends on the bond state. This is consistent with the parallel and anit-parallel correlation of the virtual qubits in the bonds, as seen in Table~\ref{table:bond_state}; in particular,  $|\phi^+\rangle=|0_x0_x\rangle+|1_x0_x\rangle$, whereas $|\phi^-\rangle=|0_y0_y\rangle+|1_y1_y\rangle$.

\section{Deformed AKLT states for universal quantum computation}\label{sec:MBQC}
Here we describe how the various deformed AKLT states from various bond states on square-octagon and cross lattices can be used for universal quantum computation. One key result is that the transition at which the capability of supporting universal quantum computation as $a$ increases seems to coincide with the transition to the ordered phase (either ferromagnetic or antiferromagnetic one, depending on the bond states). However, for the small $a$ regime, we do not know where the boundary is.

The universal quantum computation can be achieved by performing only local measurements on certain entangled states, which are referred to as {\it universal resource states\/}. This measurement-based model of quantum computation exploits entanglement as the resource. The first discovered resource state is the so-called cluster state on the square lattice by Raussendorf and Briegel~\cite{Raussendeorf2001}, and as the entanglement is consumed irreversibly during the computation, it was called the one-way quantum computer.
Generalization of cluster states to graph states~\cite{NestMBQC2006,DanielQC2008,ChenMBQC2010,2DAKLT_wei_2013} and alternative formulation using tensor-network formulation ~\cite{VerstraeteVBC2004,GrossMBQC2007} gives rise to further understanding of the universal measurement-based quantum computation. The key ingredient in establishing the universality for quantum computation requires that (i) the set of universal  gates can be simulated by local measurement and (ii) the effect of the randomness in the measurement outcome can be deal with, e.g., by adjusting later measurement bases, so that  simulation of deterministic quantum circuits can be achieved. However, it is still an open question as to what the complete characterization is of all possible universal resource states.  Exploration of other families of universal resource states and understanding of their enabling physical properties may give further insight.

The hint that the family of AKLT states may provide a useful playground for the exploration comes from the recognition that the 1D spin-1 AKLT chain can be used to simulate any sequence of one-qubit quantum gates~\cite{GrossMBQC2007,BrennenMBQC2008}. 
However, universal measurement-based quantum computation can only be achieved by using states of two-dimensionality or higher, and it was first shown that the spin-3/2 AKLT state on the hexagonal lattice does indeed provide the resource for implementing universal quantum computation in the measurement-based model~\cite{WeiQCR2011,Miyake20111656}.
This has been subsequently extended to other lattices such as the square-octagon and cross lattices~\cite{2Dbeyond_wei_2013}, as well as the spin-2 AKLT state on the square and diamond lattices~\cite{squae_wei_2015}.
The universality of the AKLT state on the hexagonal lattice was extended by Darmawan, Brennen and Bartlett~\cite{Darmawan_2012} for a deformed family, previously constructed by NKZ~\cite{Hexagon_Niggemann}, to an extended region of the phase diagram of the VBS phase up to the transition to the N\'eel ordered phase. Given the construction of the (deformed) AKLT states with various bond states and on various lattices, it is natural to inquire whether the picture holds in these other cases. 

\subsection{Procedure for showing universality}
The approach we take here to show universality for various AKLT states is based on that in Ref.~\cite{WeiQCR2011} i.e., by first constructing a local generalized measurement on all sites, usually called POVM, that reduces the AKLT states to graph states, and then showing that the associated graphs reside in the supercritical phase of percolation (as the system size increases). The appropriate POVM was constructed by Darmawan, Brennen and Bartlett to apply to the deformed AKLT states on the hexagonal lattice~\cite{Darmawan_2012}. It turns out this also applies to other trivalent lattices. The POVM consists of three elements $\{F_x(a)^\dagger F_x(a),F_y(a)^\dagger F_y(a),F_z(a)^\dagger F_z(a)\}$ and each $F_\alpha(a)\equiv_\alpha(a) \tilde{F}_\alpha D^{-1}(a)$ is associated with the measurement outcome labeled by $\alpha=x, y,z$ and is the operator that acts on the local spin that is measured, where   $q_x(a)=q_y(a)=1$, $q_z(a)=\sqrt{(a^2-1)/2}$, and
\begin{subequations}
\label{POVM2}
  \begin{eqnarray}
\tilde{F}_{x}&=&\sqrt{\frac{2}{3}}(\ket{3/2}_x\bra{3/2}+\ket{-3/2}_x\bra{-3/2}) \\
\tilde{F}_{y}&=&\sqrt{\frac{2}{3}}(\ket{3/2}_y\bra{3/2}+\ket{-3/2}_y\bra{-3/2})\\
\tilde{F}_{z}&=&\sqrt{\frac{2}{3}}(\ket{3/2}_z\bra{3/2}+\ket{-3/2}_z\bra{-3/2}),
\end{eqnarray}
\end{subequations}
where $\ket{\pm 3/2}_\alpha$'s are eigenstates of the spin operator $S_\alpha$ with eigenvalues $\pm 3/2$ (setting $\hbar=1$). 
Since the AKLT states are entangled, the outcomes $\{\alpha_v\}$ at all sites $v$'s are not independent, but Monte Carlo simulations can be used to sample them from the exact distribution~\cite{WeiQCR2011,Darmawan_2012,2Dbeyond_wei_2013}. The key ingredient in the sampling of the outcome configurations $\{\alpha_v\}$ is the acceptance probability from changing from a set $\sigma\equiv\{\alpha_v\}$ to another one $\sigma'\equiv\{\alpha_v'\}$: $p_{\rm acc}(\sigma\rightarrow \sigma') = \min[
1, r(a)]$, with $r(a)$ to be defined below in Eq.~(\ref{eqn:r}).

But to describe this probability we need to introduce an important concept: {\it domains\/}. 
A domain is a set of neighboring sites that contain the same type of outcome $\alpha=x,y,$ or $z$. 
An $\alpha$-domain can be regarded as a result of site percolation according to a given configuration $\{\alpha_v\}$, randomly generated from the POVM outcomes. 
From the result of Ref.~\cite{WeiQCR2011}, the meaning of a domain is an effective qubit, consisting of possibly many sites. 
As is necessary for carrying universal quantum computation, the number of domains should be macroscopic, i.e., proportional to total number of original sites. 
Otherwise, there would not be sufficient number of qubits that can be used. 
Moreover, the POVM on all sites projects the (deformed) AKLT states to a graph state, whose qubits are the domains. Whether the resulting graph state is useful for universal quantum computation depends entirely on its graph properties, such as connectivity and number of domains~\cite{DanielQC2008,2DAKLT_wei_2013}. Two important quantities are $|V|$ and $|E|$. $V$ denotes the set of all domains and $|V|$ their total number, and $E$ denotes the set of all inter-domain edges and $|E|$ their total number. With these notions defined, we can describe the ratio $r(a)$,
\begin{equation}
\label{eqn:r}
r(a) =\left(\frac{a^2- 1}{2}\right)^{N_z'-N_{z}}2^{|V'|-|E'|-|V|+|E|},
\end{equation}
where $N_z$ is the total number of $z$-type domain from configuration $\sigma$ and $|V|$ and $|E|$ are the total numbers of domains and inter-domain edges, respectively, associated with $\sigma$; $N_z'$, $|V'|$ and $|E'|$ are the corresponding quantities associated with configuration $\sigma'$ that the Metropolis sampling attemps to flip to from $\sigma$. The above probability ratio was derived in Ref.~\cite{Darmawan_2012}, which reduces to the original AKLT case at $a=\sqrt{3}$, derived in Ref.~\cite{WeiQCR2011}.

We note that the above POVM is valid only for $a\ge 1$; for $a<1$, a different set of POVM can be used, but it is not known how to simulate the exact distribution. Namely, no corresponding expression of $r(a)$ is known. Therefore, our discussions below will be restricted to $a\ge 1$.

Let us summarize the key criteria to check for universality.
(i) First, the domain size always has to be microscopic, i.e. it can be at most logarithmic in the system size. More
precisely, the maximum size of a domain should at most depend logarithmically on the total
number of lattice sites. This ensures that we can realize arbitrary numbers of qubits in the
graph state. As we shall see below that, close to the transition between VBS and the ordered phases, the domain size increases and become macroscopic beyond the transition. (ii)  Second, we have to make sure that the graph formed from the domains are percolated. This means
we can always find paths of connected qubits from one boundary to another if the lattice is
large. This ensures that we have enough quantum wires to perform quantum computation.
The fact that the post-measurement state is a graph state with its graph residing in the supercritical phase of percolation ensures that the state possesses
sufficient entanglement. These two criteria involve two different percolation objects: the first being the domain itself and the second being the cluster formed by domains. The quantum computational universality requires the former to be microscopic but the latter to be macroscopic.

We remark that even though the above discussions, strictly speaking, assume that  the deformed AKLT states are constructed from the singlet bond $|\psi^-\rangle$, as we shall be mostly concerned with the bi-partite lattices and as AKLT states of different bond states can be transformed to one another by local unitary transformation, the argument applies to these other bond constructions as well. The unitaries apply to sites of one sublattice can transform the deformed AKLT with $|\psi^-\rangle$ bond to those with other bonds: $|\psi^+\rangle$, $|\phi^+\rangle$, $|\phi^-\rangle$, respectively, and these unitaries are, respectively,
\begin{align}
& U_z= 
 \begin{pmatrix}
  1&0 &0&0\\
  0&-1 &0&0\\
  0&0 &1&0\\
  0&0 &0&-1\\
 \end{pmatrix}, \\ 
& U_y= 
 \begin{pmatrix}
  0&0 &0&i\\
  0&0 &-i&0\\
  0&i &0&0\\
  -i&0 &0&0\\
 \end{pmatrix}, \\
& U_x= 
 \begin{pmatrix}
  0&0 &0&1\\
  0&0 &1&0\\
  0&1 &0&0\\
  1&0 &0&0\\
 \end{pmatrix}.
 \end{align}
 Furthermore, for non-bipartite lattices, even though AKLT states of different bond states cannot be locally transformed to one another, the  argument of graph states and their Monte Carlo sampling can be supplemented by other constraints resulting from geometric frustration, e.g. see Ref.~\cite{2Dbeyond_wei_2013} for the AKLT state with $|\psi^-\rangle$ on the star lattice.

\subsection{Square-octagon lattice}
\begin{figure}[t]
  \includegraphics[width=0.5\textwidth] {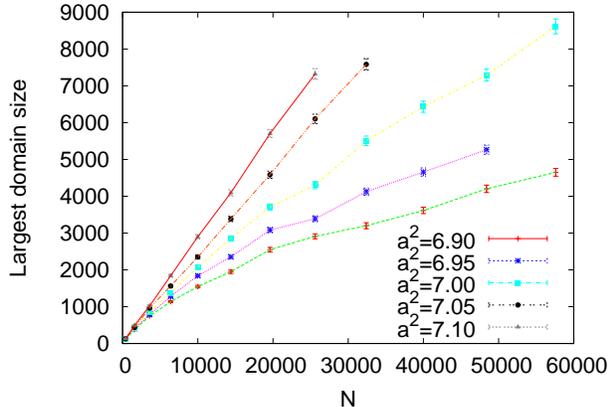}
\caption{The size of the largest domain vs. the system size. For $a^2\gtrsim 7.05$ the largest domain size is linear in $N$, the total number of sites, whereas for $a^2\lesssim 7.05$ it is logarithmic in $N$, i.e., proportional to  $\ln N$.}
  \label{fig:largestdomain}
\end{figure}

\begin{figure}[t]
  \includegraphics[width=0.5\textwidth] {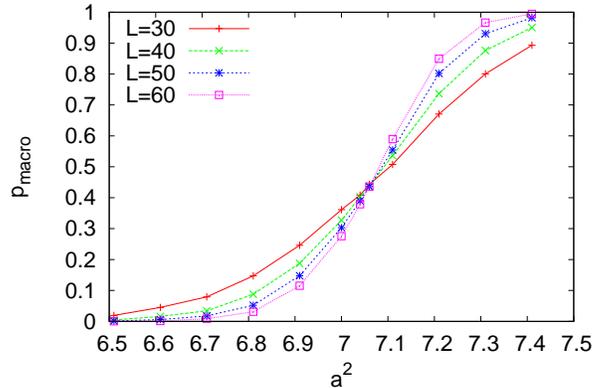}
\caption{The probability $p_{\rm macro}$ of having a macroscopic-size domain vs. the parameter $a^2$ on the square-octagon lattice with $N=4L^2$ sites. The crossing occurs approximately at $a^2\approx 7.06$, which signifies a phase transition.}
  \label{fig:largestdomainSO}
\end{figure}

First we present the largest domain size vs. the total number $N=4L^2$ of sites in Fig.~\ref{fig:largestdomain}. 
This confirms that the maximum size of a domain remains microscopic for 
$1\le a^2 \lesssim 7.05$ of the deformation parameter, verifying the  condition (i). (But to determine more precisely the transition; see Fig.~\ref{fig:largestdomainSO} and discussions below.) We also check the condition (ii) that the graphs of domains resulting from the POVM in this parameter region do form a spanning cluster with probability close to unity in this parameter regime, even for a modest size of $L=20$. To locate more precisely the transition point where the universal computational power vanishes, we examine the probability of finding a macroscopic domain via its percolation property, i.e., an alternative quantification of condition (i). In Fig.~\ref{fig:largestdomainSO}. we show the probability $p_{\rm macro}$ that there exists a domain that spans macroscopically, i.e., reaching from one boundary to
another versus the deformation parameter $a$ for various linear sizes $L$. This confirms that
there exists a phase transition, and that the transition point extracted from the percolation $a_c^2\approx 7.06(1)$ roughly agrees with the transition point
obtained from the magnetization, i.e.
$a_{c2}^2=(2.6547)^2\approx 7.0474$. The formation of a macroscopic domain is consistent with the formation of ferrormagnetic or antiferromatic order.

Our results here extend previous findings~\cite{WeiQCR2011,Darmawan_2012,2Dbeyond_wei_2013} to a wide region of the disordered VBS phase on the square-octagon lattice (for all four bond-state constructions).

\subsection{Cross lattice}
\begin{figure}[t]
  \includegraphics[width=0.5\textwidth] {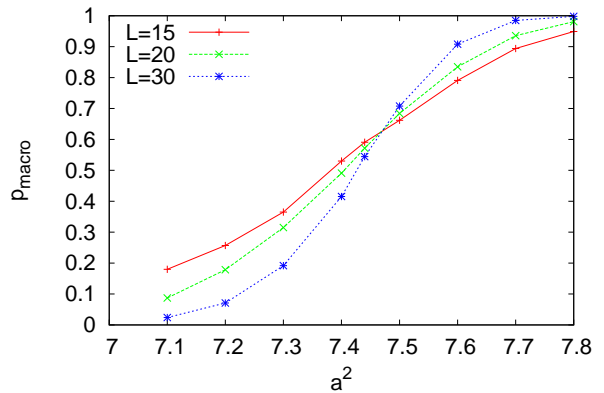}
\caption{The probability $p_{\rm macro}$ of having a macroscopic-size domain vs. the parameter $a^2$ on the cross lattice with $N=24L^2$ sites. The crossing occurs approximately at $a^2\approx 7.47$, which signifies a phase transition.}
  \label{fig:largestdomainCross}
\end{figure}
The  deformed AKLT states on the cross lattice have a similar behavior on their utility as a universal resource for MBQC. We find that in the range $1\le a <a_c$ they are universal resource states. But as $a$ approaches $a_c$ from below, the size of an effective qubit grows and becomes macroscopic at and above the transition. The largest domain size vs. the total number $N=24L^2$ of sites is shown in Fig.~\ref{fig:largestdomainCross} for different  linear sizes $L$'s.  The phase transition occurs at $a_c^2\approx 7.47(2)$, consistent with the VBS-N\'eel transition. We have also checked that for $1\le a < a_c$, the domain size remains microscopic and the graphs formed by the domains do reside in the supercritical phase of the percolation.
Thus the family of the deformed AKLT states on the cross lattice also can be used for universal MBQC for this wide range of parameter $a$ in the VBS phase, similar to that on the square-octagon lattice.

\subsection{Star lattice}

For the star lattice, the two deformed families of AKLT states with $|\psi^\pm\rangle$ are not universal, at least for $a\ge 1$. For the $|\phi^\pm\rangle$, the deformed AKLT states become ferromagnetically ordered for $a>a_c$, similar to the cases on the honeycomb, square octagon and the cross. Do we expect that the two families on the star lattice might turn out be universal in the regime $a<a_c$?  Indeed for $a$ smaller but close to $a_c$, we find that the graphs of the corresponding graph states obtained after the POVM do, with probably close to one in the large system size limit, possess a spanning cluster. However, these are graphs are tree-like, possessing a very small fraction of local loops. They are therefore not universal for quantum computation, but can be used for simulating independent qubit rotations with relatively small number of entangling gates.

\section{Conclusion} \label{sec:conclude} 

\begin{table}[h!]
\centering
\begin{tabularx} { 0.45\textwidth}{ l c c c c } 
\hline  \hline 
lattice & base &     $a_{c_1}$ &    $a_{c_2}$     &  $a_{c_2}$(solvable) \\  \hline 
Hexagon & $ \; \phi^{\pm},\psi^{\pm} \;$   & 0.421  &  2.5425 &   2.54246\\  
%
Square-octagon & $ \; \phi^{\pm},\psi^{\pm} \;$& $\times$   &  2.6547   & 2.65158  \\  
%
Cross & $\phi^{\pm},\psi^{\pm}$&  $\times$  &  2.7280  & $\times$   \\  %
Star & $\phi^{\pm}$& 0.580    &  3.0243  &   3.02438 \\  
Star & $\psi^{\pm}$&   $\times$ &  $\times$ & $\times$     \\  \hline  \hline
\end{tabularx}
 \caption{ The  summarizing phase diagrams of all cases. The phase transition from a VBS to a ordered phase occurs at $a_{c_2}$.  The last column shows the critical values obtained from mapping to exactly solvable classical models. }
 \label{table:CPsummary}
\end{table}

We investigate deformed AKLT states constructing from various bond states and on various trivalent lattices. 
The summary of phase diagram is shown in Table~\ref{table:CPsummary}. 
For the hexagonal lattice we find that there appears to be a Berezinskii-Kosterlitz-Thouless transition from the VBS phase to an quantum XY phase in the small $a$ parameter regime. 
Such transition also occurs in the deformation of other types of AKLT states with triplet-bond constructions on the hexagonal lattice.
However, we do not find such an XY phase in the deformed spin-3/2 AKLT models on other trivalent lattices such as square-octagon, cross and star lattices. 

When the parameter $a$ is sufficiently large, there is a transition from a VBS phase to a magnetically ordered phase for the deformed AKLT families on the honeycomb, square octagon and cross lattice.  Whether the ordered phase is ferromagnetic or antiferromagnetic depends on the bond states used in constructing the AKLT families.  Such a transition can be obtained approximately yet fairly accurately by mapping to a solvable classical eight-vertex model, as was done by Niggemann, Kl\"umper and Zittartz~\cite{Hexagon_Niggemann} in the hexaongal case and by Niggemann and Zittartz~\cite{SO_Niggemann} in the square-octagon case. 
We have obtained a slight improvement on the transition point in the latter case with a modified eight-vertex model, albeit it is not exactly solvable. We have verified these various cases by using numerical tensor-network methods. 

For the star lattice, the phase diagram depends on the bond states.
When constructed using singlet bonds ($\psi^-$) and the other antiparallel bonds ($\psi^+$),  the deformed family of AKLT states  remains in the same phase as the respective AKLT state (with respective $\psi^-$ or $\psi^+$ bond) throughout the whole region of the parameter. The mapping to an eight-vertex model is solvable but no transition is found.
However, for the two triplet-bond constructions, the triplet VBS phase is sandwiched between  two ferromagnetic phases (for large and small parameters, respectively), which are characterized by spontaneous magnetizations along different axes. The VBS-FM transition at large $a$ parameter regime can be obtained fairly accurately by a similar mapping to a solvable eight-vertex model, agreeing well with results from tensor-network methods.

The evidence that we found supporting the emergence of the XY phase is the essential singularity in the correlation length.  The exponent, however, is not identical to that of the BKT transition in the classical XY model, where the tuning parameter is the temperature. The parameter $a$ in our case appears in the deformation of the AKLT wave function and the parent Hamiltonian but not linearly. Moreover, our system is composed of spins with a magntidue 3/2, not the spin-1/2 case
studied previously nor the classical case. In this phase we have checked that the correlation functions decay in a power-law fashion. We also study the magnetic properties in the XY phase. There is no spontaneous magnetization. The induced magnetization has a magnitude that is independent of the field direction, but 
 its magnitude does depend on the magnitude of the field. 
This is also consistent with the XY phase. 
However, a much stronger evidence would be obtained from the so-called spin stiffness or helicity modulus that is the response of the ground-state energy under the change of boundary condition. 
This is usually accessible in Monte Carlo simulations, but as to how to calculate it in the tensor-network framework, we leave it as a future investigation.  Our results also show that as opposed to the hexagonal lattice, on either the square-octagon or cross lattice, there is no such an XY phase for the families of deformed AKLT states. At present we do not have a clear understanding, but we suspect that it may be due to multiple loop structure in these other lattices. However, we also leave this for future investigation. 

The transitions from various VBS phases to the corresponding ordered phases can be characterized by disorder-order transition in classical 8-vertex and Ising models. The associated classical order parameter matches very well the expectation value of the effective spin operator  $\tilde{S}_\alpha$ (in the $\alpha$-basis where the spin orders),
\begin{align}
\tilde{S}_\alpha= 
 \begin{pmatrix}
  1&0 &0&0\\
  0&1 &0&0\\
  0&0 &-1&0\\
  0&0 &0&-1\\
 \end{pmatrix},
 \end{align}
 evaluated with the deformed AKLT states under the same diagonal approximation that leads to the classical model. However, we do not have an analytic proof and list it as a conjecture.

We have also investigated the deformed AKLT states for universal quantum computation.  On the square-octagon and cross lattices, they are useful at least for the deformation parameter $a\ge 1$ in the VBS phase up to the transition to the ordered phase, similar to the hexagonal case~\cite{Darmawan_2012}.
The loss of the capability for universal quantum computation for various deformed AKLT states discussed in this paper is due to the growth of the size of an effective qubit to a macroscopic size, and this is consistent with the percolation of ferromagnetic or antiferromagnetic domains at the transition. For the deformed AKLT states on the star lattice, the effect of frustration prevents them from being universal for quantum computation, regardless of the internal bond states.

\section*{Acknowledgements}

The authors would like to thank Ying-Jer Kao for useful discussions.
This work was supported by the National Science Foundation under
Grants No. PHY 1314748 and No. PHY 1333903.


\begin{thebibliography}{53}%
\makeatletter
\providecommand \@ifxundefined [1]{%
 \@ifx{#1\undefined}
}%
\providecommand \@ifnum [1]{%
 \ifnum #1\expandafter \@firstoftwo
 \else \expandafter \@secondoftwo
 \fi
}%
\providecommand \@ifx [1]{%
 \ifx #1\expandafter \@firstoftwo
 \else \expandafter \@secondoftwo
 \fi
}%
\providecommand \natexlab [1]{#1}%
\providecommand \enquote  [1]{``#1''}%
\providecommand \bibnamefont  [1]{#1}%
\providecommand \bibfnamefont [1]{#1}%
\providecommand \citenamefont [1]{#1}%
\providecommand \href@noop [0]{\@secondoftwo}%
\providecommand \href [0]{\begingroup \@sanitize@url \@href}%
\providecommand \@href[1]{\@@startlink{#1}\@@href}%
\providecommand \@@href[1]{\endgroup#1\@@endlink}%
\providecommand \@sanitize@url [0]{\catcode `\\12\catcode `\$12\catcode
  `\&12\catcode `\#12\catcode `\^12\catcode `\_12\catcode `\%12\relax}%
\providecommand \@@startlink[1]{}%
\providecommand \@@endlink[0]{}%
\providecommand \url  [0]{\begingroup\@sanitize@url \@url }%
\providecommand \@url [1]{\endgroup\@href {#1}{\urlprefix }}%
\providecommand \urlprefix  [0]{URL }%
\providecommand \Eprint [0]{\href }%
\providecommand \doibase [0]{http://dx.doi.org/}%
\providecommand \selectlanguage [0]{\@gobble}%
\providecommand \bibinfo  [0]{\@secondoftwo}%
\providecommand \bibfield  [0]{\@secondoftwo}%
\providecommand \translation [1]{[#1]}%
\providecommand \BibitemOpen [0]{}%
\providecommand \bibitemStop [0]{}%
\providecommand \bibitemNoStop [0]{.\EOS\space}%
\providecommand \EOS [0]{\spacefactor3000\relax}%
\providecommand \BibitemShut  [1]{\csname bibitem#1\endcsname}%
\let\auto@bib@innerbib\@empty
\bibitem [{\citenamefont {Auerbach}(2012)}]{auerbach2012interacting}%
  \BibitemOpen
  \bibfield  {author} {\bibinfo {author} {\bibfnamefont {A.}~\bibnamefont
  {Auerbach}},\ }\href {https://books.google.com/books?id=d-sHCAAAQBAJ} {\emph
  {\bibinfo {title} {Interacting Electrons and Quantum Magnetism}}},\ Graduate
  Texts in Contemporary Physics\ (\bibinfo  {publisher} {Springer New York},\
  \bibinfo {year} {2012})\BibitemShut {NoStop}%
\bibitem [{\citenamefont {Bethe}(1931)}]{Bethe1931}%
  \BibitemOpen
  \bibfield  {author} {\bibinfo {author} {\bibfnamefont {H.}~\bibnamefont
  {Bethe}},\ }\href {\doibase 10.1007/BF01341708} {\bibfield  {journal}
  {\bibinfo  {journal} {Zeitschrift f{\"u}r Physik}\ }\textbf {\bibinfo
  {volume} {71}},\ \bibinfo {pages} {205} (\bibinfo {year} {1931})}\BibitemShut
  {NoStop}%
\bibitem [{\citenamefont {Sachdev}(2001)}]{sachdev2001quantum}%
  \BibitemOpen
  \bibfield  {author} {\bibinfo {author} {\bibfnamefont {S.}~\bibnamefont
  {Sachdev}},\ }\href {https://books.google.com/books?id=Ih\_E05N5TZQC} {\emph
  {\bibinfo {title} {Quantum Phase Transitions}}}\ (\bibinfo  {publisher}
  {Cambridge University Press},\ \bibinfo {year} {2001})\BibitemShut {NoStop}%
\bibitem [{\citenamefont {Balents}(2010)}]{Balents2010}%
  \BibitemOpen
  \bibfield  {author} {\bibinfo {author} {\bibfnamefont {L.}~\bibnamefont
  {Balents}},\ }\href {\doibase doi:10.1038/nature08917} {\bibfield  {journal}
  {\bibinfo  {journal} {Nature}\ }\textbf {\bibinfo {volume} {464}},\ \bibinfo
  {pages} {199} (\bibinfo {year} {2010})}\BibitemShut {NoStop}%
\bibitem [{\citenamefont {Wen}(1990)}]{Wen_TO1990}%
  \BibitemOpen
  \bibfield  {author} {\bibinfo {author} {\bibfnamefont {X.~G.}\ \bibnamefont
  {Wen}},\ }\href {\doibase 10.1142/S0217979290000139} {\bibfield  {journal}
  {\bibinfo  {journal} {International Journal of Modern Physics B}\ }\textbf
  {\bibinfo {volume} {04}},\ \bibinfo {pages} {239} (\bibinfo {year}
  {1990})}\BibitemShut {NoStop}%
\bibitem [{\citenamefont {Kitaev}(2003)}]{Kitaev20032}%
  \BibitemOpen
  \bibfield  {author} {\bibinfo {author} {\bibfnamefont {A.}~\bibnamefont
  {Kitaev}},\ }\href {\doibase http://dx.doi.org/10.1016/S0003-4916(02)00018-0}
  {\bibfield  {journal} {\bibinfo  {journal} {Annals of Physics}\ }\textbf
  {\bibinfo {volume} {303}},\ \bibinfo {pages} {2 } (\bibinfo {year}
  {2003})}\BibitemShut {NoStop}%
\bibitem [{\citenamefont {Yan}\ \emph {et~al.}(2011)\citenamefont {Yan},
  \citenamefont {Huse},\ and\ \citenamefont {White}}]{Yan2011}%
  \BibitemOpen
  \bibfield  {author} {\bibinfo {author} {\bibfnamefont {S.}~\bibnamefont
  {Yan}}, \bibinfo {author} {\bibfnamefont {D.~A.}\ \bibnamefont {Huse}}, \
  and\ \bibinfo {author} {\bibfnamefont {S.~R.}\ \bibnamefont {White}},\ }\href
  {\doibase 10.1126/science.1201080} {\ \textbf {\bibinfo {volume} {332}},\
  \bibinfo {pages} {1173} (\bibinfo {year} {2011})}\BibitemShut {NoStop}%
\bibitem [{\citenamefont {Depenbrock}\ \emph {et~al.}(2012)\citenamefont
  {Depenbrock}, \citenamefont {McCulloch},\ and\ \citenamefont
  {Schollw\"ock}}]{Stefan_SL2012}%
  \BibitemOpen
  \bibfield  {author} {\bibinfo {author} {\bibfnamefont {S.}~\bibnamefont
  {Depenbrock}}, \bibinfo {author} {\bibfnamefont {I.~P.}\ \bibnamefont
  {McCulloch}}, \ and\ \bibinfo {author} {\bibfnamefont {U.}~\bibnamefont
  {Schollw\"ock}},\ }\href {\doibase 10.1103/PhysRevLett.109.067201} {\bibfield
   {journal} {\bibinfo  {journal} {Phys. Rev. Lett.}\ }\textbf {\bibinfo
  {volume} {109}},\ \bibinfo {pages} {067201} (\bibinfo {year}
  {2012})}\BibitemShut {NoStop}%
\bibitem [{\citenamefont {Affleck}\ \emph {et~al.}()\citenamefont {Affleck},
  \citenamefont {Kennedy}, \citenamefont {Lieb},\ and\ \citenamefont
  {Tasaki}}]{AKLT_1988}%
  \BibitemOpen
  \bibfield  {author} {\bibinfo {author} {\bibfnamefont {I.}~\bibnamefont
  {Affleck}}, \bibinfo {author} {\bibfnamefont {T.}~\bibnamefont {Kennedy}},
  \bibinfo {author} {\bibfnamefont {E.~H.}\ \bibnamefont {Lieb}}, \ and\
  \bibinfo {author} {\bibfnamefont {H.}~\bibnamefont {Tasaki}},\ }\href
  {\doibase 10.1007/BF01218021} {\bibfield  {journal} {\bibinfo  {journal}
  {Communications in Mathematical Physics}\ }\textbf {\bibinfo {volume}
  {115}},\ \bibinfo {pages} {477}}\BibitemShut {NoStop}%
\bibitem [{\citenamefont {Pollmann}\ \emph {et~al.}(2012)\citenamefont
  {Pollmann}, \citenamefont {Berg}, \citenamefont {Turner},\ and\ \citenamefont
  {Oshikawa}}]{Pollmann2012}%
  \BibitemOpen
  \bibfield  {author} {\bibinfo {author} {\bibfnamefont {F.}~\bibnamefont
  {Pollmann}}, \bibinfo {author} {\bibfnamefont {E.}~\bibnamefont {Berg}},
  \bibinfo {author} {\bibfnamefont {A.~M.}\ \bibnamefont {Turner}}, \ and\
  \bibinfo {author} {\bibfnamefont {M.}~\bibnamefont {Oshikawa}},\ }\href
  {\doibase 10.1103/PhysRevB.85.075125} {\bibfield  {journal} {\bibinfo
  {journal} {Phys. Rev. B}\ }\textbf {\bibinfo {volume} {85}},\ \bibinfo
  {pages} {075125} (\bibinfo {year} {2012})}\BibitemShut {NoStop}%
\bibitem [{\citenamefont {Chen}\ \emph {et~al.}(2013)\citenamefont {Chen},
  \citenamefont {Gu}, \citenamefont {Liu},\ and\ \citenamefont
  {Wen}}]{ChenSPT2013}%
  \BibitemOpen
  \bibfield  {author} {\bibinfo {author} {\bibfnamefont {X.}~\bibnamefont
  {Chen}}, \bibinfo {author} {\bibfnamefont {Z.-C.}\ \bibnamefont {Gu}},
  \bibinfo {author} {\bibfnamefont {Z.-X.}\ \bibnamefont {Liu}}, \ and\
  \bibinfo {author} {\bibfnamefont {X.-G.}\ \bibnamefont {Wen}},\ }\href
  {\doibase 10.1103/PhysRevB.87.155114} {\bibfield  {journal} {\bibinfo
  {journal} {Phys. Rev. B}\ }\textbf {\bibinfo {volume} {87}},\ \bibinfo
  {pages} {155114} (\bibinfo {year} {2013})}\BibitemShut {NoStop}%
\bibitem [{\citenamefont {Chen}\ \emph {et~al.}(2012)\citenamefont {Chen},
  \citenamefont {Gu}, \citenamefont {Liu},\ and\ \citenamefont
  {Wen}}]{Chen1604}%
  \BibitemOpen
  \bibfield  {author} {\bibinfo {author} {\bibfnamefont {X.}~\bibnamefont
  {Chen}}, \bibinfo {author} {\bibfnamefont {Z.-C.}\ \bibnamefont {Gu}},
  \bibinfo {author} {\bibfnamefont {Z.-X.}\ \bibnamefont {Liu}}, \ and\
  \bibinfo {author} {\bibfnamefont {X.-G.}\ \bibnamefont {Wen}},\ }\href
  {\doibase 10.1126/science.1227224} {\ \textbf {\bibinfo {volume} {338}},\
  \bibinfo {pages} {1604} (\bibinfo {year} {2012})}\BibitemShut {NoStop}%
\bibitem [{\citenamefont {Wei}\ \emph {et~al.}(2011)\citenamefont {Wei},
  \citenamefont {Affleck},\ and\ \citenamefont {Raussendorf}}]{WeiQCR2011}%
  \BibitemOpen
  \bibfield  {author} {\bibinfo {author} {\bibfnamefont {T.-C.}\ \bibnamefont
  {Wei}}, \bibinfo {author} {\bibfnamefont {I.}~\bibnamefont {Affleck}}, \ and\
  \bibinfo {author} {\bibfnamefont {R.}~\bibnamefont {Raussendorf}},\ }\href
  {\doibase 10.1103/PhysRevLett.106.070501} {\bibfield  {journal} {\bibinfo
  {journal} {Phys. Rev. Lett.}\ }\textbf {\bibinfo {volume} {106}},\ \bibinfo
  {pages} {070501} (\bibinfo {year} {2011})}\BibitemShut {NoStop}%
\bibitem [{\citenamefont {Miyake}(2011)}]{Miyake20111656}%
  \BibitemOpen
  \bibfield  {author} {\bibinfo {author} {\bibfnamefont {A.}~\bibnamefont
  {Miyake}},\ }\href {\doibase http://dx.doi.org/10.1016/j.aop.2011.03.006}
  {\bibfield  {journal} {\bibinfo  {journal} {Annals of Physics}\ }\textbf
  {\bibinfo {volume} {326}},\ \bibinfo {pages} {1656 } (\bibinfo {year}
  {2011})},\ \bibinfo {note} {july 2011 Special Issue}\BibitemShut {NoStop}%
\bibitem [{\citenamefont {Wei}(2013)}]{2Dbeyond_wei_2013}%
  \BibitemOpen
  \bibfield  {author} {\bibinfo {author} {\bibfnamefont {T.-C.}\ \bibnamefont
  {Wei}},\ }\href {\doibase 10.1103/PhysRevA.88.062307} {\bibfield  {journal}
  {\bibinfo  {journal} {Phys. Rev. A}\ }\textbf {\bibinfo {volume} {88}},\
  \bibinfo {pages} {062307} (\bibinfo {year} {2013})}\BibitemShut {NoStop}%
\bibitem [{\citenamefont {Raussendorf}\ and\ \citenamefont
  {Briegel}(2001)}]{Raussendeorf2001}%
  \BibitemOpen
  \bibfield  {author} {\bibinfo {author} {\bibfnamefont {R.}~\bibnamefont
  {Raussendorf}}\ and\ \bibinfo {author} {\bibfnamefont {H.~J.}\ \bibnamefont
  {Briegel}},\ }\href {\doibase 10.1103/PhysRevLett.86.5188} {\bibfield
  {journal} {\bibinfo  {journal} {Phys. Rev. Lett.}\ }\textbf {\bibinfo
  {volume} {86}},\ \bibinfo {pages} {5188} (\bibinfo {year}
  {2001})}\BibitemShut {NoStop}%
\bibitem [{\citenamefont {Niggemann}\ \emph {et~al.}(1997)\citenamefont
  {Niggemann}, \citenamefont {Kl\"umper},\ and\ \citenamefont
  {Zittartz}}]{Hexagon_Niggemann}%
  \BibitemOpen
  \bibfield  {author} {\bibinfo {author} {\bibfnamefont {H.}~\bibnamefont
  {Niggemann}}, \bibinfo {author} {\bibfnamefont {A.}~\bibnamefont
  {Kl\"umper}}, \ and\ \bibinfo {author} {\bibfnamefont {J.}~\bibnamefont
  {Zittartz}},\ }\href {\doibase 10.1007/s002570050425} {\bibfield  {journal}
  {\bibinfo  {journal} {Zeitschrift f\"ur Physik B Condensed Matter}\ }\textbf
  {\bibinfo {volume} {104}},\ \bibinfo {pages} {103} (\bibinfo {year}
  {1997})}\BibitemShut {NoStop}%
\bibitem [{\citenamefont {Darmawan}\ \emph {et~al.}(2012)\citenamefont
  {Darmawan}, \citenamefont {Brennen},\ and\ \citenamefont
  {Bartlett}}]{Darmawan_2012}%
  \BibitemOpen
  \bibfield  {author} {\bibinfo {author} {\bibfnamefont {A.~S.}\ \bibnamefont
  {Darmawan}}, \bibinfo {author} {\bibfnamefont {G.~K.}\ \bibnamefont
  {Brennen}}, \ and\ \bibinfo {author} {\bibfnamefont {S.~D.}\ \bibnamefont
  {Bartlett}},\ }\href {http://stacks.iop.org/1367-2630/14/i=1/a=013023}
  {\bibfield  {journal} {\bibinfo  {journal} {New Journal of Physics}\ }\textbf
  {\bibinfo {volume} {14}},\ \bibinfo {pages} {013023} (\bibinfo {year}
  {2012})}\BibitemShut {NoStop}%
\bibitem [{\citenamefont {Verstraete}\ and\ \citenamefont
  {Cirac}(2004{\natexlab{a}})}]{PEPS_2004}%
  \BibitemOpen
  \bibfield  {author} {\bibinfo {author} {\bibfnamefont {F.}~\bibnamefont
  {Verstraete}}\ and\ \bibinfo {author} {\bibfnamefont {J.~I.}\ \bibnamefont
  {Cirac}},\ }\href@noop {} {\bibfield  {journal} {\bibinfo  {journal}
  {arXiv:cond-mat/0407066}\ } (\bibinfo {year}
  {2004}{\natexlab{a}})}\BibitemShut {NoStop}%
\bibitem [{\citenamefont {Levin}\ and\ \citenamefont
  {Nave}(2007)}]{Levin_TRG2007}%
  \BibitemOpen
  \bibfield  {author} {\bibinfo {author} {\bibfnamefont {M.}~\bibnamefont
  {Levin}}\ and\ \bibinfo {author} {\bibfnamefont {C.~P.}\ \bibnamefont
  {Nave}},\ }\href {\doibase 10.1103/PhysRevLett.99.120601} {\bibfield
  {journal} {\bibinfo  {journal} {Phys. Rev. Lett.}\ }\textbf {\bibinfo
  {volume} {99}},\ \bibinfo {pages} {120601} (\bibinfo {year}
  {2007})}\BibitemShut {NoStop}%
\bibitem [{\citenamefont {McCoy}(2011)}]{mccoy2011advanced}%
  \BibitemOpen
  \bibfield  {author} {\bibinfo {author} {\bibfnamefont {B.}~\bibnamefont
  {McCoy}},\ }\href {https://books.google.com/books?id=0u9ZrgEACAAJ} {\emph
  {\bibinfo {title} {Advanced Statistical Mechanics}}},\ International Series
  of Monographs on Physics\ (\bibinfo  {publisher} {Oxford University Press},\
  \bibinfo {year} {2011})\BibitemShut {NoStop}%
\bibitem [{\citenamefont {Baxter}(2013)}]{baxter2013exactly}%
  \BibitemOpen
  \bibfield  {author} {\bibinfo {author} {\bibfnamefont {R.}~\bibnamefont
  {Baxter}},\ }\href {https://books.google.com/books?id=eQzCAgAAQBAJ} {\emph
  {\bibinfo {title} {Exactly Solved Models in Statistical Mechanics}}},\ Dover
  Books on Physics\ (\bibinfo  {publisher} {Dover Publications},\ \bibinfo
  {year} {2013})\BibitemShut {NoStop}%
\bibitem [{\citenamefont {Huang}\ \emph {et~al.}(2013)\citenamefont {Huang},
  \citenamefont {Chen},\ and\ \citenamefont {Lin}}]{HuangSPT2013}%
  \BibitemOpen
  \bibfield  {author} {\bibinfo {author} {\bibfnamefont {C.-Y.}\ \bibnamefont
  {Huang}}, \bibinfo {author} {\bibfnamefont {X.}~\bibnamefont {Chen}}, \ and\
  \bibinfo {author} {\bibfnamefont {F.-L.}\ \bibnamefont {Lin}},\ }\href
  {\doibase 10.1103/PhysRevB.88.205124} {\bibfield  {journal} {\bibinfo
  {journal} {Phys. Rev. B}\ }\textbf {\bibinfo {volume} {88}},\ \bibinfo
  {pages} {205124} (\bibinfo {year} {2013})}\BibitemShut {NoStop}%
\bibitem [{\citenamefont {Berezinskii}()}]{Berezinskii1970}%
  \BibitemOpen
  \bibfield  {author} {\bibinfo {author} {\bibfnamefont {V.~L.}\ \bibnamefont
  {Berezinskii}},\ }\href@noop {} {\bibinfo  {journal} {Zh. Eksp. Teor. Fiz.
  {\bf 59}, 907 (1970) [Sov. Phys. JETP {\bf 42}, 493 (1971)]}\ }\BibitemShut
  {NoStop}%
\bibitem [{\citenamefont {Kosterlitz}\ and\ \citenamefont
  {Thouless}(1973)}]{KT_1973}%
  \BibitemOpen
\bibfield  {journal} {  }\bibfield  {author} {\bibinfo {author} {\bibfnamefont
  {J.~M.}\ \bibnamefont {Kosterlitz}}\ and\ \bibinfo {author} {\bibfnamefont
  {D.~J.}\ \bibnamefont {Thouless}},\ }\href
  {http://stacks.iop.org/0022-3719/6/i=7/a=010} {\bibfield  {journal} {\bibinfo
   {journal} {Journal of Physics C: Solid State Physics}\ }\textbf {\bibinfo
  {volume} {6}},\ \bibinfo {pages} {1181} (\bibinfo {year} {1973})}\BibitemShut
  {NoStop}%
\bibitem [{\citenamefont {Jos\'e}\ \emph {et~al.}(1977)\citenamefont {Jos\'e},
  \citenamefont {Kadanoff}, \citenamefont {Kirkpatrick},\ and\ \citenamefont
  {Nelson}}]{2DXYmodel_1977}%
  \BibitemOpen
  \bibfield  {author} {\bibinfo {author} {\bibfnamefont {J.~V.}\ \bibnamefont
  {Jos\'e}}, \bibinfo {author} {\bibfnamefont {L.~P.}\ \bibnamefont
  {Kadanoff}}, \bibinfo {author} {\bibfnamefont {S.}~\bibnamefont
  {Kirkpatrick}}, \ and\ \bibinfo {author} {\bibfnamefont {D.~R.}\ \bibnamefont
  {Nelson}},\ }\href {\doibase 10.1103/PhysRevB.16.1217} {\bibfield  {journal}
  {\bibinfo  {journal} {Phys. Rev. B}\ }\textbf {\bibinfo {volume} {16}},\
  \bibinfo {pages} {1217} (\bibinfo {year} {1977})}\BibitemShut {NoStop}%
\bibitem [{\citenamefont {Ding}\ and\ \citenamefont
  {Makivi\ifmmode~\acute{c}\else \'{c}\fi{}}(1990)}]{Ding_1990}%
  \BibitemOpen
  \bibfield  {author} {\bibinfo {author} {\bibfnamefont {H.-Q.}\ \bibnamefont
  {Ding}}\ and\ \bibinfo {author} {\bibfnamefont {M.~S.}\ \bibnamefont
  {Makivi\ifmmode~\acute{c}\else \'{c}\fi{}}},\ }\href {\doibase
  10.1103/PhysRevB.42.6827} {\bibfield  {journal} {\bibinfo  {journal} {Phys.
  Rev. B}\ }\textbf {\bibinfo {volume} {42}},\ \bibinfo {pages} {6827}
  (\bibinfo {year} {1990})}\BibitemShut {NoStop}%
\bibitem [{\citenamefont {Ding}(1992)}]{Ding_1992}%
  \BibitemOpen
  \bibfield  {author} {\bibinfo {author} {\bibfnamefont {H.-Q.}\ \bibnamefont
  {Ding}},\ }\href {\doibase 10.1103/PhysRevB.45.230} {\bibfield  {journal}
  {\bibinfo  {journal} {Phys. Rev. B}\ }\textbf {\bibinfo {volume} {45}},\
  \bibinfo {pages} {230} (\bibinfo {year} {1992})}\BibitemShut {NoStop}%
\bibitem [{\citenamefont {Cuccoli}\ \emph {et~al.}(1995)\citenamefont
  {Cuccoli}, \citenamefont {Tognetti},\ and\ \citenamefont
  {Vaia}}]{1995_Cuccoli}%
  \BibitemOpen
  \bibfield  {author} {\bibinfo {author} {\bibfnamefont {A.}~\bibnamefont
  {Cuccoli}}, \bibinfo {author} {\bibfnamefont {V.}~\bibnamefont {Tognetti}}, \
  and\ \bibinfo {author} {\bibfnamefont {R.}~\bibnamefont {Vaia}},\ }\href
  {\doibase 10.1103/PhysRevB.52.10221} {\bibfield  {journal} {\bibinfo
  {journal} {Phys. Rev. B}\ }\textbf {\bibinfo {volume} {52}},\ \bibinfo
  {pages} {10221} (\bibinfo {year} {1995})}\BibitemShut {NoStop}%
\bibitem [{\citenamefont {Bauer}\ \emph {et~al.}(2009)\citenamefont {Bauer},
  \citenamefont {Vidal},\ and\ \citenamefont {Troyer}}]{iPEPS_2009}%
  \BibitemOpen
  \bibfield  {author} {\bibinfo {author} {\bibfnamefont {B.}~\bibnamefont
  {Bauer}}, \bibinfo {author} {\bibfnamefont {G.}~\bibnamefont {Vidal}}, \ and\
  \bibinfo {author} {\bibfnamefont {M.}~\bibnamefont {Troyer}},\ }\href
  {http://stacks.iop.org/1742-5468/2009/i=09/a=P09006} {\bibfield  {journal}
  {\bibinfo  {journal} {Journal of Statistical Mechanics: Theory and
  Experiment}\ }\textbf {\bibinfo {volume} {2009}},\ \bibinfo {pages} {P09006}
  (\bibinfo {year} {2009})}\BibitemShut {NoStop}%
\bibitem [{\citenamefont {Nishino}\ and\ \citenamefont
  {Okunishi}(1997)}]{CTMRG1997}%
  \BibitemOpen
  \bibfield  {author} {\bibinfo {author} {\bibfnamefont {T.}~\bibnamefont
  {Nishino}}\ and\ \bibinfo {author} {\bibfnamefont {K.}~\bibnamefont
  {Okunishi}},\ }\href {\doibase 10.1143/JPSJ.66.3040} {\bibfield  {journal}
  {\bibinfo  {journal} {Journal of the Physical Society of Japan}\ }\textbf
  {\bibinfo {volume} {66}},\ \bibinfo {pages} {3040} (\bibinfo {year}
  {1997})}\BibitemShut {NoStop}%
\bibitem [{\citenamefont {Jiang}\ \emph {et~al.}(2008)\citenamefont {Jiang},
  \citenamefont {Weng},\ and\ \citenamefont {Xiang}}]{Xiang_TRG2008}%
  \BibitemOpen
  \bibfield  {author} {\bibinfo {author} {\bibfnamefont {H.~C.}\ \bibnamefont
  {Jiang}}, \bibinfo {author} {\bibfnamefont {Z.~Y.}\ \bibnamefont {Weng}}, \
  and\ \bibinfo {author} {\bibfnamefont {T.}~\bibnamefont {Xiang}},\ }\href
  {\doibase 10.1103/PhysRevLett.101.090603} {\bibfield  {journal} {\bibinfo
  {journal} {Phys. Rev. Lett.}\ }\textbf {\bibinfo {volume} {101}},\ \bibinfo
  {pages} {090603} (\bibinfo {year} {2008})}\BibitemShut {NoStop}%
\bibitem [{\citenamefont {Chen}\ \emph
  {et~al.}(2010{\natexlab{a}})\citenamefont {Chen}, \citenamefont {Gu},\ and\
  \citenamefont {Wen}}]{Xie_LU}%
  \BibitemOpen
  \bibfield  {author} {\bibinfo {author} {\bibfnamefont {X.}~\bibnamefont
  {Chen}}, \bibinfo {author} {\bibfnamefont {Z.-C.}\ \bibnamefont {Gu}}, \ and\
  \bibinfo {author} {\bibfnamefont {X.-G.}\ \bibnamefont {Wen}},\ }\href
  {\doibase 10.1103/PhysRevB.82.155138} {\bibfield  {journal} {\bibinfo
  {journal} {Phys. Rev. B}\ }\textbf {\bibinfo {volume} {82}},\ \bibinfo
  {pages} {155138} (\bibinfo {year} {2010}{\natexlab{a}})}\BibitemShut
  {NoStop}%
\bibitem [{\citenamefont {Huang}\ and\ \citenamefont
  {Wei}(2015)}]{HuangTO2015}%
  \BibitemOpen
  \bibfield  {author} {\bibinfo {author} {\bibfnamefont {C.-Y.}\ \bibnamefont
  {Huang}}\ and\ \bibinfo {author} {\bibfnamefont {T.-C.}\ \bibnamefont
  {Wei}},\ }\href {\doibase 10.1103/PhysRevB.92.085405} {\bibfield  {journal}
  {\bibinfo  {journal} {Phys. Rev. B}\ }\textbf {\bibinfo {volume} {92}},\
  \bibinfo {pages} {085405} (\bibinfo {year} {2015})}\BibitemShut {NoStop}%
\bibitem [{\citenamefont {Xie}\ \emph {et~al.}(2009)\citenamefont {Xie},
  \citenamefont {Jiang}, \citenamefont {Chen}, \citenamefont {Weng},\ and\
  \citenamefont {Xiang}}]{Xiang_SRG2009}%
  \BibitemOpen
  \bibfield  {author} {\bibinfo {author} {\bibfnamefont {Z.~Y.}\ \bibnamefont
  {Xie}}, \bibinfo {author} {\bibfnamefont {H.~C.}\ \bibnamefont {Jiang}},
  \bibinfo {author} {\bibfnamefont {Q.~N.}\ \bibnamefont {Chen}}, \bibinfo
  {author} {\bibfnamefont {Z.~Y.}\ \bibnamefont {Weng}}, \ and\ \bibinfo
  {author} {\bibfnamefont {T.}~\bibnamefont {Xiang}},\ }\href {\doibase
  10.1103/PhysRevLett.103.160601} {\bibfield  {journal} {\bibinfo  {journal}
  {Phys. Rev. Lett.}\ }\textbf {\bibinfo {volume} {103}},\ \bibinfo {pages}
  {160601} (\bibinfo {year} {2009})}\BibitemShut {NoStop}%
\bibitem [{\citenamefont {Kosterlitz}(1974)}]{Kosterlitz1974}%
  \BibitemOpen
  \bibfield  {author} {\bibinfo {author} {\bibfnamefont {J.~M.}\ \bibnamefont
  {Kosterlitz}},\ }\href {http://stacks.iop.org/0022-3719/7/i=6/a=005}
  {\bibfield  {journal} {\bibinfo  {journal} {Journal of Physics C: Solid State
  Physics}\ }\textbf {\bibinfo {volume} {7}},\ \bibinfo {pages} {1046}
  (\bibinfo {year} {1974})}\BibitemShut {NoStop}%
\bibitem [{\citenamefont {Niggemann}\ and\ \citenamefont
  {Zittartz}(2000)}]{SO_Niggemann}%
  \BibitemOpen
  \bibfield  {author} {\bibinfo {author} {\bibfnamefont {H.}~\bibnamefont
  {Niggemann}}\ and\ \bibinfo {author} {\bibfnamefont {J.}~\bibnamefont
  {Zittartz}},\ }\href {\doibase 10.1007/s100510050044} {\bibfield  {journal}
  {\bibinfo  {journal} {The European Physical Journal B - Condensed Matter and
  Complex Systems}\ }\textbf {\bibinfo {volume} {13}},\ \bibinfo {pages} {377}
  (\bibinfo {year} {2000})}\BibitemShut {NoStop}%
\bibitem [{\citenamefont {Baxter}\ and\ \citenamefont
  {Choy}(1988)}]{Baxter1988}%
  \BibitemOpen
  \bibfield  {author} {\bibinfo {author} {\bibfnamefont {R.~J.}\ \bibnamefont
  {Baxter}}\ and\ \bibinfo {author} {\bibfnamefont {T.~C.}\ \bibnamefont
  {Choy}},\ }\href {http://stacks.iop.org/0305-4470/21/i=9/a=027} {\bibfield
  {journal} {\bibinfo  {journal} {Journal of Physics A: Mathematical and
  General}\ }\textbf {\bibinfo {volume} {21}},\ \bibinfo {pages} {2143}
  (\bibinfo {year} {1988})}\BibitemShut {NoStop}%
\bibitem [{\citenamefont {Choy}\ and\ \citenamefont {Baxter}(1987)}]{Choy1987}%
  \BibitemOpen
  \bibfield  {author} {\bibinfo {author} {\bibfnamefont {T.}~\bibnamefont
  {Choy}}\ and\ \bibinfo {author} {\bibfnamefont {R.}~\bibnamefont {Baxter}},\
  }\href {\doibase http://dx.doi.org/10.1016/0375-9601(87)90162-9} {\bibfield
  {journal} {\bibinfo  {journal} {Physics Letters A}\ }\textbf {\bibinfo
  {volume} {125}},\ \bibinfo {pages} {365 } (\bibinfo {year}
  {1987})}\BibitemShut {NoStop}%
\bibitem [{\citenamefont {Kramers}\ and\ \citenamefont
  {Wannier}(1941)}]{Kramers1941}%
  \BibitemOpen
  \bibfield  {author} {\bibinfo {author} {\bibfnamefont {H.~A.}\ \bibnamefont
  {Kramers}}\ and\ \bibinfo {author} {\bibfnamefont {G.~H.}\ \bibnamefont
  {Wannier}},\ }\href {\doibase 10.1103/PhysRev.60.252} {\bibfield  {journal}
  {\bibinfo  {journal} {Phys. Rev.}\ }\textbf {\bibinfo {volume} {60}},\
  \bibinfo {pages} {252} (\bibinfo {year} {1941})}\BibitemShut {NoStop}%
\bibitem [{\citenamefont {Baxter}(1986)}]{Baxter1986}%
  \BibitemOpen
  \bibfield  {author} {\bibinfo {author} {\bibfnamefont {R.~J.}\ \bibnamefont
  {Baxter}},\ }\href {http://www.jstor.org/stable/2397830} {\bibfield
  {journal} {\bibinfo  {journal} {Proceedings of the Royal Society of London.
  Series A, Mathematical and Physical Sciences}\ }\textbf {\bibinfo {volume}
  {404}} (\bibinfo {year} {1986})}\BibitemShut {NoStop}%
\bibitem [{\citenamefont {Van~den Nest}\ \emph {et~al.}(2006)\citenamefont
  {Van~den Nest}, \citenamefont {Miyake}, \citenamefont {D\"ur},\ and\
  \citenamefont {Briegel}}]{NestMBQC2006}%
  \BibitemOpen
  \bibfield  {author} {\bibinfo {author} {\bibfnamefont {M.}~\bibnamefont
  {Van~den Nest}}, \bibinfo {author} {\bibfnamefont {A.}~\bibnamefont
  {Miyake}}, \bibinfo {author} {\bibfnamefont {W.}~\bibnamefont {D\"ur}}, \
  and\ \bibinfo {author} {\bibfnamefont {H.~J.}\ \bibnamefont {Briegel}},\
  }\href {\doibase 10.1103/PhysRevLett.97.150504} {\bibfield  {journal}
  {\bibinfo  {journal} {Phys. Rev. Lett.}\ }\textbf {\bibinfo {volume} {97}},\
  \bibinfo {pages} {150504} (\bibinfo {year} {2006})}\BibitemShut {NoStop}%
\bibitem [{\citenamefont {Browne}\ \emph {et~al.}(2008)\citenamefont {Browne},
  \citenamefont {Elliott}, \citenamefont {Flammia}, \citenamefont {Merkel},
  \citenamefont {Miyake},\ and\ \citenamefont {Short}}]{DanielQC2008}%
  \BibitemOpen
  \bibfield  {author} {\bibinfo {author} {\bibfnamefont {D.~E.}\ \bibnamefont
  {Browne}}, \bibinfo {author} {\bibfnamefont {M.~B.}\ \bibnamefont {Elliott}},
  \bibinfo {author} {\bibfnamefont {S.~T.}\ \bibnamefont {Flammia}}, \bibinfo
  {author} {\bibfnamefont {S.~T.}\ \bibnamefont {Merkel}}, \bibinfo {author}
  {\bibfnamefont {A.}~\bibnamefont {Miyake}}, \ and\ \bibinfo {author}
  {\bibfnamefont {A.~J.}\ \bibnamefont {Short}},\ }\href
  {http://stacks.iop.org/1367-2630/10/i=2/a=023010} {\bibfield  {journal}
  {\bibinfo  {journal} {New Journal of Physics}\ }\textbf {\bibinfo {volume}
  {10}},\ \bibinfo {pages} {023010} (\bibinfo {year} {2008})}\BibitemShut
  {NoStop}%
\bibitem [{\citenamefont {Chen}\ \emph
  {et~al.}(2010{\natexlab{b}})\citenamefont {Chen}, \citenamefont {Duan},
  \citenamefont {Ji},\ and\ \citenamefont {Zeng}}]{ChenMBQC2010}%
  \BibitemOpen
  \bibfield  {author} {\bibinfo {author} {\bibfnamefont {X.}~\bibnamefont
  {Chen}}, \bibinfo {author} {\bibfnamefont {R.}~\bibnamefont {Duan}}, \bibinfo
  {author} {\bibfnamefont {Z.}~\bibnamefont {Ji}}, \ and\ \bibinfo {author}
  {\bibfnamefont {B.}~\bibnamefont {Zeng}},\ }\href {\doibase
  10.1103/PhysRevLett.105.020502} {\bibfield  {journal} {\bibinfo  {journal}
  {Phys. Rev. Lett.}\ }\textbf {\bibinfo {volume} {105}},\ \bibinfo {pages}
  {020502} (\bibinfo {year} {2010}{\natexlab{b}})}\BibitemShut {NoStop}%
\bibitem [{\citenamefont {Wei}\ \emph {et~al.}(2012)\citenamefont {Wei},
  \citenamefont {Affleck},\ and\ \citenamefont
  {Raussendorf}}]{2DAKLT_wei_2013}%
  \BibitemOpen
  \bibfield  {author} {\bibinfo {author} {\bibfnamefont {T.-C.}\ \bibnamefont
  {Wei}}, \bibinfo {author} {\bibfnamefont {I.}~\bibnamefont {Affleck}}, \ and\
  \bibinfo {author} {\bibfnamefont {R.}~\bibnamefont {Raussendorf}},\ }\href
  {\doibase 10.1103/PhysRevA.86.032328} {\bibfield  {journal} {\bibinfo
  {journal} {Phys. Rev. A}\ }\textbf {\bibinfo {volume} {86}},\ \bibinfo
  {pages} {032328} (\bibinfo {year} {2012})}\BibitemShut {NoStop}%
\bibitem [{\citenamefont {Verstraete}\ and\ \citenamefont
  {Cirac}(2004{\natexlab{b}})}]{VerstraeteVBC2004}%
  \BibitemOpen
  \bibfield  {author} {\bibinfo {author} {\bibfnamefont {F.}~\bibnamefont
  {Verstraete}}\ and\ \bibinfo {author} {\bibfnamefont {J.~I.}\ \bibnamefont
  {Cirac}},\ }\href {\doibase 10.1103/PhysRevA.70.060302} {\bibfield  {journal}
  {\bibinfo  {journal} {Phys. Rev. A}\ }\textbf {\bibinfo {volume} {70}},\
  \bibinfo {pages} {060302} (\bibinfo {year} {2004}{\natexlab{b}})}\BibitemShut
  {NoStop}%
\bibitem [{\citenamefont {Gross}\ and\ \citenamefont
  {Eisert}(2007)}]{GrossMBQC2007}%
  \BibitemOpen
  \bibfield  {author} {\bibinfo {author} {\bibfnamefont {D.}~\bibnamefont
  {Gross}}\ and\ \bibinfo {author} {\bibfnamefont {J.}~\bibnamefont {Eisert}},\
  }\href {\doibase 10.1103/PhysRevLett.98.220503} {\bibfield  {journal}
  {\bibinfo  {journal} {Phys. Rev. Lett.}\ }\textbf {\bibinfo {volume} {98}},\
  \bibinfo {pages} {220503} (\bibinfo {year} {2007})}\BibitemShut {NoStop}%
\bibitem [{\citenamefont {Brennen}\ and\ \citenamefont
  {Miyake}(2008)}]{BrennenMBQC2008}%
  \BibitemOpen
  \bibfield  {author} {\bibinfo {author} {\bibfnamefont {G.~K.}\ \bibnamefont
  {Brennen}}\ and\ \bibinfo {author} {\bibfnamefont {A.}~\bibnamefont
  {Miyake}},\ }\href {\doibase 10.1103/PhysRevLett.101.010502} {\bibfield
  {journal} {\bibinfo  {journal} {Phys. Rev. Lett.}\ }\textbf {\bibinfo
  {volume} {101}},\ \bibinfo {pages} {010502} (\bibinfo {year}
  {2008})}\BibitemShut {NoStop}%
\bibitem [{\citenamefont {Wei}\ and\ \citenamefont
  {Raussendorf}(2015)}]{squae_wei_2015}%
  \BibitemOpen
  \bibfield  {author} {\bibinfo {author} {\bibfnamefont {T.-C.}\ \bibnamefont
  {Wei}}\ and\ \bibinfo {author} {\bibfnamefont {R.}~\bibnamefont
  {Raussendorf}},\ }\href {\doibase 10.1103/PhysRevA.92.012310} {\bibfield
  {journal} {\bibinfo  {journal} {Phys. Rev. A}\ }\textbf {\bibinfo {volume}
  {92}},\ \bibinfo {pages} {012310} (\bibinfo {year} {2015})}\BibitemShut
  {NoStop}%
\bibitem [{\citenamefont {Fan}\ and\ \citenamefont {Wu}(1969)}]{Fan1969}%
  \BibitemOpen
  \bibfield  {author} {\bibinfo {author} {\bibfnamefont {C.}~\bibnamefont
  {Fan}}\ and\ \bibinfo {author} {\bibfnamefont {F.~Y.}\ \bibnamefont {Wu}},\
  }\href {\doibase 10.1103/PhysRev.179.560} {\bibfield  {journal} {\bibinfo
  {journal} {Phys. Rev.}\ }\textbf {\bibinfo {volume} {179}},\ \bibinfo {pages}
  {560} (\bibinfo {year} {1969})}\BibitemShut {NoStop}%
\bibitem [{\citenamefont {Fan}\ and\ \citenamefont {Wu}(1970)}]{Fan1970}%
  \BibitemOpen
  \bibfield  {author} {\bibinfo {author} {\bibfnamefont {C.}~\bibnamefont
  {Fan}}\ and\ \bibinfo {author} {\bibfnamefont {F.~Y.}\ \bibnamefont {Wu}},\
  }\href {\doibase 10.1103/PhysRevB.2.723} {\bibfield  {journal} {\bibinfo
  {journal} {Phys. Rev. B}\ }\textbf {\bibinfo {volume} {2}},\ \bibinfo {pages}
  {723} (\bibinfo {year} {1970})}\BibitemShut {NoStop}%
\bibitem [{\citenamefont {Wu}\ and\ \citenamefont
  {Lin}(1987{\natexlab{a}})}]{Lin1987}%
  \BibitemOpen
  \bibfield  {author} {\bibinfo {author} {\bibfnamefont {F.~Y.}\ \bibnamefont
  {Wu}}\ and\ \bibinfo {author} {\bibfnamefont {K.~Y.}\ \bibnamefont {Lin}},\
  }\href {http://stacks.iop.org/0305-4470/20/i=16/a=049} {\bibfield  {journal}
  {\bibinfo  {journal} {Journal of Physics A: Mathematical and General}\
  }\textbf {\bibinfo {volume} {20}},\ \bibinfo {pages} {5737} (\bibinfo {year}
  {1987}{\natexlab{a}})}\BibitemShut {NoStop}%
\bibitem [{\citenamefont {Wu}\ and\ \citenamefont
  {Lin}(1987{\natexlab{b}})}]{Wu_Lin1987}%
  \BibitemOpen
  \bibfield  {author} {\bibinfo {author} {\bibfnamefont {F.~Y.}\ \bibnamefont
  {Wu}}\ and\ \bibinfo {author} {\bibfnamefont {K.~Y.}\ \bibnamefont {Lin}},\
  }\href {http://stacks.iop.org/0305-4470/20/i=16/a=049} {\bibfield  {journal}
  {\bibinfo  {journal} {Journal of Physics A: Mathematical and General}\
  }\textbf {\bibinfo {volume} {20}},\ \bibinfo {pages} {5737} (\bibinfo {year}
  {1987}{\natexlab{b}})}\BibitemShut {NoStop}%
\end{thebibliography}

\appendix
\section{The classical vertex model} \label{App:classicalvertex}

\begin{figure}[ht]
\includegraphics[width=0.1\textwidth] {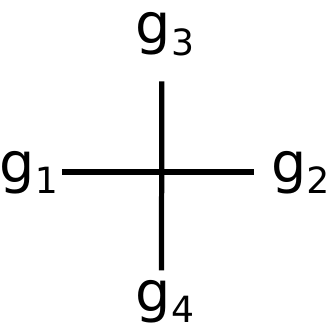}
\caption{ Vertex with interaction energy $E(g_1, g_2, g_3, g_4; u)$ }
  \label{fig:one_weight}
\end{figure}

A general vertex model is a lattice model which has classical state variables $g$ associated with the links of the lattice~\cite{mccoy2011advanced,baxter2013exactly}. 
The interactions between those state variables are characterized by an interaction energy $E$ lying on the vertices (hence the name vertex model) where $E$ depends on the state variables of the adjacent bonds and maybe one additional independent parameter $u$ (usually referred to as the spectral parameter); see Fig.~\ref{fig:one_weight}.

Solving such a model usually means computing the partition function or equivalently the free energy (per site). 
Therefore, we assign a Boltzmann weight $W$ to each vertex by
\begin{align}
W(g_1,g_2,g_3,g_4;u) = e^{-\beta E(g_1,g_2,g_3,g_4;u)}.
\end{align}
The partition function is defined by 
\begin{align}
Z = \sum_G \prod_i  W_i(g_1,g_2,g_3,g_4;u),
\end{align}
where $G=\{g\}$ runs over all allowed bond configurations and $i$ runs over all vertices (assuming a finite lattice to begin with).  
The free energy per site in the thermodynamic limit is then given by
\begin{align}
f= -\beta^{-1} \lim_{N \to \infty } \frac{1}{N} \ln (Z).
\end{align}

\begin{figure}[ht]
 \includegraphics[width=0.5\textwidth]{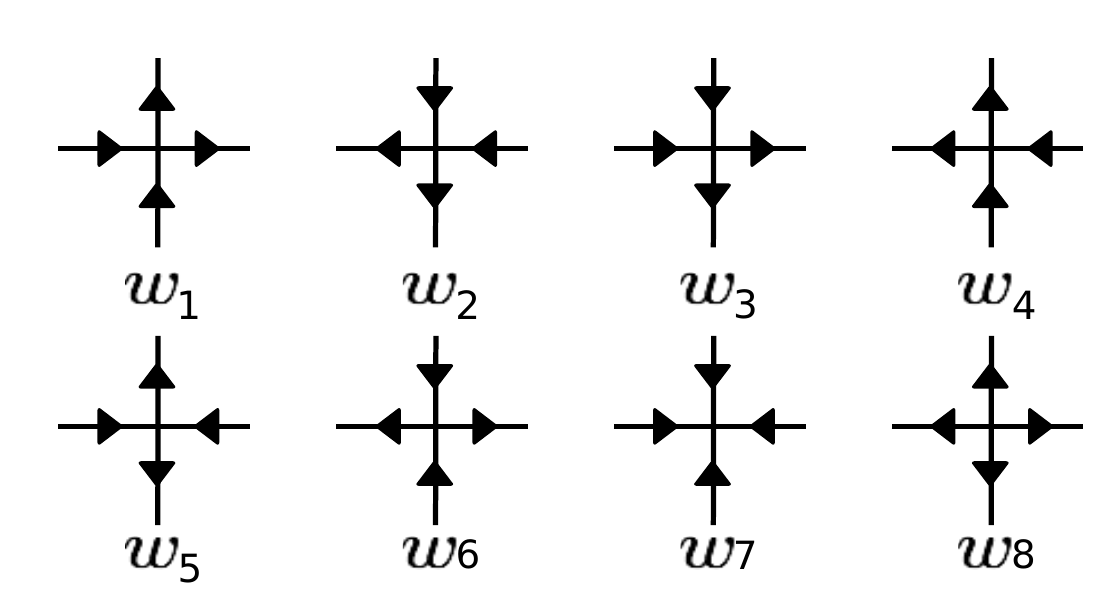}
\caption{The allowed vertices of the eight-vertex model.}
  \label{fig:weight}
\end{figure}

In the following we shall restrict ourselves to the eight-vertex model. 
The state variables $g$ are restricted to two discrete values, which
 we choose to be 0 and 1. 
In the eight-vertex model, the only allowed bond configurations have an even number of $0's$ adjacent to each vertex. 
This, of course, also means that each vertex has an even amount of $1's$ adjacent to it. 
We can now introduce a graphical representation of the model. 
An horizontal bond occupied by $g = 0$ is represented by an arrow pointing left while an vertical bond occupied by $g = 0$ is represented by an arrow pointing upward. 
Bonds with a state variable $g = 1$ assigned to it are represented by arrows pointing to the right and downward respectively. 
The allowed vertices of the square lattice are shown in Fig.~\ref{fig:weight}.

Their local Boltzmann weights are denoted by $\omega_i$. 
Thus, the second restriction can be rephrased as follows: an allowed vertex has an even number of arrows pointing towards it.

\section{Solution of free fermion} \label{App:freefermion}

In order to find a phase transition,  we have to solve the model. 
The free fermion condition restricts once again the possible Boltzmann weights. 
They have to obey the equation
\begin{align}
w_1w_2+w_3w_4=w_5w_6+w_7w_8.
\end{align}
Fan and Wu solved the eight-vertex model on the square lattice when this condition holds in \cite{Fan1969} and in more detail in \cite{Fan1970} using dimer methods. 
The closed expression for the free energy per site reads
\begin{align}
 \beta f  = & - \frac{1}{8\pi^2}  \int_0^{2\pi} d \theta    \int_0^{2\pi} d \phi \ln \big(  2a+2b \cos(\theta)+ 2c \cos(\phi)  \notag \\ 
 & +2d \cos (\theta-\phi) + 2e \cos(\theta+\phi) \big),
\end{align}
where $a$, $b$, $c$, $d$, and $e$ are functions of the Boltzmann weights, i.e.
\begin{align}
& 2a =   w_1^2+ w_2^2 +w_3^2+ w_4^2 , \notag \\ 
&b =  w_1w_3-w_2w_4  \notag \\ 
&c =  w_1w_4-w_2w_3   \notag \\ 
&d =  w_3w_4-w_7w_8   \notag \\ 
&e =  w_3w_4-w_5w_6 
\label{smalla}
\end{align}
Simple trigonometric expansions lead to
\begin{align}
 \beta f  = & - \frac{1}{8\pi^2}  \int_0^{2\pi} d \theta    \int_0^{2\pi} d \phi 
  \ln \big(  2A(\phi)  \notag \\ 
 &+2B(\phi) \cos(\theta) + 2C(\phi)\cos(\theta) \big),
\end{align}
where 
\begin{align}
& A(\phi) = a+ c \cos(\phi),   \notag \\ 
& B(\phi) =b+(d+e) \cos(\phi),   \notag \\ 
& C(\phi) = (d-e) \sin(\phi)
\label{capA}
\end{align}
The integration over $\theta$ can be done easily using the formula
\begin{align}
 &  \int_0^{2\pi} d \theta \ln \big(  2x+  2y \cos(\theta)+ 2z \cos(\theta)   \big)  \notag \\ 
 &  = 2\pi \ln \Big(x + \sqrt{ x^2-y^2-z^2} \Big). 
\end{align}
Thus, the free energy per site becomes
\begin{align}
 \beta f  =  - \frac{1}{4\pi^2}   \int_0^{2\pi} d \phi 
  \ln \Big(  A(\phi) +   \sqrt{Q(\phi)}   \Big),
\end{align}
where 
\begin{align}
Q(\phi) = A(\phi)^2-B(\phi)^2-C(\phi)^2. 
\end{align}
Finding the phase transition means finding non-analytic points of the free-energy per site. 
In Ref.~\cite{Fan1970}  the authors argued that this translates into detecting non-analytic points of $Q(\phi)$. 
Assuming that all Botzmann weights are non-zero they showed furthermore that this can only happen if $Q(\phi)$ is not a perfect square. 
Then, either $Q(0)=0$ or $Q(\pm \pi)=0$ marks the critical point where some derivatives of $Q$ diverge.

\smallskip
\noindent {\bf Ferromagnetic transitions on the star lattice}. 
Luckily, the Boltzmann weights from the various deformed AKLT wavefunctions  in the main text satisfy the free-fermion condition, such as those on the hexagonal lattice and those via on-site diagonal approximations on the square-octagon, as well those with $|\phi^\pm\rangle$ bond states on the  star lattice. The latter ones, shown in Eq.~(\ref{eqn:star8}), are among the new results in this paper, and we check explicitly that  the free-fermion condition is satisfied:
\begin{align}
w_1w_2+w_3w_4 = \frac{1}{2}\big( (a^4-1)^4+(a^8+18a^4+24a^2+85)^2  \big) 
\end{align} 
and 
\begin{align}
w_5w_6+w_7w_8  = \frac{1}{2}\big( (a^4-1)^4+(a^8+18a^4+24a^2+85)^2  \big) .
\end{align} 
Thus, the solution presented in the early part  can be applied. Using  Eqs.~(\ref{smalla}) and (\ref{capA}), we get
\begin{align}
&A(\phi) = 2(a^2+3)(a^4+7)(a^4-1)^4\cos(\phi)  \notag \\
&    \quad +\frac{1}{8}\big( 3(a^2+1)^4(a^2-1)^8+(a^6+3a^4+15a^2+14)^4   \big),   \notag \\
& B(\phi) = -2(a^2+3)(a^4-1)^4(a^4+7)\big(  \cos(\phi)-1  \big)    \notag \\
&C(\phi) = 2(a^2+3)(a^4-1)^4(a^4+7)  \sin(\phi).
\end{align}
Therefore, 
\begin{align}
&Q(\phi)=  -4(a^2+3)(a^4+7)^2(a^4-1)^8\sin^2(\phi )  \notag \\
           &-4(a^2+3)(a^4+7)^2(a^4-1)^8  \big(\cos^2(\phi )-1 \big)  \notag \\
           &+\Big(    2(a^2+3)(a^4+7)(a^4-1)^4 \cos(\phi)    \notag \\
           &+ \frac{1}{8}\big( 3(a^2+1)^4(a^2-1)^8+(a^6+3a^4+15a^2+13)^4  \big)   \Big)^2.
\end{align}

\begin{figure}[ht]
\includegraphics[width=0.5\textwidth]{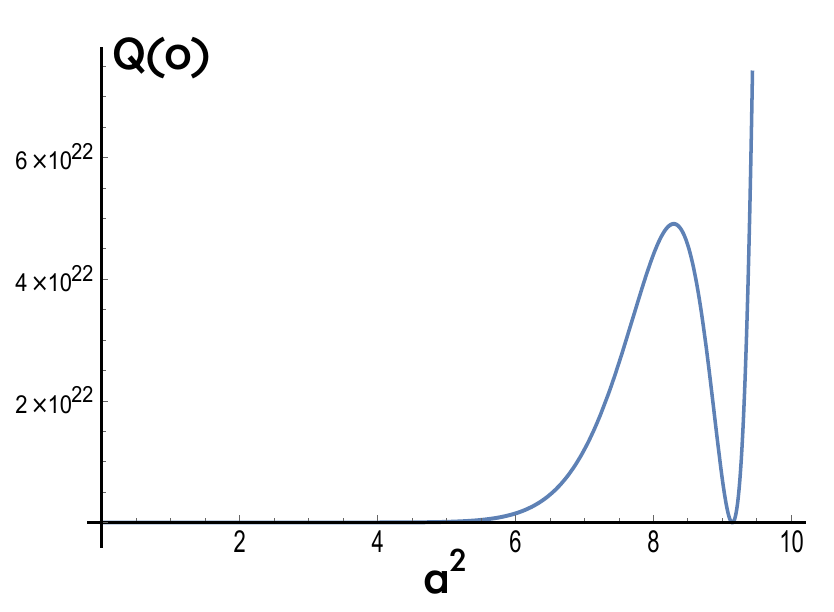}
\caption{ $Q(0)$ as a function of $a^2$.}
\label{fig:solutionQ}
\end{figure}

The transition point is given by the roots of $Q(0/\pm\pi)$ under the condition that  $a\in \mathbb{R}$. 
In this case, we can find such a point, namely
\begin{align}
& a_c = \big(2+\sqrt{3}+ \sqrt{2(6+5\sqrt{3})}   \big)^{\frac{1}{2}} \approx 3.02438,    \notag \\
& \phi = 0.
\end{align}
The transitions on other free-fermion models mentioned above on other lattices can be obtained similarly, as were done in Refs.~\cite{Hexagon_Niggemann,SO_Niggemann}.

\section{ The Ising model on the union jack and checkerboard lattices} \label{App:Isingasunionjack}

\begin{figure}[ht]
\includegraphics[width=0.5\textwidth] {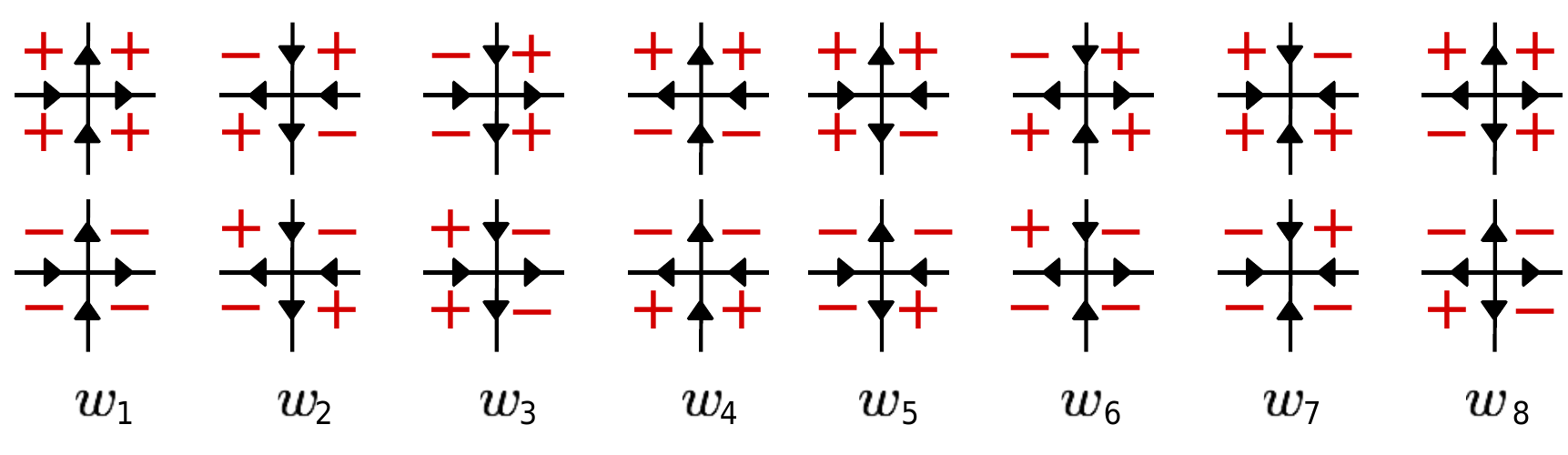}
\caption{ Spin-vertex correspondence: The plus and minus signs represent the Ising spins on the original lattice, the arrows correspond to the state variables of the dual lattice. }
\label{fig:spin_vertex}
\end{figure}

The Ising model belongs to the class of spin models which have state variables sitting on the vertices and interaction energies associated with the bonds of the lattice~\cite{Choy1987,Lin1987}. 
Its partition function is
\begin{align}
Z = \sum_{\{\sigma\}} e ^{\Sigma_{<i,j>} K_r \sigma_i \sigma_j},
\end{align}
where the outer sum runs over all spin configurations of the lattice $\sigma = {\sigma_1, ..., \sigma_N }$,
$\sigma_i \in \{1,-1\}$, the inner sum runs over all nearest neighbor pairs, and r denotes the type of the nearest neighbor bond.
 As discussed in~\cite{Lin1987}, we can obtain a free fermion eight vertex model by performing the outer sum over all fourfold connected vertices.
This effectively removes those spins from the lattice and we obtain a square lattice.
\begin{figure}[ht]
\includegraphics[width=0.5\textwidth]{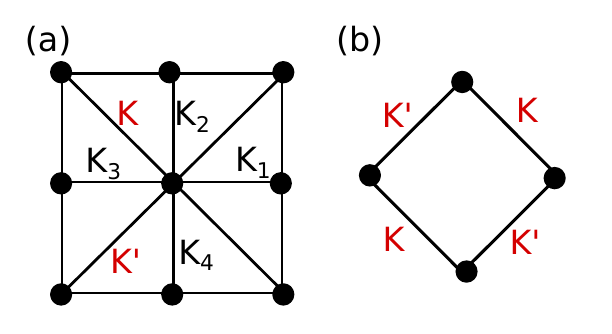}
\caption{ (a) Unit cell of the Ising model on the union jack lattice and its interaction energies. The dots represent the classical spins.(b)Unit cell of the effective square lattice.}
\label{fig:union_jack}
\end{figure}
The partition function  can be rewritten as follows, 
\begin{align}
Z = \sum_{\sigma'}   & 2\times \prod  e^{\frac{1}{2}  \big(  K'(\sigma_j \sigma_k+ \sigma_i \sigma_l) +  K(\sigma_i \sigma_j+ \sigma_k\sigma_l)  \big)  } \notag\\
& \cosh(K_1\sigma_i + K_2\sigma_j + K_3\sigma_k+ K_4\sigma_l),
\label{patition_uni}
\end{align}
where the product runs over all plaquettes shown in Fig.~\ref{fig:union_jack}(b), the outer sum runs over all spin configurations of the eightfold connected spins in Fig.~\ref{fig:union_jack}(a), and the factor 1/2 in the exponential negates the double counting of bonds of the type $K$ and $K'$.
We can draw the dual lattice which is also a square lattice, and identify its bonds with state variables according to the spin configuration of the original lattice.
This two-to-one map is shown in Fig.~\ref{fig:spin_vertex}. 
The Boltzmann weights associated with the vertices of the dual lattice are precisely the factors of the product in Eq.~\ref{patition_uni} when we plug in the corresponding values of the Ising spins.
Hence, the Boltzmann weights can be expressed in terms of the Ising interaction energies, i.e 
\begin{align}
&w_1= 2 \rho e^{K+K'} \cosh (K_1+K_2+K_3+K_4) \notag \\
&w_2=2 \rho e^{-K-K'} \cosh (K_1-K_2+K_3-K_4) \notag \\
&w_3=2 \rho e^{-K+K'} \cosh (K_1-K_2-K_3+K_4) \notag \\
&w_4=2 \rho e^{K-K'} \cosh (K_1+K_2-K_3-K_4) \notag \\
&w_5=2 \rho   \cosh (K_1-K_2+K_3+K_4) \notag \\
&w_6=2 \rho  \cosh (K_1+K_2+K_3-K_4) \notag \\
&w_7=2 \rho  \cosh (K_1+K_2-K_3+K_4) \notag \\
&w_8=2 \rho  \cosh (-K_1+K_2+K_3+K_4) ,
\label{uni_wight}
\end{align}
which satisfy the free fermion condition~\cite{Choy1987,Lin1987}. 
We introduced an additional free parameter $\rho$. 
Hence, Eq.~(\ref{patition_uni})  is proportional to the partition function of the free fermion eight vertex model defined by Eq.~(\ref{uni_wight}). 

We now return to  the  square-octagon lattice. 
In particular, we are going to compute the spontaneous magnetization of the free fermion eight-vertex model generated by the quantum state with bonds $|\omega \rangle = |00\rangle \pm |11\rangle$.
The  corresponding Boltzmann weights  can be given by
\begin{align}
&w_1 = \frac{1}{2} ( a^8+4a^6+ 30a^4 + 52 a^2  + 41 )  \notag \\
&w_2 = \frac{1}{2}  (a^1-1)^4   \notag \\
&w_3 =  w_4 =  \frac{1}{2} ( a^2-1 )^3    ( a^2+3 )      \notag \\
&w_5 =w_6 =  w_7 =  w_8=  \frac{1}{2} (a^2-1)^2(a^4+2a^2+5).
\end{align} 
For the interaction energies of the Ising model, it can be  given
\begin{align}
&k = K_1=K_2=K_3=K_4 = \frac{1}{2}\ln \big(  \frac{a^2 -1}{a^2+3} \big)     \notag \\
& \kappa = K=K'=- \ln \big(  \frac{a^2 -1}{a^2+3} \big) \notag \\
&\rho = \frac{1}{2}(a^2-1)^3 (a^2+3)  
\end{align}

Thus, the interaction energies are physical, meaning real, for $a > 1$. 
As argued in~\cite{Choy1987}, the spontaneous magnetization of the free fermion eight vertex model is then given by the spontaneous magnetization of an eightfold connected Ising spin. 
In terms of the Boltzmann weights of the free fermion model, this means
\begin{align}
 M =
  \begin{cases}
    (1-\Omega ^{-2})^{1/8}      & \quad \text{if }   \Omega^2 >1\\
    0  & \quad  \text{ others},\\
  \end{cases}
\end{align}
where 
\begin{align}
\Omega^2 =  \frac{(w_1^2+w_2^2-w_3^2-w_4^2)^2-4(w_5w_6-w_7w_8)^2}{ 16 \times w_5 w_6 w_7 w_8 }.
\end{align}
 The Ising model undergoes a phase transition at $\Omega ^2 =1$ and $a_c = \Big(\sqrt{2(5+4\sqrt(2)} +\sqrt{2}+1 )    \Big)^{1/2} \approx 2.65158. $
 This is exactly the same transition point which we extracted from the free energy per site of the eight-vertex model.
 In~\cite{Wu_Lin1987}, they also provided a criterion which  can be used to characterize the ordered phase. We have to calculate the four quantities
\begin{align}
& E_1 = -(K+K' +|K_1 +K_2 +K_3 +K_4| )  \notag \\
&E_2 =  - ( -K-K' +|K_1 -K_2 +K_3 -K_4|)  \notag \\
&E_3 = - (-K+K' +|K_1 -K_2 -K_3 +K_4|) \notag \\
&E_4 =- (K-K- +|K_1 +K_2 -K_3 -K_4|).
\end{align}
The system is either in a ferromagnetic phase if $E_1 < E_2,E_3,E_4$, or in a antiferromagnetic phase if $E_2 < E_1,E_3,E_4$, or in a metamagnetic phase if $E_3 < E_1,E_2,E_4$ or $E_4 < E_1,E_2,E_3$.

From this, we find that for $1 < a < a_c$ the free fermion eight-vertex model is in a ferromagnetic phase.
However, from numerical results  of the deformed AKLT states we find that the the phase is ferromagnetic for $a > a_c$ instead. 
This seemingly contradiction can be explained through the Hadamard transformation used in reducing the 16 vertices to 8 vertices. 
As noted in~\cite{Hexagon_Niggemann}, this transformation maps the low temperature region of the quantum state to the high temperature region of the eight-vertex model and vice versa.

To obtain the correct temperature behavior of the magnetization in the classical model, we can employ the Kramers-Wannier duality. 
This duality maps the original Ising model to an Ising model on the dual lattice and the low-temperature regime of the original model to the high-temperature regime of the dual model and vice versa~\cite{Kramers1941}. 
The union jack lattice is dual to the square-octagon lattice~\cite{Baxter1988}. 
However, instead of working directly on the union jack lattice, we will use an alternative lattice.
In ~\cite{Baxter1986} , the free fermion model is also equivalent to the Ising model on the checkerboard lattice. 
This can be seen using the same trick, namely by performing the sum over every second row of spins in Fig.~\ref{fig:checker} and rewriting the partition function. 

Then drawing the dual lattice and performing the map shown in Fig.~\ref{fig:spin_vertex} gives the equivalent eight-vertex model. 
In terms of Ising interactions $\{J_i\}$ the free fermion Boltzmann weights are
\begin{align}
&w_1 =  2 \rho \cosh (J_1+J_2+J_3+J_4)  \notag \\
&w_2 =  2 \rho \cosh (J_1-J_2+J_3-J_4)  \notag \\
&w_3 =  2 \rho \cosh (J_1-J_2-J_3+J_4)  \notag \\
&w_4 = 2 \rho \cosh (J_1+J_2-J_3-J_4)  \notag \\
&w_5 = 2 \rho  e^{M+P}   \cosh (J_1-J_2+J_3+J_4)  \notag \\
& w_6 = 2 \rho  e^{-(M+P)}   \cosh (J_1+J_2+J_3-J_4)  \notag \\
& w_7 =   2 \rho e^{M-P}   \cosh (J_1+J_2-J_3+J_4)  \notag \\
&w_8=   2 \rho e^{-M+P}  \cosh (-J_1+J_2+J_3+J_4)  
\end{align} 
where $M$ and $P$ are free parameters~\cite{Baxter1986}.
The union jack Ising model with $K = K'$ and $K_1 = K_2 = K_3 = K_4$ is equivalent to a checkerboard Ising model with $J_1 = J_3$ ~\cite{Choy1987} .
The solutions of the Ising interactions in terms of the free fermion Boltzmann weights are known~\cite{Baxter1986}. 
 Furthermore it was shown that the sublattice spontaneous magnetization of the union jack Ising model is equivalent to the spontaneous magnetization of the checkerboard Ising model~\cite{Choy1987}.

The checkerboard lattice is self-dual and the Ising interactions of the dual model are given by $e^{-2J_i'}=\tanh (J_i)$.
The spontaneous magnetization of the dual model is
\begin{align}
 M =
  \begin{cases}
    (1-\Omega ^{-2})^{1/8}      & \quad \text{if }   \Omega^2 >1\\
    0  & \quad  \text{ others}.\\
  \end{cases}
\end{align}
The dual magnetization is plotted in  Fig.~\ref{fig:dual_M} and it can be regarded as the disorder parameter of the free fermion model~\cite{Baxter1988}.
Supported by numerical evidence provided  in  Fig.~\ref{fig:dual_M}, we conjecture that the spontaneous magnetization of dual Ising model corresponds to the ground state expectation value of the effective spin operator
\begin{align}
\tilde{S}_z= 
 \begin{pmatrix}
  1&0 &0&0\\
  0&1 &0&0\\
  0&0 &-1&0\\
  0&0 &0&-1\\
 \end{pmatrix}.
\end{align}
The transition points, which can be deduced by solving $\Omega^2 = 1$, are equal to the transition points which we obtained from the corresponding eight-vertex models.

 \end{document}